\input pipi.sty
\input epsf.sty
\input psfig.sty
\magnification1000
\def\ntightboxit{\relax}
\raggedbottom

\nopagenumbers
\rightline\timestamp
\rightline{FTUAM 04-14}
\rightline{hep-ph/0411334}

\bigskip
\hrule height .3mm
\vskip.6cm
\centerline{{\bigfib The pion-pion scattering amplitude}}
\medskip
\centerrule{.7cm}
\vskip1cm
\setbox8=\vbox{\hsize65mm {\noindent\fib J. R. Pel\'aez} 
\vskip .1cm
\noindent{\addressfont Departamento de F\'{\i}sica Te\'orica,~II\hb
 (M\'etodos Matem\'aticos),\hb
Facultad de Ciencias F\'{\i}sicas,\hb
Universidad Complutense de Madrid,\hb
E-28040, Madrid, Spain}}
\centerline{\box8}
\smallskip
\setbox7=\vbox{\hsize65mm \fib and} 
\centerline{\box7}
\smallskip
\setbox9=\vbox{\hsize65mm {\noindent\fib F. J. 
Yndur\'ain} 
\vskip .1cm
\noindent{\addressfont Departamento de F\'{\i}sica Te\'orica, C-XI\hb
 Universidad Aut\'onoma de Madrid,\hb
 Canto Blanco,\hb
E-28049, Madrid, Spain.}\hb}
\smallskip
\centerline{\box9}
\bigskip

\setbox0=\vbox{\abstracttype{Abstract}
We obtain reliable  $\pi\pi$
scattering amplitudes consistent with  experimental data, both at low and high energies,
and  fulfilling appropriate analyticity properties. 
We do this by first fitting experimental low energy ($s^{1/2}\leq1.42\,{\rm GeV}$) phase
shifts and inelasticities  with expressions that incorporate analyticity and unitarity.
In particular, for the  S wave with isospin~0, we discuss in detail several sets of 
experimental data. This provides  low energy partial wave amplitudes that 
summarize the known experimental information. Then, we  
 impose Regge behaviour as follows from factorization and experimental data 
for the imaginary parts of the scattering amplitudes 
at higher energy, and   check fulfillment of dispersion relations up to 0.925~GeV. This 
allows us  to improve  our fits. The ensuing  $\pi\pi$ scattering
amplitudes  are then shown to verify dispersion relations up to 1.42~GeV, as well as 
$s\,-\,t\,-\,u$ crossing sum rules and other consistency conditions. 
The improved parametrizations therefore provide a reliable representation of
pion-pion  amplitudes with which one can test chiral perturbation
theory calculations, pionium decays,  or use as input  for CP-violating $K$
decays.  In this respect, we find  $[a_0^{(0)}-a_0^{(2)}]^2=(0.077\pm0.008)\,M^{-2}_\pi$
 and $\delta_0^{(0)}(m^2_K)-\delta_0^{(2)}(m^2_K)=52.9\pm1.6^{\rm o}$.
}
\centerline{\box0}
\brochureendcover{Typeset with \physmatex}
\brochureb{\smallsc j. r. pel\'aez and f. j.  yndur\'ain}{\smallsc 
the pion-pion scattering amplitude}{1}

\brochuresection{1. Introduction}

\noindent
A precise and unbiased knowledge of the $\pi\pi$ scattering amplitude 
has become increasingly important in the last years. 
This is so, in particular, because  $\pi\pi$ scattering
 is one of the few places where one has more
observables than unknown constants in  a chiral perturbation theory (ch.p.t.) analysis 
to one loop, so it provides a test of ch.p.t. in this approximation, as well as 
a window to higher order terms. 
Beside this, an accurate determination of the S wave scattering lengths,
 and of the phase shifts, provide 
essential information for three subjects under intensive experimental investigation 
at present, viz., light scalar spectroscopy, pionic atom decays
and  CP violation in the kaonic system.

In two  recent papers, Ananthanarayan, Colangelo, Gasser and Leutwyler\ref{1} 
and Colangelo, Gasser and Leutwyler\ref{2}  have used 
experimental information, analyticity and unitarity (in the form of  
the Roy equations\ref{3}) and, in the second paper, also chiral perturbation theory, 
to construct a precise $\pi\pi$ 
scattering amplitude at low energy, 
$s^{1/2}\leq0.8\,\gev$.
Unfortunately, however, the analysis of refs.~1,~2 presents some weak points. 
First of all,  the input scattering amplitude at high energy
 ($s^{1/2}\gsim1.5\,\gev$)
which these authors use 
 is  incompatible
 with
experimental data\ref{4} on $\pi\pi$ cross sections and 
also contradicts known properties of standard Regge theory; see a 
detailed analysis in refs.~5,~6,~7 and Appendix~B here. 
Moreover,  the errors these authors take 
for some of their experimental input data are 
 optimistic, as shown in ref.~7. As a consequence  of all this 
the  $\pi\pi$ amplitudes of ref.~2 
do not satisfy well a number of consistency tests, as we show in  \sect~7 here 
(see also refs.~6,~7 for more details). 

Some of the shortcomings of the articles in refs.~1,~2, notably incorrect Regge
behaviour,  are also reproduced in
the  papers of Descotes et al., and Kami\'nski, Le\'sniak and Loiseau,\ref{8}
who also base their analysis in the Roy equations but, 
 since the errors given by these authors are substantially 
larger than those of ref.~2, their effects
 are now less pronounced. 
Therefore, we still need to find reliable pion-pion scattering 
amplitudes compatible with physical data both at high and low energy, as well as
to verify to what extent 
such  amplitudes agree with ch.p.t.

In the present paper we address ourselves to the first question; 
that is to say, we try to find 
what {\sl experiment} implies for the 
$\pi\pi$ amplitudes. 
To avoid biases, we will start by performing fits to experimental data
 on phase shifts and inelasticities, incorporating only   
 the highly safe requirements of analyticity and unitarity, in the low energy region 
$s^{1/2}\leq1.42\,\gev$.   
In particular, for the S0 wave below 0.95~\gev, where the 
experimental situation is confused, we perform a global fit, as well 
as individual fits to various sets of data. 
These fits are described in \sect~2 and, in the following Sections, we
investigate to which extent 
 the ensuing scattering
amplitudes  are consistent, in particular with 
high energy information. To do so, we 
 assume  Regge behaviour, as given in ref.~5 (slightly improved; see 
Appendix~B), above 1.42~\gev: using this
we check in \sect~3  fulfillment of
 forward dispersion relations, 
 for the three independent $\pi\pi$ scattering amplitudes.
This, in particular, permits us to discriminate among the 
various sets of phase shifts for the S0 wave, leaving only 
a few solutions which are consistent with dispersion relations (and, as it turns out, 
very similar one to the other, as discussed in \sect~4).

When dealing with different data sets one has to
weight not only the data on a single experiment but one has to take into account 
the 
reliability of the experiments themselves. So we have done
for many waves, where some clearly faulty experimental 
data have only been considered only to conservatively enlarge the uncertainties.
Concerning the most controversial S0 wave, we have used the very reliable 
data coming from  $K_{l4}$
and $K\rightarrow\pi\pi$ decays; to this we add 
 the results from other experimental analyses of $\pi\pi$ scattering
available in the literature. We then use  forward dispersion
to test consistency of the several sets of data.

The present study should therefore be considered, in particular, 
 as a guideline to the consistency (especially  
with forward dispersion relations) of the various
 data sets.
 
Next, we use these dispersion relations to improve the central values of 
the parameters of the fits given in \sect~2. 
The result of such analysis  (\sect~4) is that one can get a precise 
 description for all 
waves, consistent with forward 
dispersion relations up to $s^{1/2}\sim0.95\,\gev$ and a bit less so  
($\lsim1.5\,\sigma$ level) in the 
whole energy range, $2M_\pi\leq s^{1/2}\leq1.42\,\gev$, and 
even below threshold, down to $s^{1/2}=\sqrt{2}M_\pi$. 
  The greater uncertainties affect the S0 
wave for $s^{1/2}>0.95\,\gev$, a not unexpected feature, 
and, to a lesser extent, the P wave 
above 1.15~\gev.  

In \sect~5 we verify that the scattering amplitudes 
we have obtained,  which were  shown to   
satisfy $s\,-\,u$ crossing (by checking the dispersion relations), 
also verify $s\,-\,t$ crossing, in that they satisfy two typical 
crossing sum rules. In \sect~6 we use the scattering
amplitudes we have determined  and the method of the Froissart--Gribov
representation to calculate  a number of low energy
 parameters for P, D and some higher waves which, 
in particular, provides further consistency tests. 
We also evaluate, in \sect~7, the important quantities 
$[a_0^{(0)}-a_0^{(2)}]^2$ and  $\delta_0^{(0)}(m^2_K)-\delta_0^{(2)}(m^2_K)$
for which we find
$$[a_0^{(0)}-a_0^{(2)}]^2=(0.077\pm0.008)\,M^{-2}_\pi,\quad
\delta_0^{(0)}(m^2_K)-\delta_0^{(2)}(m^2_K)=52.9\pm1.6^{\rm o}.
$$ 
Also in \sect~7 we compare our results with those obtained by
 other authors using Roy equations 
and ch.p.t. 
However, in the present paper we will not address ourselves to the 
question of the chiral perturbation theory analysis of our $\pi\pi$ amplitudes.

Our paper is finished in \sect~8 with a brief Summary, as well as 
with a few Appendices. 
In Appendix~A, we collect the formulas obtained 
with our best fits. In Appendix~B we 
 give a brief discussion of the Regge formulas used; 
in particular, we present an improved evaluation of the parameters for rho exchange. 
Appendix~C is devoted to a discussion of the shortcomings of 
experimental
phase shift analyses above $\sim1.4\,\gev$, which justifies our preference for 
using Regge formulas in this energy region.

We end this Introduction with a few words on notation and normalization conventions. 
We will here denote amplitudes with a fixed value of isospin, say $I$, in channel 
$s$, simply by $F^{(I)}$, $f_l^{(I)}$; we will specify the channel,  $F^{(I_s)}$, 
when there is danger of confusion. 
For amplitudes with fixed isospin in channel $t$, we write explicitly  $F^{(I_t)}$.

For scattering amplitudes with well defined isospin in channel $s$, $I_s$, we write
$$\eqalign{
F^{(I_s)}(s,t)=&\,2\times\sum_{l={\rm even}}(2l+1)P_l(\cos\theta)f_l^{(I_s)}(s),
\quad I_s=\hbox{even},\cr
F^{(I_s)}(s,t)=&\,2\times\sum_{l={\rm odd}}(2l+1)P_l(\cos\theta)f_l^{(I_s)}(s),
\quad I_s=\hbox{odd,}\cr
f_l^{(I)}(s)=&\,\dfrac{2s^{1/2}}{\pi k}\hat{f}_l^{(I)},\quad
 \hat{f}_l^{(I)}=\sin\delta_l^{(I)}(s)\ee^{\ii\delta_l^{(I)}(s)}.\cr 
\cr}
\equn{(1.1a)}$$
The last formula is only valid 
when only the elastic channel is open. 
When inelastic channels  open this equation is no more valid, 
but one can still write
$$\hat{f}_l(s)=
\left[\dfrac{\eta_l\,\ee^{2\ii\delta_l}-1}{2\ii}\right].
\equn{(1.1b)}$$ 
The factor 2 in the first formulas in (1.1a) is due 
 to Bose statistics. Because of this, 
even waves only exist with isospin $I=0,\,2$ and odd waves
must  necessarily have isospin $I=1$. 
For this reason, we will often omit the 
isospin index for odd waves, 
writing e.g. $f_1$, $f_3$ instead of $f_1^{(1)}$, 
$f_3^{(1)}$.
Another  convenient simplification that we  use here 
is to denote the $\pi\pi$ partial waves by S0, S2, P, D0, D2, F, etc., 
in self-explanatory notation.

The quantity $\eta_l$, called the {\sl inelasticity parameter} for wave $l$, 
 is  positive and smaller than or equal to unity. 
The elastic and inelastic cross sections, for a given wave, are given in terms of 
$\delta_l$ and $\eta_l$ by
$$\sigma^{\rm el.}_l=\tfrac{1}{2}\left\{\dfrac{1+\eta_l^2}{2}-
\eta\cos2\delta_l\right\},\quad
\sigma^{\rm inel.}_l=\dfrac{1-\eta_l^2}{4};
\equn{(1.2)}$$
$\sigma^{\rm el.}_l,\;\sigma^{\rm inel.}_l$ are defined so that, 
for collision of particles $A$, $B$ (assumed distinguishable), 
$$\sigma_{\rm tot.}=\dfrac{4\pi^2}{\lambda^{1/2}(s,m_A,m_B)}\,
\dfrac{2s^{1/2}}{\pi k}\sum_l(2l+1)
\left[\sigma^{\rm el.}_l+\sigma^{\rm inel.}_l\right].
\equn{(1.3)}$$

When neglecting isospin violations (which we  do unless 
explicitly stated otherwise) we will take the 
 convention of approximating the pion mass by 
$M_\pi=m_{\pi^\pm}\simeq139.57\,\mev$.  
We  also define  scattering lengths, $a_l^{(I)}$, and effective range parameters,
 $b_l^{(I)}$, by
$$\dfrac{\pi}{4 M_\pi k^{2l}}\real f_l(s)\simeqsub_{k\to0}
a_l^{(I)}+b_l^{(I)} k^2+\cdots\,,\quad k=\sqrt{s/4-M^2_\pi}.
\equn{(1.4)}$$

\booksection{2. The $\pi\pi$ scattering amplitudes at low energy ($s^{1/2}\leq1.42\,\gev$)}

\noindent
We present in this  section
parametrizations of the S, P, D,  F and G waves in $\pi\pi$ scattering 
obtained fitting
experimental data at low energy, $s^{1/2}\leq1.42\,\gev$, 
which will provide a representation of the 
{\sl experimental} $\pi\pi$ scattering amplitudes in this energy
range: this may be considered as an {\sl energy-dependent} phase 
shift analysis.  Above 1.42~\gev, we will use the Regge
expressions obtained in ref.~5,  which we reproduce and discuss in Appendix~B in the present paper.
In following Sections we will 
verify to which extent the scattering amplitudes that one finds in this way 
are consistent with dispersion relations or crossing symmetry, 
within the errors given.

Before entering into the actual fits, a few words are due on the choice of the 
energy 1.42~\gev\ as limit between the 
regions where we use phase shift analyses and Regge representations. 
Experimental phase shift analyses exist up to $\sim2~\gev$. 
However, as is known, phase shift analyses become ambiguous as soon as 
{\sl inelastic} processes become important. 
As we will discuss in some detail in Appendix~C, the existing experimental
 phase shift analyses become suspect for energies above $\sim1.4$~\gev:
 in particular, they produce cross sections that deviate from experimentally measured
total cross sections. 

As for the higher energy range, we have shown in ref.~5 that a Regge description of 
the imaginary part of the $\pi\pi$ scattering amplitudes 
agrees well with $\pi\pi$ cross section data (and, through factorization, also 
with $\pi N$ and $NN$ data) 
for kinetic energies above $\sim1\,\gev$; 
for $\pi\pi$ scattering,  
down to $s^{1/2}\simeq1.4$, or even to $1.3$~\gev, depending on the process.
We have chosen the limiting energy to be 1.42~\gev\ as a 
reasonable compromise; but one could have taken  
lower limits, or slightly higher ones, as well. 
In fact, it is not possible to go below 1.4~\gev\ with a Regge description for 
the $I_t=1$ amplitude or for $\pi^0\pi^0$ scattering, 
since both contain the isospin~zero amplitudes, which show two resonances 
just below that region [the $f_2(1270)$ and the $f_0(1370)$]. 
But it is possible to choose a lower junction point between 
 the phase shift analyses and the Regge formulas 
for $\pi^0\pi^+$ scattering. 
The influence of this is negligible for $\pi^0\pi^+$, as we will show in \subsect~3.1.2, 
provided one stays in the range 
$1.32\,\gev\leq s^{1/2}\leq 1.42\,\gev$, with the larger values favoured.

We now turn to the parametrizations. In writing them we will  use the
requirements of  analyticity and elastic unitarity. 
Extra information has been added to help stabilize the fits
in those channels where the low energy data is scarce; in particular,
information 
on Adler zeros or scattering lengths. For the latter we impose values obtained
from their Froissart-Gribov representation, but with very
large error bars to cover the different values available in the literature.
The effect on this of the high energy representation is negligible.
 
The method used to take into account unitarity and analyticity will be 
the standard one of the effective range formalism, 
supplemented by a conformal expansion. 
To be precise, for a given partial wave 
$f_l^{(I)}(s)$ we write,
 for any value (complex or real) of $s$,
 $$f_l^{(I)}(s)=\dfrac{2s^{1/2}}{\pi k}\dfrac{1}{2s^{1/2}k^{-2l-1}\phiv_l^{(I)}(s)-\ii}
=\dfrac{k^{2l}}{\pi}\dfrac{1}{\phiv_l^{(I)}(s)-\ii k^{2l+1}/2s^{1/2}} .
\equn{(2.1)}$$
The effective range function $\phiv_l^{(I)}(s)$ is real on the segment $0\leq s\leq s_0$, but it
 will be {\sl complex} above  $s_0$, and also for $s\leq0$. 
Here $s_0$ is the energy squared above which inelastic processes are nonnegligible. 
Using only the 
 requirements of causality and conservation of probability, 
it can be shown that $\phiv_l^{(I)}(s)$ is
analytic in the complex
$s$-plane, cut from
$-\infty$ to 0 and from $s_0$ to $+\infty$; 
see \fig~1.

We can profit from these analytical properties as follows. 
We map the cut plane into the unit disk (as in \fig~1) by 
means of the conformal transformation
$$w(s)=
\dfrac{\sqrt{s}-\sqrt{s_0-s}}{\sqrt{s}+\sqrt{s_0-s}}.
$$
The properties of analyticity and elastic unitarity of 
$f_l^{(I)}(s)$ are then strictly equivalent 
to uniform convergence (in the variable $w$) in the disk $|w|<1$ of the series 
$$\phiv_l^{(I)}(s)=\sum_{n=0}^\infty B_n w(s)^n.
$$
This series thus converges, in the variable $s$, uniformly in the interior of 
the whole cut plane; 
in general very quickly. It will turn
out that  we will need only two or three terms in the expansions for the partial wave
amplitudes.
 
\topinsert{
\bigskip
\setbox0=\vbox{{\psfig{figure=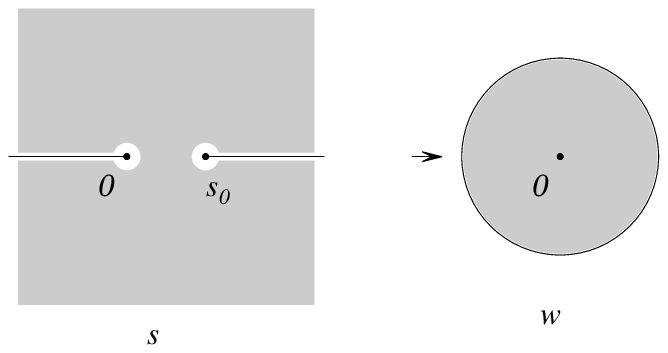,width=9.truecm,angle=-0}}} 
\setbox6=\vbox{\hsize 6truecm\captiontype\figurasc{Figure 1. }{\hb The  
mapping $s\to w$.\hb
\phantom{XX}}\hb
\vskip.1cm} 
\medskip
\line{\tightboxit{\box0}\hfil\box6}
\medskip
}\endinsert

Before starting with the actual fits, 
we here say a few words on the values of $s_0$ (i.e., the energy at 
which we consider inelasticity  not negligible) that we will take for the 
various waves. 
For the S0 wave, the $\bar{K}K$ channel is strongly coupled, so 
here $s_0=4m^2_K$. 
For other waves the $\bar{K}K$ channel is weakly coupled. 
In the cases where we have sufficiently precise data
 (that is, for the P wave and the low energy S2 wave),  
we will take  $s^{1/2}_0=1.05\,\gev$ which is, approximately, the $2\pi\rho$ threshold. 

For the D0 wave, and for the S2 wave at intermediate  energy, 
$1.05\,\gev\lsim s^{1/2}\lsim1.4\,\gev$,  
inelasticity is detected. 
It is small (but not negligible) while, unfortunately, the data are not 
precise enough to perform a fully consistent analysis.
 We then follow a
different strategy:  we fit the experimental phase shifts using formulas 
that  neglect
inelasticity below 1.45
\gev\
  (which is, approximately, the $2\rho$ threshold). 
The experimental inelasticity between 1 \gev\ and 1.42 \gev\ is then added by hand.

Finally, for the D2, F waves, for which experiments detect no inelasticity below 
1.42 \gev, $s_0$ is taken equal to 1.45~\gev.

\bigskip
\goodbreak
\booksubsection{2.1. The P  wave}
\vskip-0.4truecm
\booksubsection{2.1.1. Parametrization of the P wave below 1 \gev}

\noindent
We will consider first the P wave for $\pi\pi$ scattering below 1 \gev, 
to exemplify the methods, because it is the one for which more precise 
results are obtained. 
We start thus considering the region of energies 
 where the inelasticity is below the  2\% level; say,  
$s_0^{1/2}\leq 1.05\,\gev^2$. 

We then expand $\phiv_1(s)$ in powers of $w$, and, reexpressing 
$w$ in terms of $s$, the expansion will be convergent over all the cut $s$-plane. 
Actually, and because we know that the P wave resonates at $s=M^2_\rho$, 
it is more convenient to expand not  $\phiv_1(s)$ 
itself, but ${\psi}(s)$ given by
$$\phiv_1(s)=(s-M^2_\rho){\psi}(s)/4;
\equn{(2.2a)}$$
so we write
$${\psi}(s)=\left\{B_0+B_1w+
\cdots\right\};\quad.
\equn{(2.2b)}$$

In terms of  $\phiv_1(s)$ we thus find the expression for the cotangent of the phase shift, 
 keeping two terms in the expansion,
$$\cot\delta_1(s)=\dfrac{s^{1/2}}{2k^3}
(M^2_\rho-s)\left\{B_0+B_1\dfrac{\sqrt{s}-\sqrt{s_0-s}}{\sqrt{s}+\sqrt{s_0-s}}
\right\};\quad s_0^{1/2}=1.05\;\gev.
\equn{(2.3)}$$
$M_\rho,\,B_0,\,B_1$ are free parameters to be fitted to experiment.
In terms of $\phiv_1,\,\psi$ we have, for the rho width, and the scattering length, $a_1$,
$$\eqalign{
\gammav_\rho=&\,\dfrac{2k^3_\rho}{M^2_\rho{\psi}(M^2_\rho)}, 
\quad
k_\rho=\tfrac{1}{2}\sqrt{M^2_\rho-4M^2_\pi},\cr
a_1=&\,\dfrac{-1}{4M_\pi\phiv_1(4M_{\pi}^2)}=
\dfrac{1}{M_\pi(M^2_\rho-4M^2_\pi)\psi(4M^2_\pi)}.
\cr}
\equn{(2.4)}$$

The values $B_0=\hbox{const.}$, $B_{i\geq 1}=0$ would 
correspond to a perfect Breit--Wigner. 
Actually, it is known that the $\rho$ deviates from a pure Breit--Wigner 
and for a precision parametrization 
 two terms, $B_0$ and $B_1$, have to be kept in (2.3). 
Note that the parametrization holds not only on the physical region 
$4M_{\pi}^2\leq s\leq s_0$, but on the unphysical region $0\leq s\leq 4M_{\pi}^2$ and 
also over the 
whole region of the complex $s$ plane with $\imag s\neq 0$.
The parametrization given now is the one that has less biases, in the sense 
that no model has been used: we have imposed only the highly safe requirements of 
analyticity and unitarity, which follow from  
causality and conservation of probability.

The best values for our parameters are actually obtained not 
from fits to $\pi\pi$ scattering data, but from fits to the 
pion form factor.\ref{9} 
This is obtained from $e\pi$ scattering, from $e^+e^-\to\pi^+\pi^-$ and from 
$\tau\to\nu\pi\pi$ decay, where we have a large number of
precise data, and pions  are on their mass shells. 
Including systematic experimental errors in the fits, and including
 in the fit also the constraint 
 $a_1=(38\pm3)\times10^{-3}\,M_{\pi}^{-3}$ for the scattering length, we have
$$\matrix{
{\rm from}\;\pi^+\pi^-:\;&B_0=1.074\pm0.006,\quad B_1=0.13\pm0.04;\quad
 M_\rho=774.0\pm0.4\,\mev
\cr
{\rm from}\;\pi^+\pi^0:\;&B_0=1.064\pm0.006,\quad B_1=0.13\pm0.04;\quad
 M_\rho=773.1\pm0.6\,\mev.\cr
}
\equn{(2.5a)}$$
The fit is excellent; one has $\chidof=245/(244-13)$, and the slight 
excess over unity is due to the well-known incompatibility of  OPAL and other 
data for $\tau$ decay at low invariant mass.

The corresponding values 
 for the rho width, and P wave scattering length and effective range parameter, are
$$\eqalign{
a_1=&\,(37.8\pm0.8)\times10^{-3}M_{\pi}^{-3},
\quad b_1=(4.74\pm0.09)\times10^{-3}M_{\pi}^{-5},\quad
\gammav_{\rho^0}=146.0\pm0.8;\quad{\rm from}\;\pi^+\pi^-\cr
a_1=&\,(37.8\pm0.8)\times10^{-3}M_{\pi}^{-3},
\quad b_1=(4.78\pm0.09)\times10^{-3}M_{\pi}^{-5},\quad
\gammav_{\rho^+}=147.7\pm0.7;\quad{\rm from}\;\pi^+\pi^0.\cr
}
\equn{(2.5b)}$$
Although the values of the experimental $\pi\pi$ 
phase shifts were {\sl not} included in the fit, the 
phase shifts that (2.5a) implies are in very good agreement with them, 
as shown in \fig~2.

\topinsert{
\setbox0=\vbox{{\psfig{figure=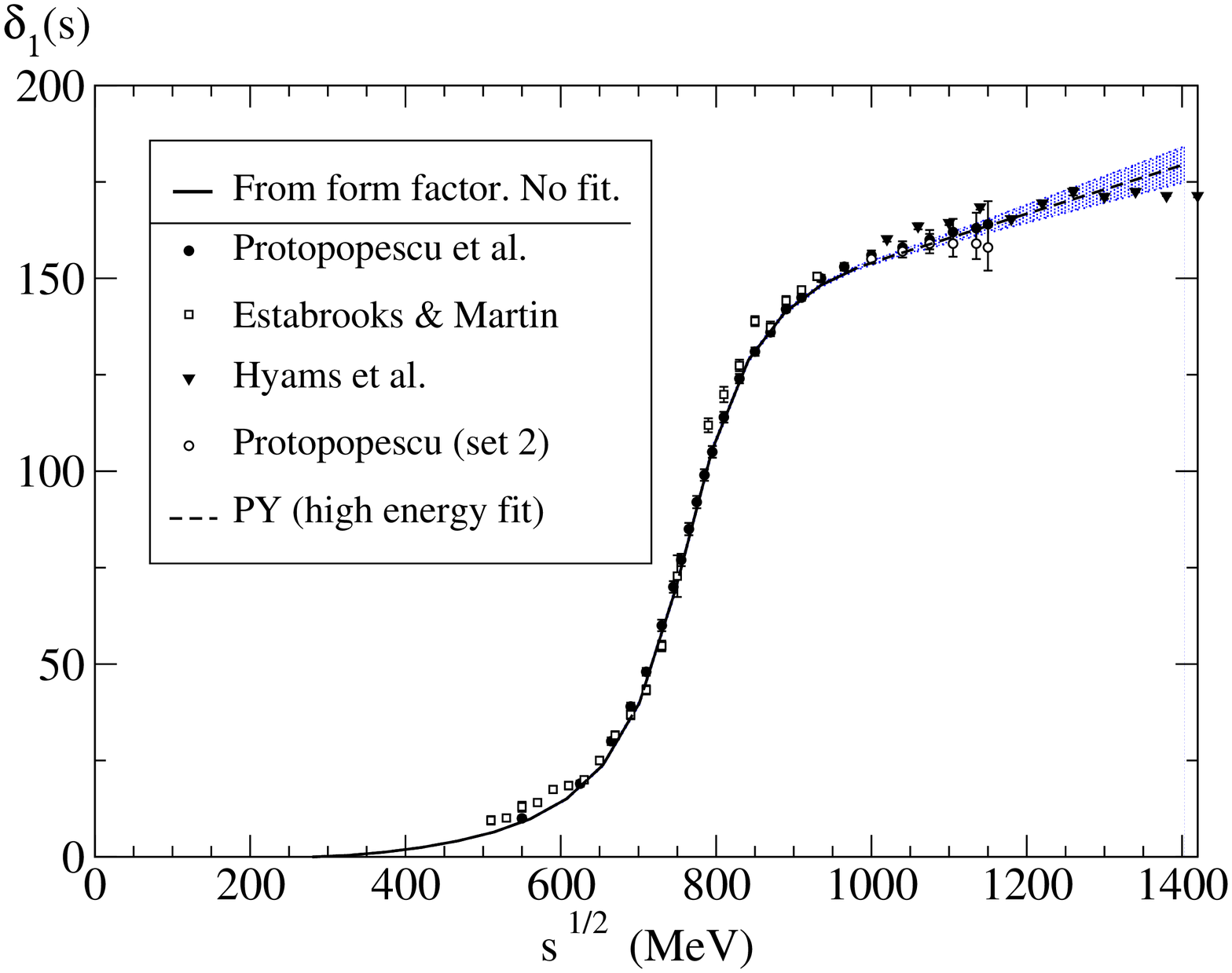,width=11.2truecm,angle=-90}}}
\setbox6=\vbox{\hsize 15truecm\captiontype\figurasc{Figure 2. }{The  
phase shifts of solution 1 from Protopopescu et al.,\ref{10} 
Hyams et al.\ref{11a} and 
Estabrooks and Martin\ref{11a} compared with the prediction with 
the parameters (2.6)   (solid line below 1~\gev). 
We emphasize that this is {\sl not} a fit 
to these experimental data, but is obtained 
from the pion form factor.\ref{9} The error here is like the thickness of the line.
\hb
\quad Above 1~\gev, the dotted line and error (PY) are as 
follows from the fit (2.7).}
} 
\medskip
\centerline{\tightboxit{\box0}}
\bigskip
\centerline{\box6}
\bigskip
}\endinsert

If  neglecting  violations of isospin invariance, 
 one would expect numbers equal to the 
average of both sets. If we also increase the error, 
so as to take into account that due to isospin breaking, by adding linearly 
half the difference between both determinations (2.5a), we find 
$$\eqalign{
B_0=&\,1.069\pm0.011,\quad B_1=0.13\pm0.05,\quad M_{\rho}=773.6\pm0.9,\cr 
a_1=&\,(37.6\pm1.1)\times10^{-3}M_{\pi}^{-3},\quad
  b_1=(4.73\pm0.26)\times10^{-3}M_{\pi}^{-5};
}
\equn{(2.6)}$$
the corresponding parametrization 
 we take to be valid up to $s^{1/2}=1\,\gev$.
\vskip0.5truecm

\booksubsection{2.1.2. The P wave for $1\gev\leq s^{1/2}\leq 1.42\gev$}

\noindent
In the range $1\,\gev\leq s^{1/2}\leq 1.42\,\gev$ one is sufficiently
 far away from thresholds to neglect 
their influence (the coupling to $\bar{K}K$ is negligible). 
A purely empirical parametrization that agrees with the data in 
 Protopopescu et al.\ref{10} and Hyams et al.,\ref{11a} 
 up to 
1.42 \gev, within errors, is obtained   with a linear fit to the phase and inelasticity:
$$\eqalign{
\delta_1(s)=&\,\lambda_0+\lambda_1(\sqrt{s/\hat{s}}-1),\quad\eta_1(s)=
1-\epsilon(\sqrt{s/\hat{s}}-1);\cr
\epsilon=&\,0.30\pm0.15,\quad \lambda_0=2.69\pm0.01,\quad\lambda_1=1.1\pm0.2.\cr }
\equn{(2.7)}$$
Here $\hat{s}=1\,{\gev}^2$.
The value of $\lambda_0$ ensures the 
agreement of the phase with the value given in the previous Subsection at
 $s=\hat{s}\equiv1\,\gev^2$.
 This fit is good (Fig.~2). 

It should be remarked, however, that there are other solutions for the P wave  
above $1.15\,\gev$. 
In the first analysis of  Hyams et al.,\ref{11a} that we have followed here, 
a resonance  occurs with a mass 
 $\sim1.6\,\gev$,
 and its effect is 
only felt above $\sim1.4\,\gev$; 
but in the second analysis by the same group,  Hyams et al.,\ref{11b} 
a broad, highly 
inelastic resonance (actually, more a spike than a resonance) 
with a mass $\sim1.35\,\gev$ appears, 
 or does
not appear,  depending on the solution chosen. 
Finally, the Particle Data Group (based mostly on evidence from 
$e^+e^-$ annihilation and $\tau$ decay) report a resonance at 1.45~\gev:
one has to admit that the situation for the P 
wave above $1.15\,\gev$ is not clear. We will return to this later.

\booksubsection{2.2. The S waves}
\vskip-0.5truecm
\booksubsection{2.2.1. Parametrization of the S wave for  $I=2$}

\noindent
We consider three sets of experimental data.
 The first 
 corresponds to solution A in the paper by 
Hoogland et al.,\ref{12} who use the reaction $\pi^+ p\to\pi^+\pi^+n$.
We will not consider the so-called solution B in this paper; 
while it produces results similar to the other, its errors 
are clearly underestimated.  
The second set corresponds 
 to  the work of Losty
et al.,\ref{12} who analyze instead  $\pi^- p\to\pi^-\pi^-\Deltav$. 
The third set are the data of Cohen et al.,\ref{4} who consider  $\pi^- n\to\pi^-\pi^-p$. 
 These  three sets represent a substantial improvement over 
other determinations; since they produce two like charge pions, only isospin 2 
contributes, and one gets rid of the large  S0 wave  contamination. 
However, they still present the problem that 
one does not have scattering of real pions.

\topinsert{
\setbox0=\vbox{{\psfig{figure=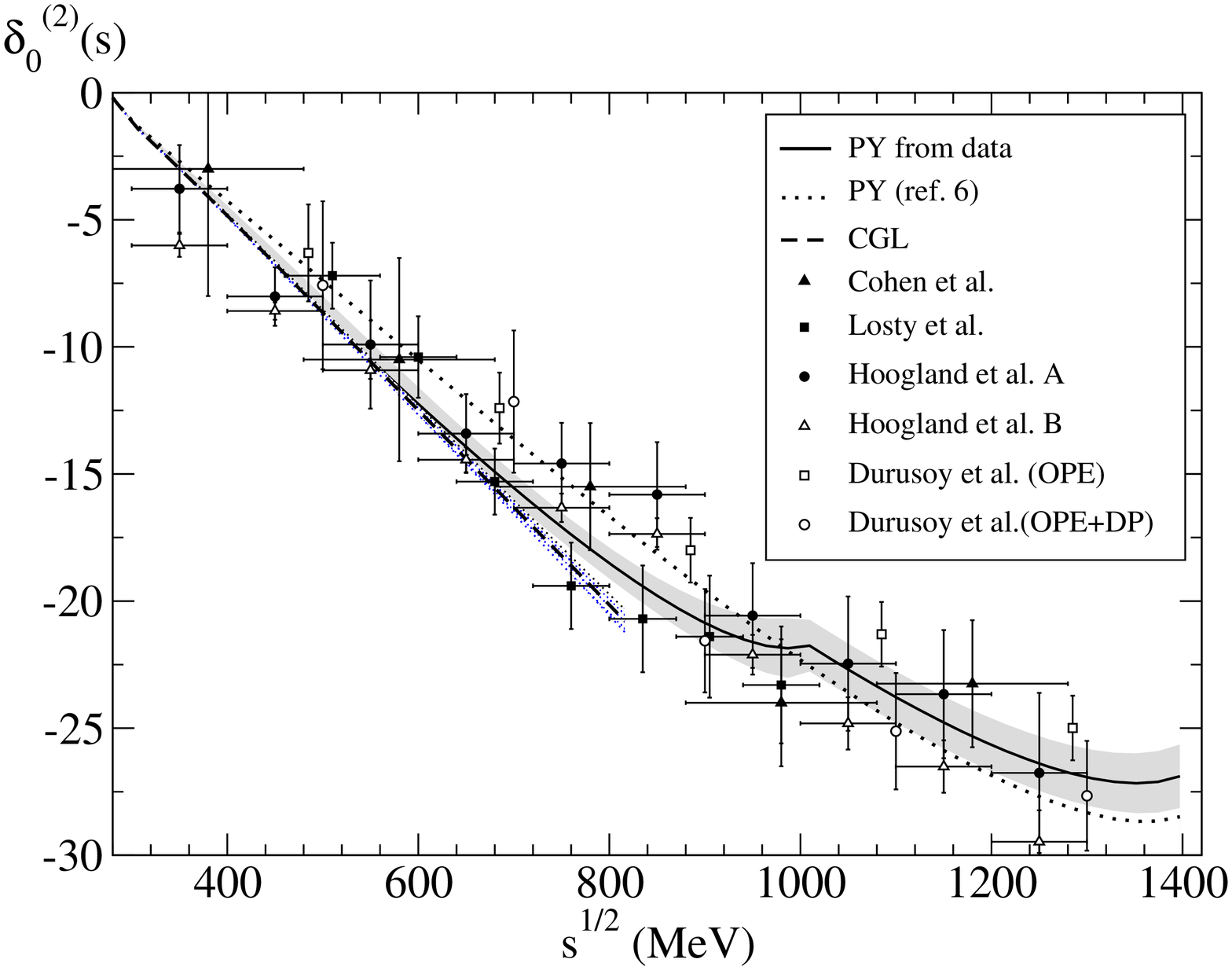,width=11.2truecm,angle=-90}}} 
\setbox6=\vbox{\hsize 15truecm\captiontype\figurasc{Figure 3. }{
Continuous line: The  
$I=2$, $S$-wave phase shifts, and error bands, corresponding to (2.7), 
(2.8) and (2.9), 
denoted by PY. 
Dotted line: the fit from ref.~6. 
The experimental points are from Losty et al. (black squares), Hoogland et al.,\ref{12}
 solution A 
(black dots), and Cohen et al.\ref{4} (black triangles). 
We have also included data from Durusoy et al. and from Solution~B of 
Hoogland et al., although they were not used in the fits.
The dashed line, which lies below 
our fit, is the S2 phase of Colangelo, Gasser and
Leutwyler,\ref{2}  with error band attached. 
}\hb} 
\centerline{\tightboxit{\box0}}
\bigskip
\centerline{\box6}
\medskip
}\endinsert

For isospin 2, there is no low energy resonance, but $f_0^{(2)}(s)$
 presents the feature that a zero 
is expected (and, indeed, confirmed by the fits) 
in the region $0<s<4M_{\pi}^2$. 
This zero of $f_0^{(2)}(s)$ is related to the so-called Adler zeros 
and, to lowest order in chiral perturbation theory,  
occurs at $s=2z_2^2$ with $z_2=M_\pi$. 
We note that, unlike the corresponding zero for the S0 wave, 
$2z_2^2$ is inside the region where the conformal expansion is expected to converge.

In ref.~6, we  neglected inelasticity below 1.45~\gev, and fitted  
 all experimental data, for $s^{1/2}\leq1380\,\mev$. 
A more precise determination (and, above all, with more 
realistic error estimates)
 is obtained if we realize that 
 inelasticity is detected 
 experimentally 
 above the $2\pi\rho$ threshold, $\hat{s}^{1/2}=1.05\,\gev$, 
so we should fit separately the low and high energy regions (\fig~3).

In the low energy region,  we  fix $z_2=M_\pi$ and fit only the low energy data, 
$s^{1/2}<1.0\,\gev$; later, in \sect~4, we will allow $z_2$ to vary. We 
write
$$\cot\delta_0^{(2)}(s)=\dfrac{s^{1/2}}{2k}\,\dfrac{M_{\pi}^2}{s-2z_2^2}\,
\left\{B_0+B_1\dfrac{\sqrt{s}-\sqrt{\hat{s}-s}}{\sqrt{s}+\sqrt{\hat{s}-s}}\right\},
\quad z_2\equiv M_\pi;\quad \hat{s}^{1/2}=1.05\;\gev.
\equn{(2.8a)}$$
Then we get $\chidof=13.0/(25-2)$ and  
$$B_0=-80.4\pm2.8,\quad B_1=-73.6\pm12.6;\quad a_0^{(2)}=(-0.052\pm0.012)\,M_{\pi}^{-1};\quad
b_0^{(2)}=(-0.085\pm0.011)\,M_{\pi}^{-3}.\equn{(2.8b)}$$

For the high energy region we  neglect the inelasticity below $1.45$ \gev,  
and then add inelasticity by hand. 
We consider two extreme possibilities: fitting the whole range, 
or fitting only high energy data ($s^{1/2}\geq0.91\,\gev$), 
requiring agreement of the 
central value with the low energy determination at $s^{1/2}=1\,\gev$.  
We accept as the best result an average of the two, and thus have
$$\eqalign{
\cot\delta_0^{(2)}(s)=&\,\dfrac{s^{1/2}}{2k}\,\dfrac{M_{\pi}^2}{s-2M^2_\pi}\,
\left\{B_0+B_1\dfrac{\sqrt{s}-\sqrt{s_0-s}}{\sqrt{s}+\sqrt{s_0-s}}\right\};\cr
s_0^{1/2}=&\,1.45\;\gev;\quad B_0=-123\pm6,\quad B_1=-118\pm14,\cr
}
\equn{(2.9)}$$
and we have enlarged the errors to cover  both 
extreme cases. We will not 
consider varying the position of the Adler zero for this high energy piece.

The inelasticity  may be described by the 
empirical fit
$$\eta_0^{(2)}(s)=\cases{
1-\epsilon(1-\hat{s}/s)^{3/2},\quad 
\epsilon=0.18\pm0.12;\quad s>\hat{s}=(1.05\;\gev)^2;\cr
1,\qquad s<\hat{s}=(1.05\;\gev)^2.\cr}
\equn{(2.10)}$$
These formulas are expected to hold 
from $s^{1/2}=1.0\,\gev$ up to $1.42$ \gev.
As shown in \fig~3, 
both determinations, (2.8-9) and that in ref.~6 are almost overlapping (their error bands 
actually overlap).

\booksubsection{2.2.2. Parametrization of the S wave for  $I=0$ 
below  $0.95\,\gev$ (global fit)}

\noindent
The S wave with isospin zero is  difficult to deal with.
 Here we have a very broad 
enhancement, variously denoted as $\epsilon,\,\sigma,\,
f_0$, around
$s^{1/2}= {\mu}_0\sim800\,\mev$. 
We will not discuss here whether
 this enhancement is a {\sl bona fides} 
resonance; we merely remark that in all experimental
 phase shift analyses $\delta_0^{(0)}(s)$ crosses 90\degrees\ 
somewhere between 600 and 900 \mev; we {\sl define}  ${\mu}_0$ as the 
energy at which the phase equals 90\degrees.
Moreover, we expect  a zero of $f_0^{(0)}(s)$ (Adler zero), hence a pole of 
the effective range function 
$\phiv_0^{(0)}(s)$, for $s=\tfrac{1}{2}z_0^2$ with $\tfrac{1}{2}z_0^2$
 in the region $0<
s<4M_{\pi}^2$.  Chiral perturbation theory suggests that 
$z_0\simeq M_\pi$.
 
We can distinguish two energy regions: below $s^{1/2}_0=2m_K$ we are under 
the $\bar{K}K$ threshold. 
Between $s^{1/2}_0$ and $s^{1/2}\sim1.2$ there is a nonegligible
 coupling between the 
 $\bar{K}K$ and $\pi\pi$ channels and the analysis becomes very unstable, because there 
is little information on the process $\pi\pi\to\bar{K}K$ and even less on 
$\bar{K}K\to\bar{K}K$. 
We will  present later  an empirical fit in the 
region of energies around and above 1 \gev, and 
 we will now concentrate in the low energy region.

For the theoretical formulas we 
  impose the Adler zero at $s=\tfrac{1}{2}M_{\pi}^2$
 (no attempt is made to vary this for the moment; 
see \sect~4),
 and a zero of $\cot\delta_0^{(0)}(s)$ at
$s={\mu}^2_0$, ${\mu}_0$ a free parameter.
 Then we map the $s$ plane, cut along the left hand
cut ($s\leq0$) and  the $\bar{K}K$ cut, writing
$$\cot\delta_0^{(0)}(s)=\dfrac{s^{1/2}}{2k}\,
\dfrac{M_{\pi}^2}{s-\tfrac{1}{2}z_0^2}\,
\dfrac{{\mu}^2_0-s}{{\mu}^2_0}\Big\{B_0+B_1w(s)+\cdots\Big\},
$$
and
$$
w(s)=\dfrac{\sqrt{s}-\sqrt{s_0-s}}{\sqrt{s}+\sqrt{s_0-s}},\quad
s_0=4m^2_K
$$ 
(we have taken $m_K=0.496\,{\gev}$).

This parametrization does not represent fully the 
coupling of the $\bar{K}K$ channel and we will thus only 
take it to be valid up to  $s^{1/2}=0.95\,\gev$.

On the experimental side the situation is still a bit confused,
  although it has cleared up
substantially in the last years thanks to the experimental information 
on  $K_{l4}$ and   $K_{2\pi}$ decays. The 
 information we have on this S0 wave is of
three kinds:  from phase shift analysis in collisions\ref{10,11a}
 $\pi p\to\pi\pi N,\Deltav$; 
from the decay\ref{13} $K_{l4}$; and  from  $K_{2\pi}$ decays.
The last gives the value of the combination 
$\delta_0^{(0)}-\delta_0^{(2)}$ at $s^{1/2}=m_K$; 
the decay $K_{l4}$ gives $\delta_0^{(0)}-\delta_1$ at low energies,
 $s^{1/2}\lsim380\;\mev$. 
If using  recent $K_{2\pi}$ information,\ref{14} 
combined with older determinations, and
 with  the $I=2$ phase obtained in the previous subsection,  
one finds the phase
$$\delta_0^{(0)}(m^2_K)=43.3\pm3\degrees.
\equn{(2.11)}
$$

We will here  include in the fit the low energy data from $K_{l4}$
 decay,\fnote{As a technical point, we mention that we have 
increased by 50\% the error in  the point at highest energy, $s^{1/2}=381.4\,\mev$, 
from the  $K_{e4}$ compilation of Pislak et al., because this
 experimental value represents an average
 over a long energy range that extends 
to the edge of phase space.} shown in \fig~4, 
and 
we impose the  value of $\delta_0^{(0)}(m_K^2)$ from 
$K_{2\pi}$ given in (2.11).
The main virtue of these $K$ decay data is that they 
refer to pions on their mass shell; but, unfortunately, this leaves 
too much room  at 
the upper energy range,
$s^{1/2}\gsim0.6\,\gev$. 
If we fit only  $K$ decay data we can only 
use one parameter $B_0$ in the conformal expansion: if including another 
parameter, spureous minima would appear. We get a good fit, 
albeit with rather large errors:
$$\eqalign{
\cot\delta_0^{(0)}(s)=&\,\dfrac{s^{1/2}}{2k}\,\dfrac{M_{\pi}^2}{s-\tfrac{1}{2}z_0^2}\,
\dfrac{{\mu}^2_0-s}{{\mu}^2_0}\,B_0,
\quad z_0\equiv M_\pi;\cr
{B}_0=&\,18.5\pm1.7,\quad 
{\mu}_0=766\pm95\,\mev;\quad \dfrac{\chi^2}{\rm d.o.f.}=\dfrac{5.7}{12-2}.\cr
\quad 
a_0^{(0)}=&\,(0.22\pm0.02)\times M_{\pi}^{-1};\quad\delta_0^{(0)}(m_K)= 43\pm5\degrees.
\cr    }
\equn{(2.12)}$$

\topinsert{
\setbox0=\vbox{{\psfig{figure=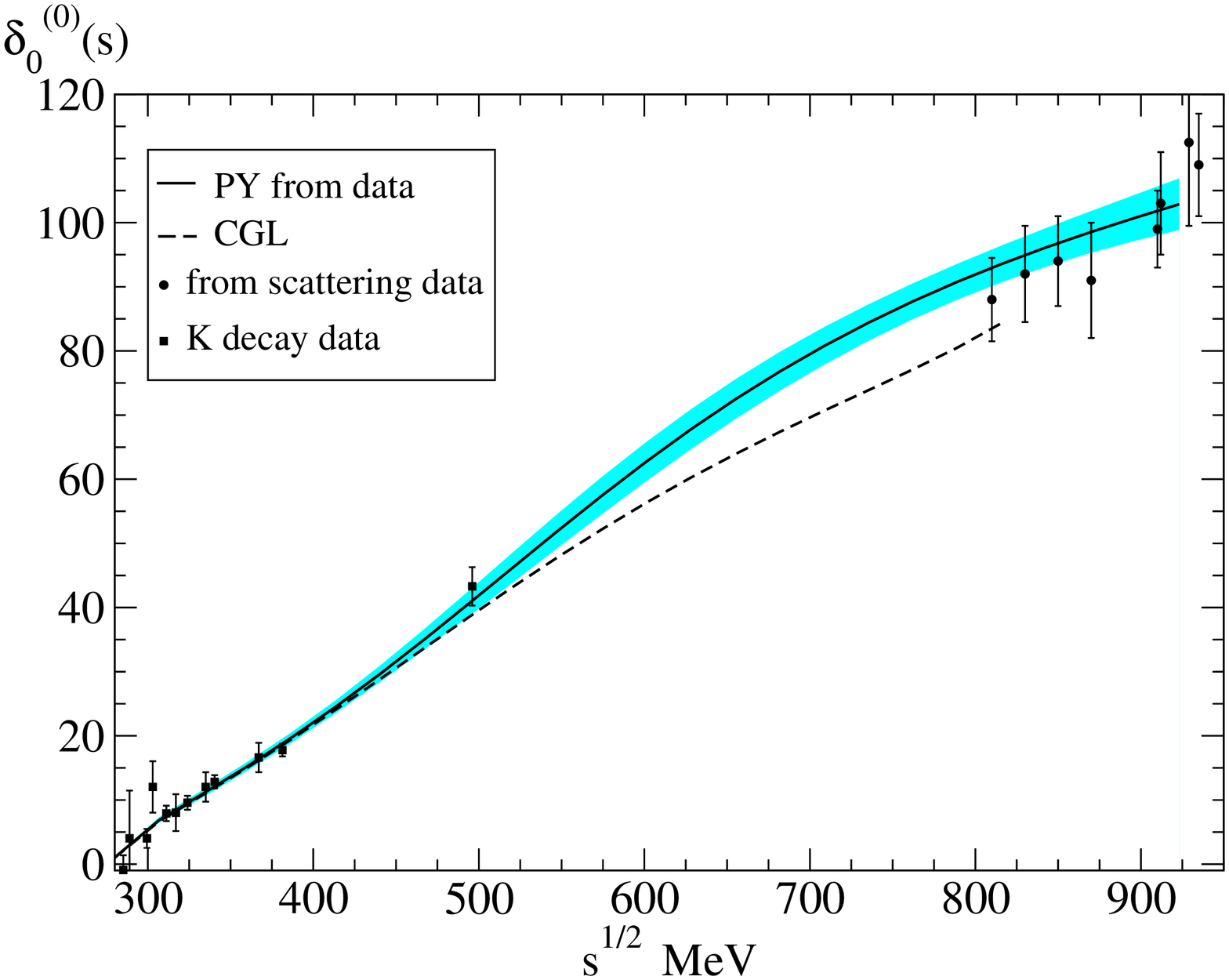,width=12.truecm,angle=-90}}}
\setbox6=\vbox{\hsize 15truecm\captiontype\figurasc{Figure 4. }{
The  
$I=0$, $S$-wave phase shifts and error bands corresponding to Eq.~(2.14) (PY, continuous
line).  Also shown (black dots) are the points from  $K_{l4}$ and $K_{2\pi}$ decays,
 and only the high energy data
 included in
the fits, as given in Eq.~(2.13). 
The dashed line is the solution of Colangelo et al.\ref{2} 
}\hb} 
\centerline{\tightboxit{\box0}}
\bigskip
\centerline{\box6}
\medskip
}\endinsert

  To improve on  this we have to add further data (and one more 
parameter $B_1$). 
To do so,  we can follow two different procedures. 
We can add to the  $K_{e4}$, $K_{2\pi}$ data various sets of 
experimental phase shifts, fitting each set 
individually; 
this we will do in \subsect~2.2.3. 
Or we can follow what we consider to be the best procedure: 
we combine in the fit data from various experiments at energies above 0.8~\gev, 
which we will call a {\sl global fit}. 
The reason to choose only data above 
0.8~\gev\ is that,  
 in the region 
between $0.81\,\gev$ and $0.97\,\gev$, the more relevant
 experimental results 
have overlapping error bars, something that does not happen at other energies 
(see Fig.~5). 
By combining several sets we may expect to average out systematic errors, 
at least to some extent.

At high energy we  thus include the following sets of data: first of all, the values  
$$\eqalign{
\delta_0^{(0)}(0.870^2\,\gev^2)=&\,91\pm9\degrees;\quad
\delta_0^{(0)}(0.910^2\,\gev^2)=\,99\pm6\degrees;\cr
\delta_0^{(0)}(0.935^2\,\gev^2)=&109\pm8\degrees;\quad 
\delta_0^{(0)}(0.965^2\,\gev^2)=134\pm14\degrees.\cr
}
\equn{(2.13a)}$$
These points are taken from solution 1 of Protopopescu et al.\ref{10} 
(both with and without modified moments), with the error 
increased by the difference between this and Solution~3 data in the same reference. 
We will also include in the fit the data, at similar energies,
of Grayer et al.:\ref{11a}
$$\eqalign{
\delta_0^{(0)}(0.912^2\,\gev^2)=&\,103\pm8\degrees;\quad
\delta_0^{(0)}(0.929^2\,\gev^2)=112.5\pm13\degrees;\cr
\delta_0^{(0)}(0.952^2\,\gev^2)=&\,126\pm16\degrees;\quad 
\delta_0^{(0)}(0.970^2\,\gev^2)=141\pm18\degrees.\cr
}
\equn{(2.13b)}$$
The central values are obtained averaging the  solutions given by 
Grayer et al., except solution~E, and the error is calculated adding quadratically
 the statistical error of the highest point, the 
statistical error of the lowest point (for each energy) and the difference 
between the central value and the farthest point. 

Finally, we add three points between 0.8 and 0.9 \gev\ obtained 
averaging the $s$-channel solution 
of Estabrooks and Martin  and solution 1 of Protopopescu et al., which
represent two extremes.  The error is obtained adding the difference between these two 
in quadrature to the largest statistical error. 
In this way we obtain the numbers,
$$\eqalign{
\delta_0^{(0)}(0.810^2\,\gev^2)=&\,88\pm6\degrees;\quad
\delta_0^{(0)}(0.830^2\,\gev^2)=92\pm7\degrees;\quad
\delta_0^{(0)}(0.850^2\,\gev^2)=94\pm6\degrees. 
\cr
}
\equn{(2.13c)}$$

\topinsert{
\medskip
\setbox0=\vbox{{\psfig{figure=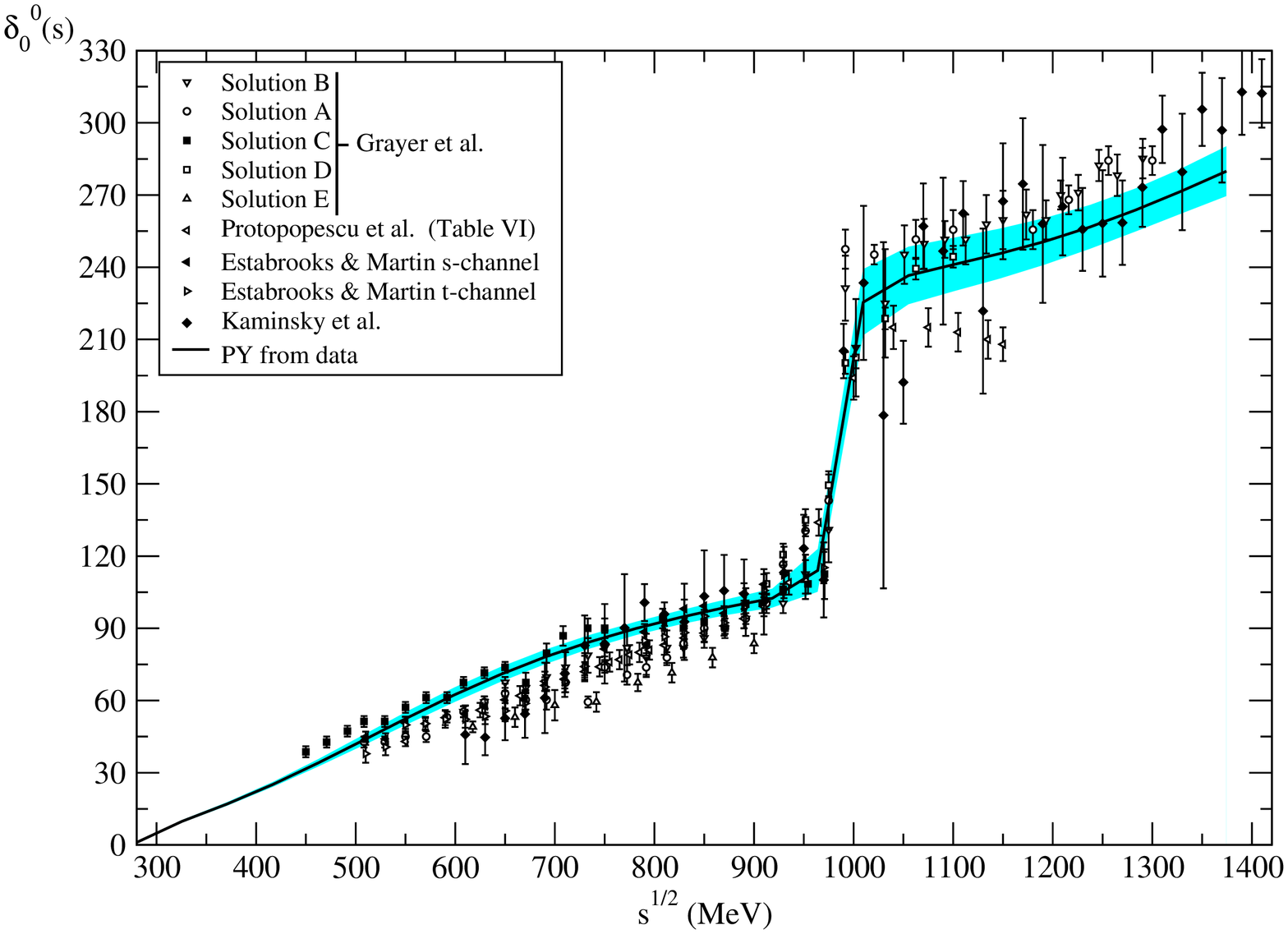,width=14truecm,angle=-90}}}
\setbox6=\vbox{\hsize 15truecm\captiontype\figurasc{Figure 5. }{
The  
$I=0$, $S$-wave phase shifts and error band in the whole energy range as 
given by our fits (PY), 
Eqs.~(2.14), (2.15).  The experimental points  from 
$K_{l4}$ and
$K_{2\pi}$ decays are not shown.}} 
\centerline{\tightboxit{\box0}}
\bigskip
\centerline{\box6}
}\endinsert

In spite of the generous errors taken, it should be noted that these 
data   could still contain 
 systematic errors, beyond those 
taken into account in (2.13), which may contaminate the 
results of the fit at the higher energies. In \sect~4 
we will use dispersion relations to 
improve the parametrization of this S0 wave.

We now have enough constraints to include  two
 parameters $B_i$ in the expansion of $\cot\delta_0^{(0)} $.
We find
$$\eqalign{
\cot\delta_0^{(0)}(s)=&\,\dfrac{s^{1/2}}{2k}\,\dfrac{M_{\pi}^2}{s-\tfrac{1}{2}z_0^2}\,
\dfrac{{\mu}^2_0-s}{{\mu}^2_0}\,
\left\{B_0+B_1\dfrac{\sqrt{s}-\sqrt{s_0-s}}{\sqrt{s}+\sqrt{s_0-s}}\right\},
\quad z_0\equiv M_\pi;\cr
{B}_0=&\,21.04,\quad {B}_1=6.62,\quad
{\mu}_0=782\pm24\,\mev;\quad {\chi^2}/{\rm d.o.f.}={15.7}/(19-3).\cr
\quad 
a_0^{(0)}=&\,(0.230\pm0.010) M_{\pi}^{-1},\quad b_0^{(0)}=(0.268\pm0.011)
M_{\pi}^{-3};\quad\delta_0^{(0)}(m_K)= 41.0\degrees\pm2.1\degrees;
\cr    }
\equn{(2.14a)}$$
this fit (shown in \fig~4) we take to be valid for $s^{1/2}\leq0.95\,\gev$. 
The errors of the $B_i$ are strongly correlated; uncorrelated errors are obtained if 
replacing the $B_i$ by the 
parameters $x,\,y$ with
$$B_0=y-x;\quad B_1=6.62-2.59 x;\quad y=21.04\pm0.70,\quad x=0\pm 2.6.
\equn{(2.14b)}$$

\booksubsection{2.2.3. Parametrization of the S wave for  $I=0$ 
below  $0.95\,\gev$ (individual fits)}

\noindent
In this Subsection we summarize, in Table~1, the results of  fits to data from 
 $K_{l4}$ and $K_{2\pi}$ decays including also, individually,  
data from various sets of  phase shift analyses. 
The method of fit is like  that used in the previous Subsection;
in particular, 
 we have fixed the Adler zero at $z_0=M_\pi$ in all these fits.

\midinsert
{\medskip
\setbox0=\vbox{
\setbox1=\vbox{\petit \offinterlineskip\hrule
\halign{
&\vrule#&\strut\hfil\ #\ \hfil&\vrule#&\strut\hfil\ #\ \hfil&
\vrule#&\strut\hfil\ #\ \hfil&
\vrule#&\strut\hfil\ #\ \hfil&
\vrule#&\strut\hfil\ #\ \hfil&
#&\strut\hfil\ #\ \hfil&#&\strut\hfil\ #\ \hfil\cr
 height2mm&\omit&&\omit&&\omit&&\omit&&\omit&&\omit&&\omit&\cr 
&\hfil \hfil&&\hfil $B_0$ \hfil&&\hfil $B_1$\hfil&&\hfil ${\mu}_0$ (\mev)\hfil&&
\hfil $M_\pi\times a_0^{(0)}$\hfil&&&&\hfil& \cr
 height1mm&\omit&&\omit&&\omit&&\omit&&\omit&&\omit&&\omit&\cr
\noalign{\hrule} 
height1mm&\omit&&\omit&&\omit&&\omit&&\omit&&\omit&&\omit&\cr
&PY, Eq. (2.14)&&\vphantom{\Big|}$21.04\;^{\rm (a)}$&&
$6.62\;^{\rm(a)}$&&$782\pm24$&&$0.230\pm0.010$& &&&
& \cr 
\noalign{\hrule}
height1mm&\omit&&\omit&&\omit&&\omit&&\omit&&\omit&&\omit&\cr
&\vphantom{\Big|}$K\; {\rm decay\;  only}$
&&$\phantom{\Big|}18.5\pm1.7$&&$\equiv0$&&$766\pm95$&&$0.218\pm0.021$&
&&&& \cr
\noalign{\hrule}
height1mm&\omit&&\omit&&\omit&&\omit&&\omit&&\omit&&\omit&\cr
&\vphantom{\Big|}
${\displaystyle{{ K\; {\rm decay\; data}}}\atop{\displaystyle +\,{\rm Grayer,\;B}}}$&&
$\phantom{\Big|}22.7\pm1.6$ &&$12.3\pm3.7 $&&$858\pm15 $&
&\hfil$0.246\pm0.042$ \hfil&
&&&& \cr
\noalign{\hrule}
height1mm&\omit&&\omit&&\omit&&\omit&&\omit&&\omit&&\omit&\cr
&\vphantom{\Big|}
${\displaystyle{{ K\; {\rm decay\; data}}}\atop{\displaystyle +\,{\rm Grayer,\;C}}}$&&
$\phantom{\Big|} 16.8\pm0.85$&&$-0.34\pm2.34 $&&$787\pm9 $&
&\hfil $0.236\pm0.023 $ \hfil&
&&&& \cr
\noalign{\hrule}
height1mm&\omit&&\omit&&\omit&&\omit&&\omit&&\omit&&\omit&\cr
&\vphantom{\Big|}
${\displaystyle{{ K\; {\rm decay\; data}}}\atop{\displaystyle +\,{\rm Grayer,\;E}}}$&&
$21.5\pm3.6 $&&$12.5\pm7.6 $&&$1084\pm110 $&
&\hfil $0.26\pm0.05$  \hfil&
&&&& \cr
\noalign{\hrule}
height1mm&\omit&&\omit&&\omit&&\omit&&\omit&&\omit&&\omit&\cr
&\vphantom{\Big|}
${\displaystyle{{ K\; {\rm decay\; data}}}\atop{\displaystyle +\,{\rm Kaminski}}}$&&
$ 27.5\pm3.0$&&$21.5\pm7.4 $&&$789\pm18 $&
&\hfil $0.25\pm0.10 $ \hfil&
&&&& \cr
\noalign{\hrule}
height1mm&\omit&&\omit&&\omit&&\omit&&\omit&&\omit&&\omit&\cr
&\vphantom{\Big|}
${\displaystyle{{ K\; {\rm decay\; data}}}\atop{\displaystyle +\,{\rm Grayer,\;A}}}$&&
$ 28.1\pm1.1$&&$26.4\pm2.8 $&&$866\pm6 $&
&\hfil $0.29\pm0.04 $ \hfil&
&&&& \cr
\noalign{\hrule}
height1mm&\omit&&\omit&&\omit&&\omit&&\omit&&\omit&&\omit&\cr
&\vphantom{\Big|}
${\displaystyle{{ K\; {\rm decay\; data}}}\atop{\displaystyle +\,{{\rm EM},\;s{\rm -channel}}}}$&&
$ 29.8\pm1.3$&&$25.1\pm3.3 $&&$811\pm7 $&
&\hfil $0.27\pm0.05 $ \hfil&
&&&& \cr
\noalign{\hrule}
height1mm&\omit&&\omit&&\omit&&\omit&&\omit&&\omit&&\omit&\cr
&\vphantom{\Big|}
${\displaystyle{{ K\; {\rm decay\; data}}}\atop{\displaystyle +\,{{\rm EM},\;t{\rm -channel}}}}$&&
$ 29.3\pm1.4$&&$26.9\pm3.4 $&&$829\pm6 $&
&\hfil $0.27\pm0.05 $ \hfil&
&&&& \cr
\noalign{\hrule}
height1mm&\omit&&\omit&&\omit&&\omit&&\omit&&\omit&&\omit&\cr
&\vphantom{\Big|}
${\displaystyle{{ K\; {\rm decay\; data}}}\atop{\displaystyle +\,{\rm Protopopescu\,VI}}}$&&
$ 27.0\pm1.7$&&$22.0\pm4.1 $&&$855\pm10 $&
&\hfil $0.26\pm0.05 $ \hfil&
&&&& \cr
\noalign{\hrule}
height1mm&\omit&&\omit&&\omit&&\omit&&\omit&&\omit&&\omit&\cr
&\vphantom{\Big|}
${\displaystyle{{ K\; {\rm decay\; data}}}\atop{\displaystyle +\,{\rm Protopopescu\,XII}}}$&&
$ 25.5\pm1.7$&&$18.5\pm4.1 $&&$866\pm14 $&
&\hfil $0.25\pm0.05 $ \hfil&
&&&& \cr
\noalign{\hrule}
height1mm&\omit&&\omit&&\omit&&\omit&&\omit&&\omit&&\omit&\cr
&\vphantom{\Big|}
${\displaystyle{{ K\; {\rm decay\; data}}}\atop{\displaystyle +\,{\rm Protopopescu\,VIII}}}$&&
$ 27.1\pm2.3$&&$23.8\pm5.0 $&&$913\pm18 $&
&\hfil $0.27\pm0.07 $ \hfil&
&&&& \cr
\noalign{\hrule}}
\vskip.05cm}
\centerline{\box1}
\bigskip
{\noindent\petit
 PY, Eq.~(2.14): our 
{\sl global} fit, \subsect~2.2.2. The rest are fits to either $K$ decay data alone, or 
combining these with various $\pi\pi$ scattering data sets.
Grayer A, B, C, E: the  solutions in the 
paper of Grayer et al.\ref{11a}  (solution ~D only concerns data above 1~\gev). 
EM: the
solutions of Estabrooks and  Martin.\ref{11a}  Kaminski: the papers of 
Kami\'nski et al.\ref{11c}  Protopopescu VI, XII and VIII:  
 the corresponding solutions in ref.~10. Solutions A, B and C of Grayer et al., as well as both 
solutions EM and Kami\'nski et al. are from 
energy-independent  analyses. The rest are from energy-dependent evaluations.\quad 
$^{\rm(a)}$ We do not give errors here as they are strongly correlated; cf.~\equn{(2.14b)}.}
\medskip
\centerline{\sc Table~1}
\smallskip
\centerrule{5truecm}}
\box0
}
\endinsert

A few comments are in order.  
First of all, we note that the 
solution~E of Grayer et al., as well as what one finds with only 
$K$ decay data, have  
 very large errors. 
Moreover,  solution~E of Grayer et al. 
 only contains eight points which clearly lie below 
all other experimental data; see \fig~5. 
Second, it is seen that the values of the parameters in the first five fits 
(which, as will be shown in \sect~4, are the more reliable ones)
cluster around our solution, labeled  PY,~Eq.~(2.14) in  Table~1. This is 
as should be expected, since our global solution sums up  
 information from  various experimental sets.
Third,  
 unlike our global solution, which includes systematic errors, 
the errors given in  Table~1 for  the other 
fits are only the {\sl statistical} ones. 
 Finally, we remark that all fits other than the first six have parameters that 
differ a lot from those 
obtained fitting only $K$ decay data. 
Since this last fit is already very good, this makes the lower six 
fits in Table~1 suspect. 
Indeed, and as we will see, they lead to scattering amplitudes 
that do not satisfy well dispersion relations.

\booksubsection{2.2.4. The $I=0$ S wave between $950\,\mev$ and $1420\,\mev$}

\noindent
The description of 
pion-pion scattering above the $\bar{K}K$ threshold would imply a full two-channel 
formalism. To determine the three independent components of the 
effective range matrix $\bf \Phi$, $\phiv_{11}$, 
$\phiv_{22}$ and $\phiv_{12}$, one requires 
measurement of  three cross sections. 
Failing this, one gets an indeterminate set, which is reflected 
very clearly in the wide variations of the effective range matrix parameters in the 
energy-dependent fits of Protopopescu et al.\ref{10} and 
Hyams et al.\ref{11a}

The raw data themselves are also incompatible; 
Protopopescu et al. find a phase shift that flattens above $s^{1/2}\simeq 1.04\,\gev$, 
while that of Hyams et al. or Grayer et al. continues to grow. 
This incompatibility is less marked if we choose the 
solution with modified higher moments by 
Protopopescu et al. (Table~XIII there). 
The inelasticities are compatible among the various determinations from 
$\pi\pi$ scattering, 
including that of Hyams et al.,\ref{11b} and 
Kami\'nski et al.,\ref{11c} although the errors
 of Protopopescu et al. appear to be much underestimated. 
They are, however, systematically, well above what one finds\ref{11d} from 
$\pi\pi\to\bar{K}K$ scattering: 
see Fig.~6. 

We will here give a purely empirical fit, using the $\pi\pi$ data. 
We write
$$\eqalign{
\cot\delta_0^{(0)}(s)=&\,c_0\,\dfrac{(s-M^2_s)(M^2_{f}-s)}{M^2_{f} s^{1/2}}\,
\dfrac{|k_2|}{k^2_2},\quad k_2=\dfrac{\sqrt{s-4m^2_K}}{2}\cr
\eta_0^{(0)}=&\,1-\left(\epsilon_1\dfrac{k_2}{s^{1/2}}+\epsilon_2\dfrac{k_2^2}{s}\right)
\,\dfrac{M'^2-s}{s};\quad M'=1.5\;\gev\; ({\rm fixed}).\cr}
\equn{(2.15a)}$$
The fit to the inelasticity gives
$$\epsilon_1=6.4\pm0.5,\quad \epsilon_2=-16.8\pm1.6;
\quad\chi^2/({\rm d.o.f.})=0.7.
\equn{(2.15b)}$$
This result is driven by the data of Protopopescu et al., 
whose accuracy is, unfortunately, much overestimated: see \fig~6.

If, instead of fitting $\eta_0^{(0)}$ to the $\pi\pi$ data of Protopopescu et al. 
and Hyams et al. we had fitted the data\ref{11d} from  
$\pi\pi\to\bar{K}K$ scattering (shown in \fig~6), we would have
 found values for the $\epsilon_i$ 
much smaller than what was given in (2.15b):
$$\epsilon_1=2.4\pm0.2,\quad \epsilon_2=-5.5\pm0.8;
\quad\chi^2/({\rm d.o.f.})=1.3.
\equn{(2.15b')}$$
 
We have checked that the influence of using (2.15b) or (2.15b$'$)
 on the dispersion relations and other 
evaluations of integrals, to be considered later, is minute, for energies below 0.95~\gev.
This is because the inelasticity affects little the {\sl imaginary} 
part of the partial wave (on the average). 
 Above 1~\gev, if we 
took the $\epsilon_i$ following from  $\pi\pi\to K\bar{K}$, Eq.~(2.15b$'$),   
the dispersion relations  would be slightly better fulfilled; see 
\subsect~4.1, at the end. 
In spite of this, we stick to (2.15b). 
Taking $\eta_0^{(0)}$ from one set of experiments and $\delta_0^{(0)}$ 
from another (incompatible with the first) would be an inconsistent procedure. 

To fit $\delta_0^{(0)}$ we also require  it to agree with the low energy determination 
we found in  \subsect~2.2.2 at $s^{1/2}=0.95\,\gev$. 
If we include the data of Protopopescu et al. in the fit we find a poor fit with 
$\chidof=39/(14-2)$ and the parameters 
$$c_0=1.72\pm0.08,\quad M_s=930\;\mev,\quad M_f=1340\;\mev.$$
The error in $c_0$,  corresponding to $3\,\sigma$,
is purely nominal. If we keep $M_s$ fixed and 
remove the data of Protopopescu et al., we get a good \chidof, and now 
$$c_0=0.79\pm0.25,\quad M_f=1270\;\mev.$$
If we want to be compatible with
 the data of Hyams et al., we must increase the errors.
We then take the numbers
$$c_0=1.3\pm0.5,\quad M_f=1320\pm50\;\mev,\quad M_s=930\;\mev\;({\rm fixed}).
\equn{(2.16)}$$
The fit is shown in \fig~6.

\topinsert{
\setbox0=\vbox{{\psfig{figure=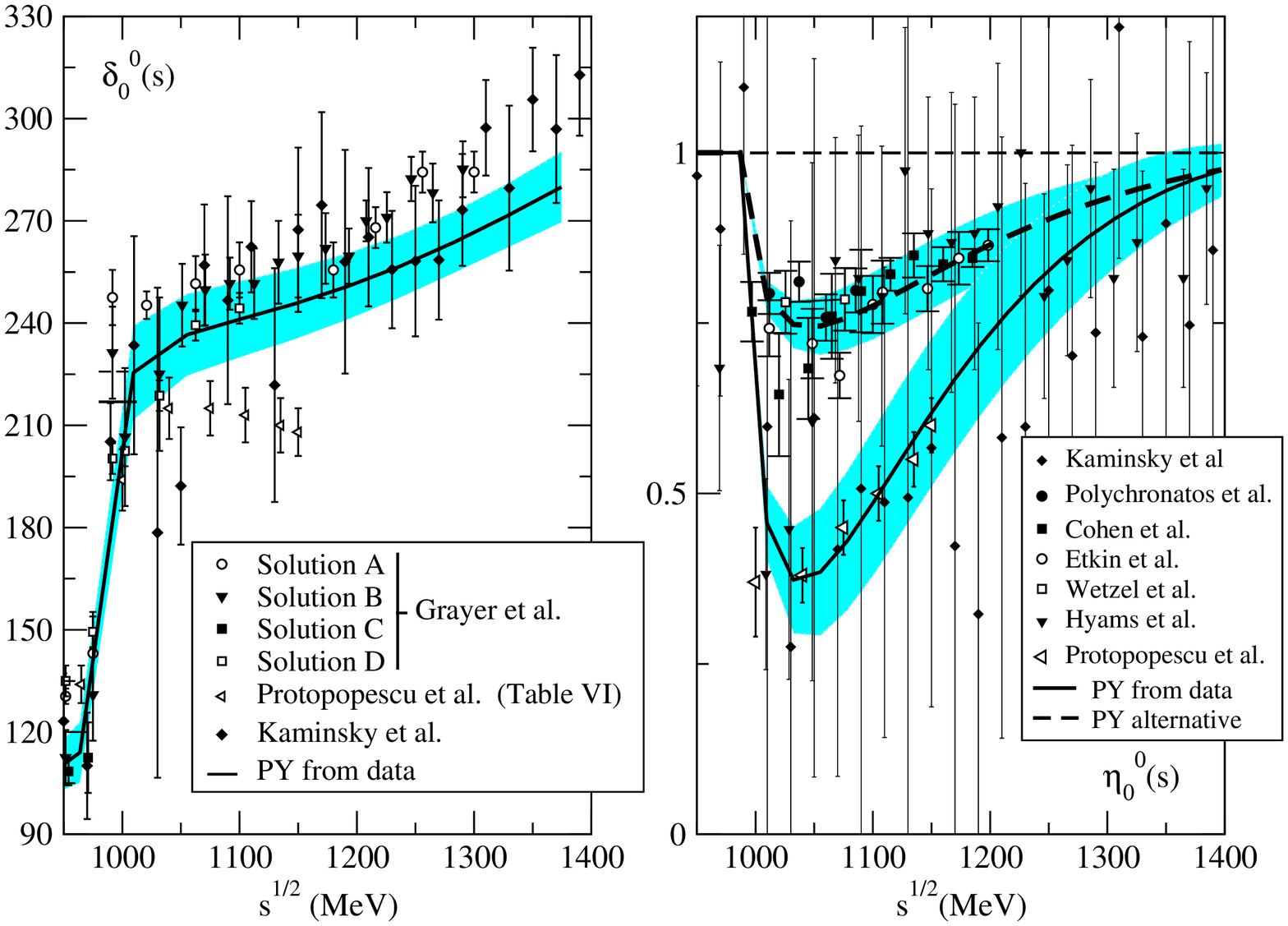,width=14truecm,angle=-90}}} 
\setbox6=\vbox{\hsize 15truecm\captiontype\figurasc{Figure 6. }{Fit to the
 $I=0$, $S$-wave  inelasticity
 and phase shift  between 
 950 and 1400 \mev. 
Data  from refs.~10,~11. The 
difference between the determinations of $\eta_0^{(0)}$ from $\pi\pi\to\pi\pi$ 
(PY from data) and 
from $\pi\pi\to\bar{K}K$ (PY alternative) is apparent here.
}\hb} 
\centerline{\tightboxit{\box0}}
\bigskip
\centerline{\box6}
\medskip
}\endinsert

 We emphasize again that these are  purely empirical fits and, moreover, 
they are fits to 
data which certainly have uncertainties well beyond their nominal errors, 
as given in  given in (2.15b) and (2.16); something that is obvious for  $\eta_0^{(0)}$
from \fig~6. 
It follows that relations such as dispersion relations 
in which the S0 wave plays an important role will be unreliable for energies 
near and, especially, above $\bar{K}K$ threshold (below these energies, however, 
both (2.15b) and (2.15b$'$) give very similar results). 
In fact, we will check that mismatches occur when $s^{1/2}>0.95\,\gev$.
A sound description of the S0 wave for $s^{1/2}>0.95\,\gev$
  in $\pi\pi$ scattering would require more refined 
parametrizations and, above all,
 use of more information than just $\pi\pi$ experimental data, 
and lies beyond the scope of the present paper.

\booksubsection{2.3. The D waves}

\noindent
The D waves cannot be described with the same accuracy as the S, P waves. 
The data are scanty, and have huge errors. 
That one can get reasonable fits at all is due to the fact that one can use low
 energy information 
from sum rules; specifically, we will impose in the fits the 
values of the scattering lengths that follow from the Froissart--Gribov representation. 
Note that this is not circular reasoning, and it only 
introduces a small correlation: the  Froissart--Gribov representation 
for the D0, D2, F waves depends mostly on the S0, S2 and P waves, and very little on the
 D0, D2, F waves themselves.

\booksubsection{2.3.1. Parametrization of the $I=2$ D wave}

\noindent
For isospin equal 2 we would only expect important inelasticity when the channel 
 $\pi\pi\to\rho\rho$  opens up, 
so we will take the value $s_0=1.45^2\,\gev^2\sim4M^2_\rho$ for the energy at which 
elasticity is not negligible.

 But life is complicated: 
a pole term is necessary to get an acceptable fit down to low energy
since we expect $\delta_2^{(2)}$ to change sign near threshold.
The experimental measurements (Losty et al.; Hoogland et al.\ref{12}) give 
negative and small values for the phase above some $500\,\mev$, while  
 chiral perturbation calculations 
and the Froissart--Gribov representation 
indicate a positive scattering length.\ref{6}
 If we want a parametrization that 
applies down to threshold, we must incorporate this  
zero of the phase shift. 
What is more, the clear inflection seen in data around 1~\gev\ implies that 
we will have to go to third order in the 
conformal expansion. So 
we write
$$\cot\delta_2^{(2)}(s)=
\dfrac{s^{1/2}}{2k^5}\,\Big\{B_0+B_1 w(s)+B_2 w(s)^2\Big\}\,
\dfrac{{M_\pi}^4 s}{4({M_\pi}^2+\deltav^2)-s}
\equn{(2.17a)}$$
with $\deltav$ a free parameter fixing the zero and
$$w(s)=\dfrac{\sqrt{s}-\sqrt{s_0-s}}{\sqrt{s}+\sqrt{s_0-s}},\quad
 s_0^{1/2}=1450\,\mev.$$
Since  the data we have on this wave are not accurate (cf.~\fig~7) 
we have to include extra information. To be precise, we 
include in the fit the value of  
 the scattering length  that follows
 from the Froissart--Gribov representation, 
$$a_2^{(2)}=(2.78\pm0.37)\times10^{-4}\,M_{\pi}^{-5},
$$
but {\sl not} that of the effective range parameter,
$$
\quad b_2^{(2)}=(-3.89\pm0.28)\times10^{-4}\,{M_\pi}^{-7}
$$
 (see below, \sect~6).

\topinsert{
\setbox0=\vbox{{\psfig{figure=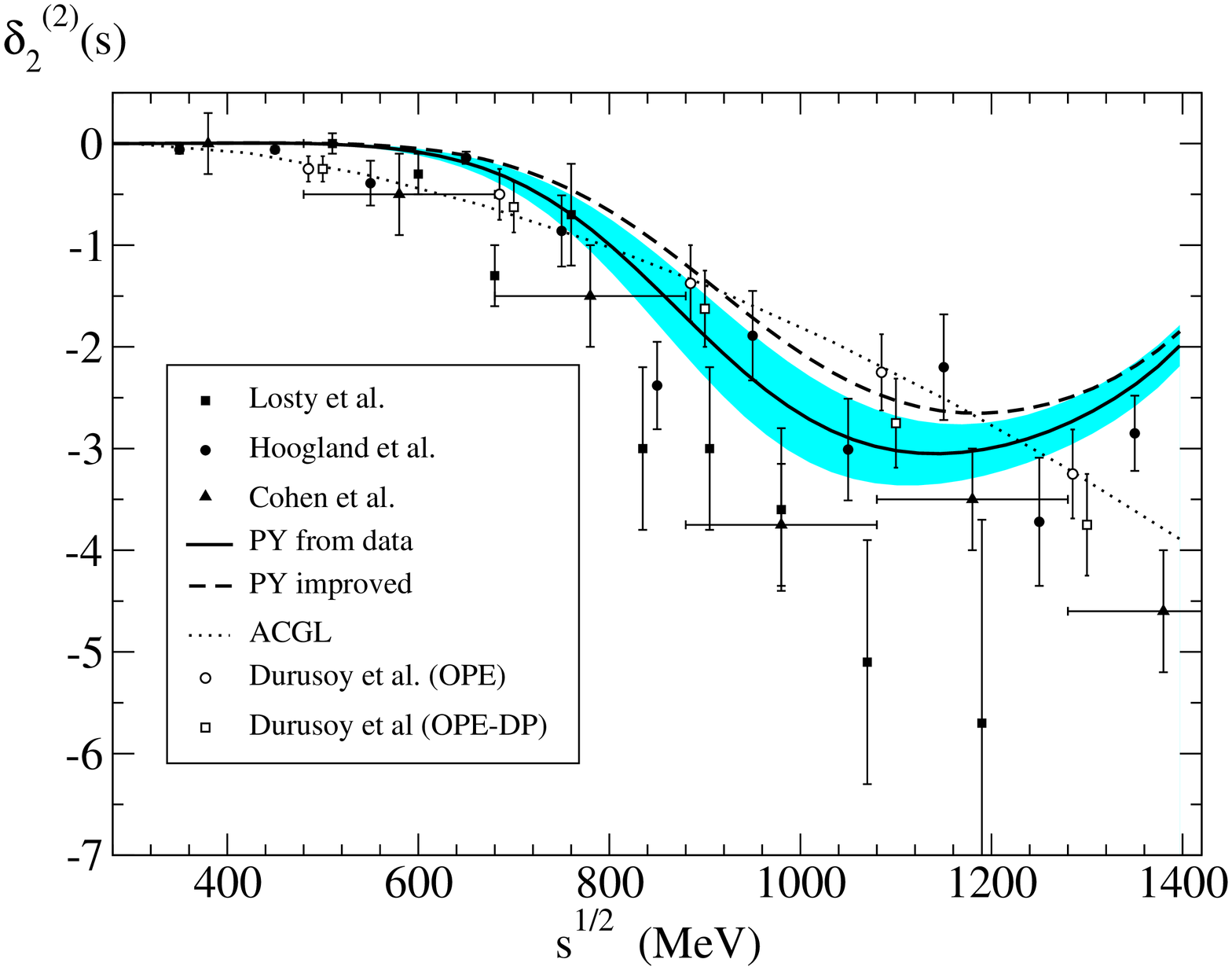,width=12truecm,angle=-90}}} 
\setbox6=\vbox{\hsize 15truecm\captiontype\figurasc{Figure 7. }{Fits to the  
$I=2$, $D$-wave phase shift. Continuous line (PY) with Eq.~(2.18).
Dashed line: PY, after improving with dispersion relations (\sect~4).
Dotted line: the fit used in refs.~1,~2 (ACGL, CGL).
Also shown are the data points of Losty et al.,
 from solution~A of Hoogland et al.,\ref{12}  
 from Cohen et al.\ref{4} and Durusoy et al.,\ref{12} the last nont included in the fit.
}} 
\centerline{\tightboxit{\box0}}
\bigskip
\centerline{\box6}
\medskip
}\endinsert

We  get a mediocre fit, $\chidof=71/(25-3)$, and the values of the parameters
 are
$$B_0=(2.4\pm0.3)\times10^3,\quad B_1=(7.8\pm0.8)\times10^3,\quad
 B_2=(23.7\pm3.8)\times10^3,\quad
\deltav=196\pm20\,\mev.
\equn{(2.17b)}$$
We have rescaled the errors by the square root of the \chidof

  The fit, which may be found in \fig~7, returns 
 reasonable
 numbers for the scattering length and  for the effective range parameter,
$b_2^{(2)}$: 
$$a_2^{(2)}=(2.5\pm0.9)\times10^{-4}\,{M_\pi}^{-5};\quad
b_2^{(2)}=(-2.7\pm0.8)\times10^{-4}\,{M_\pi}^{-7}.
\equn{(2.18)}$$

Although the twist  
of $\delta_2^{(2)}(s)$ at $s^{1/2}\sim1.05\,\gev$ is probably 
connected to the opening of the 
$2\pi\rho$ channel, 
we neglect the inelasticity of the D2 wave, since it is not detected experimentally. 
This, together with the incompatibility of 
the various sets of experimental data 
and the poor convergence of the conformal series,  indicates  that
the solution given by (2.17) is, very likely, somewhat displaced with respect 
to the  ``true" D2 wave 
at the higher energy range (say for $s^{1/2}\gsim0.7\,\gev$). 
In fact, the values of the parameters will be improved in \sect~4 
with the help of 
dispersion relations; 
the D2 phase shift one finds by so doing is slightly displaced 
with respect to that following from (2.17), 
as shown in \fig~7.

\booksubsection{2.3.2. Parametrization of the $I=0$ D wave}

\noindent
The D wave with isospin 0 in $\pi\pi$ scattering presents two resonances 
below $1.7\,\gev$: the $f_2(1270)$ and the $f_2(1525)$, 
that we will denote respectively by $f_2$, $f'_2$. 
Experimentally, 
$\gammav_{f_2}=185\pm4\,\gev$ and $\gammav_{f'_2}=76\pm10\,\gev.$ 
The first, $f_2$,  
couples mostly to $\pi\pi$, with small  couplings to 
$\bar{K}K$ ($4.6\pm0.5\,\%$), $4\pi$ ($10\pm3\,\%$) and 
$\eta\eta$. The second 
couples mostly to $\bar{K}K$, with a small coupling to $\eta\eta$ and $2\pi$, 
respectively $10\pm3\,\%$ and $0.8\pm0.2\,\%$. 
This means that the channels $\pi\pi$ and $\bar{K}K$ are 
essentially decoupled and, to
 a 15\% accuracy, we may neglect inelasticity up to $s \simeq1.45^2\gev^2$.

There are not many experimental data on the D wave which, at accessible energies, 
 is  small. 
So, the compilation of  $\delta_2^{(0)}$ 
phase shifts of Protopopescu et al.\ref{10} gives significant numbers  
for $\delta_2^{(0)}$ 
only in the range $810\,\mev\leq s^{1/2}\leq 1150\,\mev$. 
In view of this, it is impossible to get accurately the D wave scattering lengths, 
or indeed any other low energy parameter, from this information: so, 
we will  include information 
on $a_2^{(0)}$ to help stabilize the fits.

\topinsert{
\setbox0=\vbox{{\psfig{figure=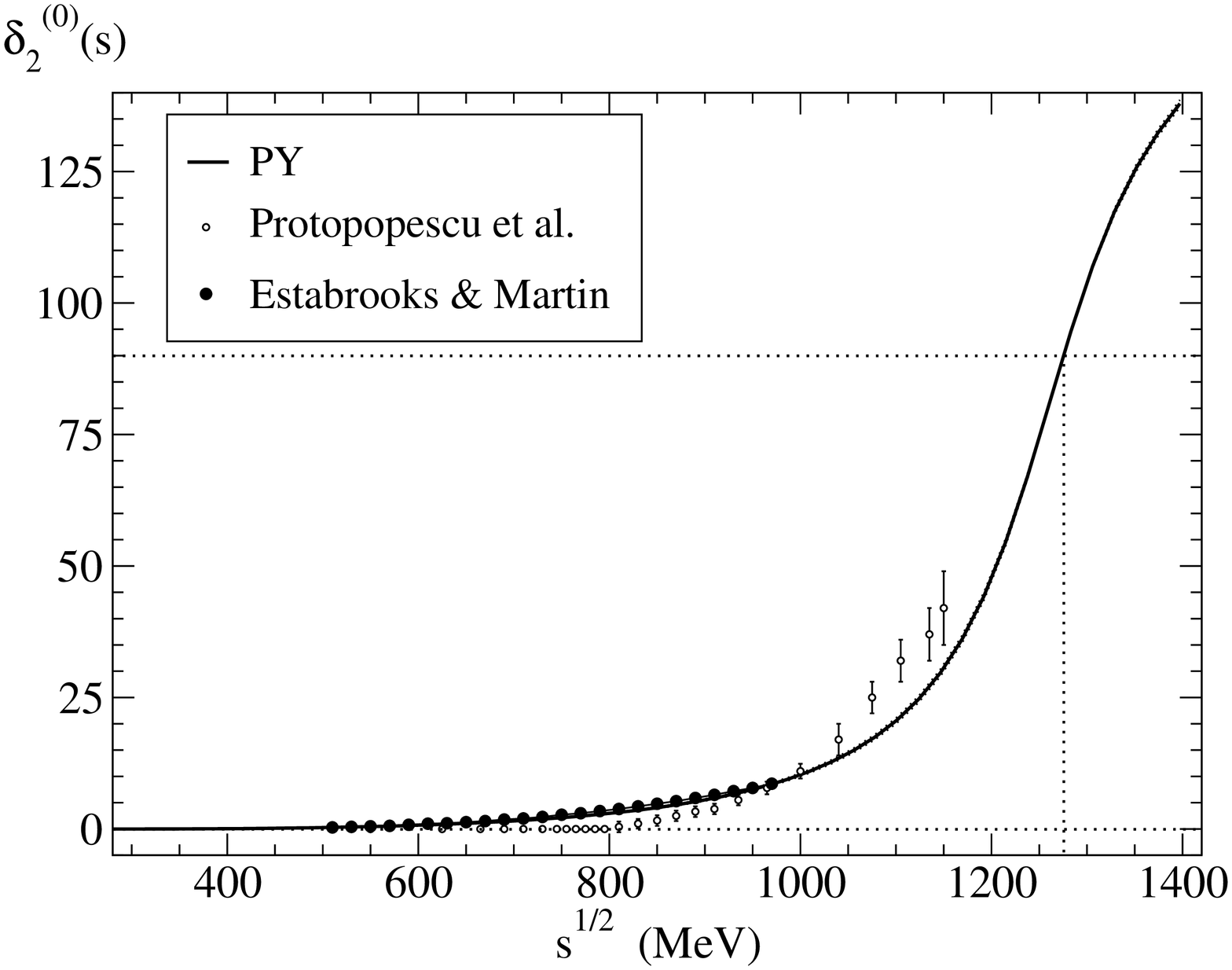,width=11.2truecm,angle=-90}}} 
\setbox6=\vbox{\hsize 15truecm\captiontype\figurasc{Figure 8. }{Fit to the  
$I=0$, $D$-wave phase shift. 
Also shown are the data points from solution~1 of Protopopescu et al. (open circles) and 
data  of  Estabrooks and Martin (black dots). 
The dotted lines mark  the $f_2(1270)$ resonance, whose 
location and width were included in the fit.
}\hb} 
\centerline{\tightboxit{\box0}}
\bigskip
\centerline{\box6}
\medskip
}\endinsert

We take the data of Protopopescu et al.\ref{10}  
and consider the so-called  ``solution 1",
with the two possibilities given in Table~VI and 
Table~XIII (with modified higher moments), in  
 the range 
mentioned before, $s^{1/2}=0.810\,\gev$ to $1.150\,\gev$.
The problem with these data points is that they are certainly biased, 
as indeed they are 
quite incompatible with
those of other experiments. 
We can stabilize the fits by 
 fitting also the points\fnote{For these data we arbitrarily 
take a common error of 10\%.} of Estabrooks and Martin,\ref{11a}
 and  imposing the value of the
 width of the
$f_2$ resonance,  with the condition $\gammav_{f_2}=185\pm10\,\mev$, 
as well as the value of the scattering
length.
 We write
$$\cot\delta_2^{(0)}(s)=\dfrac{s^{1/2}}{2k^5}\,(M^2_{f_2}-s)\,{{M_\pi}^2}\,
\Big\{B_0+B_1w(s)\big\}
\equn{(2.19a)}$$
and
$$w(s)=\dfrac{\sqrt{s}-\sqrt{s_0-s}}{\sqrt{s}+\sqrt{s_0-s}},\quad
 s_0^{1/2}=1450\,\mev;\quad
M_{f_2}=1275.4\,\mev.$$ 
We find
$$ B_0=23.6\pm0.7,\quad B_1=24.7\pm1.0.
\equn{(2.19b)}$$
The fit is not good in that we get $\chidof=300/(52-2)$. However, it improves 
substantially if we
 exclude the data of Protopopescu et al., which are strongly biased 
(as is  clearly seen in \fig~8); the central values, however, vary very little. 
We have included in the errors in (2.19b)  half the difference between the 
two possibilities (with or without the data of Protopopescu et al.)

The fit returns the values
$$\eqalign{
a_2^{(0)}=&\,(18.0\pm2.8)\times10^{-4}\,\times M_{\pi}^{-5},\quad
b_2^{(0)}=(-8.1\pm3.1)\;10^{-4}\,\times M_{\pi}^{-7};\cr
\gammav_{f_2}=&\,190\pm8\,\mev.\cr
}
\equn{(2.20)}$$
The value of $b_2^{(0)}$ agrees within  $1.3\,\sigma$ with the more 
precise result obtained with the Froissart--Gribov representation (see below), 
that gives
$$a_2^{(0)}=(18.70\pm0.41)\;10^{-4}\,\times M_{\pi}^{-5},\quad
b_2^{(0)}=(-4.16\pm0.30)\;10^{-4}\,\times M_{\pi}^{-7}.
\equn{(2.21)}$$
This agreement  is remarkable, taking into account that the D-wave
$b_2$ calculations are less reliable since they are
the ones with a {\sl relatively} large contribution of 
the {\sl derivative} of the I=2 t-channel amplitude,
 which is very uncertain (see
ref.~6).

We then take into account the inelasticity by writing
$$\eta_2^{(0)}(s)=\cases{1,\qquad  s< 4m_K^2,\;
\cr
1-\epsilon\,\dfrac{k_2(s)}{k_2(M^2_{f_2})},\quad
\epsilon=0.262\pm0.030;\quad s> 4m_K^2;\cr} 
\equn{(2.22)}$$
$ k_2=\sqrt{s/4-m^2_K}.$
We have fixed the coefficient $\epsilon$ fitting the inelasticities
 of Protopopescu et al.,\ref{10}  
and the experimental inelasticity of the $f_2$; the overall $\chidof$ of this 
fit is $\sim1.8$.

In principle, by including  $b_2^{(0)}$ in our fit 
the nominal uncertainties in the ``fit to data"
parameters could be decreased. However, we have not done so, since,
as explained before and in ref.~6, the D-wave $b_2$'s are less reliable 
than the $a_2$'s (used as input with very large errors) and, 
therefore, 
the apparent improvement would be made at the cost of reliability.

\booksubsection{2.4. The F, G waves}

\noindent
The contribution of the F wave to our sum rules and dispersion relations is very small, 
but it is interesting to evaluate it (we have included its contribution to all relations)
to check, 
precisely, the convergence of the partial wave series.
The contribution of the G0, G2 waves is completely negligible,  
for the quantities we calculate in our paper, 
but we  describe  fits to these for completeness.

\booksubsection{2.4.1. The F wave}
\noindent
 The experimental situation
for the F wave is somewhat confused. 
 According to Protopopescu et al.,\ref{10} it starts
negative
 (but compatible with zero at the $2\,\sigma$ level) and
 becomes positive around $s^{1/2}=1\,\gev$. 
Hyams et al.\ref{11a}  report a positive $\delta_3(s)$ 
when it differs from zero (above $s^{1/2}=1\,\gev$). 
In both cases no inelasticity is detected up to $s^{1/2}\sim1.4\,\gev$.

The corresponding scattering length may be calculated with the help
 of the Froissart--Gribov 
representation and one finds
$$a_3=(6.3\pm0.4)\times10^{-5}\,M_{\pi}^{-7}.
$$
It could in principle be
 possible that $\delta_3(s)$ changes sign {\sl twice}, once near
threshold  and once near $s^{1/2}=1\,\gev$.    
However,  we disregard 
this possibility and write, simply,
$$\cot\delta_3(s)=\dfrac{s^{1/2}M_{\pi}^6}{2k^7}\,\Big\{B_0+B_1w(s)\Big\},\quad
w(s)=
\dfrac{\sqrt{s}-\sqrt{s_0-s}}{\sqrt{s}+\sqrt{s_0-s}},
\equn{(2.23a)}$$
with $s_0^{1/2}=1.45\,\gev$, and {\sl impose} the value of $a_3$. 

It is to be understood
 that this parametrization provides only an empirical representation of the available data, 
and that it may not be reliable except at very low energies,  
where it is dominated by the scattering length, and for  
$s^{1/2}\gsim 1\,\gev$. We fit  data of Protopopescu et al.,\ref{10}
 for energies above 1 \gev, and data of Hyams et al.\ref{11b} 
We have estimated the errors of this last set (not given in the paper)
 as the distance from the average value
to the extreme values in the  different 
solutions given.  We find 
$$\dfrac{\chi^2}{\rm d.o.f.}\simeq\dfrac{7.7}{14-2},\quad 
B_0=(1.09\pm0.03)\times10^5,\quad B_1=(1.41\pm0.04)\times 10^5.
\equn{(2.23b)}$$
The errors have been increased by including as an error the
 variation that affects
 the central values when using only one of the two
sets of data.
 We do not include separately the effects of the $\rho_3(1690)$, 
since its tail is  incorporated in the fitted data.

\booksubsection{2.4.2.The G waves.}
\noindent
The experimental information on the G waves is very scarce. 
For the wave G2, we have two nonzero values for $\delta_4^{(2)}$ 
 from Cohen et al.,\ref{4} and four significant ones from Losty et al.;\ref{12} 
they are somewhat incompatible. 
We then fit the data separately, with a one-parameter formula;
we write
$$\cot\delta_4^{(2)}(s)=\dfrac{s^{1/2}M^8_\pi}{2k^9}\,B.$$
If we fit the data of Losty et al. we find $B=(-0.56\pm0.09)\times10^{6}$, 
while from Cohen et al. we get  $B=(-10.2\pm3.0)\times10^{6}$. 
Fitting both sets together we find $B=-9.1\times10^{6}$, 
and a very poor  $\chidof=32/(6-1)$. 
Enlarging the resulting  error 
to cover $6\,\sigma$, we  obtain our best result, 
$$\cot\delta_4^{(2)}(s)=\dfrac{s^{1/2}M^8_\pi}{2k^9}\,B,\quad B=(-9.1\pm3.3)\times 10^6.
\equn{(2.24)}$$

This formula can only be considered as 
 valid  only for a limited range, $0.8\leq s^{1/2}\leq1.5\,\gev$. 
In fact, from the Froissart--Gribov representation it follows
 that the G2 scattering length is
{\sl positive}. One has 
$a_4^{(2)}=(4.5\pm0.2)\times10^{-6}\,M_{\pi}^{-9},
$
while (2.24) would give a negative value.

For the G0 wave, the situation is similar. 
However,  we know of the existence of a very inelastic
 resonance with mass  around 2 \gev. 
An effective value for the imaginary part of the corresponding partial wave may be found 
in Appendix~A.

\brochuresection{3. Forward dispersion relations}

\noindent We expect that the scattering amplitudes that follow from the 
phase shifts (and inelasticities) at low energy ($s^{1/2}\leq1.42\,\gev$), 
and the Regge expressions 
at high energy, will satisfy dispersion relations since they 
fit well the experimental data and are therefore good approximations to the 
physical scattering amplitudes. 
In the present Section we will check that this is the case, 
at low energies ($s^{1/2}\lsim0.95\,\gev$), for 
three independent scattering amplitudes (that form a complete set), which we 
will conveniently take the following $t$-symmetric or antisymmetric combinations: 
$\pi^0\pi^0\to\pi^0\pi^0$, $\pi^0\pi^+\to\pi^0\pi^+$, and the 
amplitude $I_t=1$,  corresponding to isospin unity in the $t$ channel.
The reason for choosing these amplitudes is that 
the amplitudes for $\pi^0\pi^0$ and $\pi^0\pi^+$ 
depend  only on two isospin states, and have positivity properties: 
their imaginary parts are sums of positive terms. 
Because of this, the {\sl errors} are much reduced for them. 
This is easily verified if we compare with the errors 
in the dispersion relations for $\pi^0\pi^0$ and $\pi^0\pi^+$ 
with those for  the amplitude 
with $I_t=1$, which depends on three isospin states 
and has no positivity properties (see below, in \fig~15).

Here we will not cover the full energy range or try to improve the parameters by 
requiring fulfillment of the 
dispersion relations, something that we leave for \sect~4. 
We will start discussing  the global fit in \subsect~2.2.2; 
the results using the individual fits for S0, as in \subsect~2.2.3, 
will be discussed later, in \sect~3.3.

In our analysis one should take into account that, 
for the amplitudes that contain the S0 
wave, the uncertainties for it above 0.95~\gev\ induce large errors in the dispersive
integrals,  and the agreement between dispersive integrals and real parts of the scattering 
amplitudes evaluated directly becomes affected. 
This is particularly true for the $\pi^0\pi^0$ 
amplitude, dominated by the S0 contribution, 
where the mismatch becomes important above $\sim0.7\,\gev$.
For the $\pi^0\pi^+$ amplitude, however, since it is  not affected by the S0 problem, 
the fulfillment of the dispersion relation is good, within reasonably small errors, 
up to the very region where Regge behaviour takes over, $s^{1/2}\simeq1.42\,\gev$.

This is a good place to comment on the 
importance of the contribution of the Regge region (i.e., from energy above 
1.42~GeV) to the various dispersive integrals. 
Of course, this depends on each dispersion relation. 
As an indication, we mention that for the unsubtracted dispersion relation 
(3.7) at threshold, the contribution of the Regge region 
($s\geq1.42$~GeV) is 9\%. For the subtracted relation 
(3.1a) it is of 10\% at $s^{1/2}=0.5$~GeV and 20\% at
 $s^{1/2}=0.8$~GeV.

A final general comment is that here, as well as for the sum rules 
that we will discuss in \sect~5 and 
the Froissart--Gribov calculations (\sect~6),  
we only include the contributions of waves up to and including the F wave. 
We have checked, in a few typical cases, that the 
contributions of the G0, G2 waves are completely negligible.
\vskip0.5truecm

\brochuresubsection{3.1. Even amplitude dispersion relations (with the global fit)}
\vskip-0.5truecm
\booksubsection{3.1.1. $\pi^0\pi^0$ scattering}
 
\noindent
We  consider here the forward dispersion 
relation for  $\pi^0\pi^0$ scattering, subtracted at threshold, $4M^2_\pi$.
 We write, with $F_{00}(s)$ the forward $\pi^0\pi^0$ amplitude,
$$\real F_{00}(s)-F_{00}(4M_{\pi}^2)=
\dfrac{s(s-4M_{\pi}^2)}{\pi}\pepe\int_{4M_{\pi}^2}^\infty\dd s'\,
\dfrac{(2s'-4M^2_\pi)\imag F_{00}(s')}{s'(s'-s)(s'-4M_{\pi}^2)(s'+s-4M_{\pi}^2)}.
\equn{(3.1a)}$$
In particular, for $s=2M^2_\pi$, which will be important when we later discuss 
the Adler zeros (\sect~4), we have
$$F_{00}(4M_{\pi}^2)=F_{00}(2M_{\pi}^2)+D_{00},\quad D_{00}=
\dfrac{8M_{\pi}^4}{\pi}\int_{4M_{\pi}^2}^\infty\dd s\,
\dfrac{\imag F_{00}(s)}{s(s-2M_{\pi}^2)(s-4M_{\pi}^2)}.
\equn{(3.1b)}$$

We first check the sum rule (3.1b). 
We take  for 
$F_{00}(4M_{\pi}^2),\,F_{00}(2M_{\pi}^2)$ the values that follow from 
our fits to experimental data of \sect~2, which provide a representation 
of partial waves valid below threshold 
(provided $s>0$), and evaluate the dispersive integral with 
the parametrizations we obtained also in \sect~2. 
We find  fulfillment to less than $1\,\sigma$:
$$D_{00}=(43\pm3)\times10^{-3}
\equn{(3.2a)}$$
and
$$
F_{00}(4M^2_\pi,0)-F_{00}(2M^2_\pi,0)=(33\pm22)\times10^{-3}.
\equn{(3.2b)}$$
For the difference (which should vanish if the 
dispersion relation was exactly fulfilled, and which takes into account correlations) 
we get
$$F_{00}(4M^2_\pi,0)-F_{00}(2M^2_\pi,0)-D_{00}=
(-10\pm23)\times10^{-3}.
\equn{(3.2c)}$$

\topinsert{
\setbox0=\vbox{{\psfig{figure=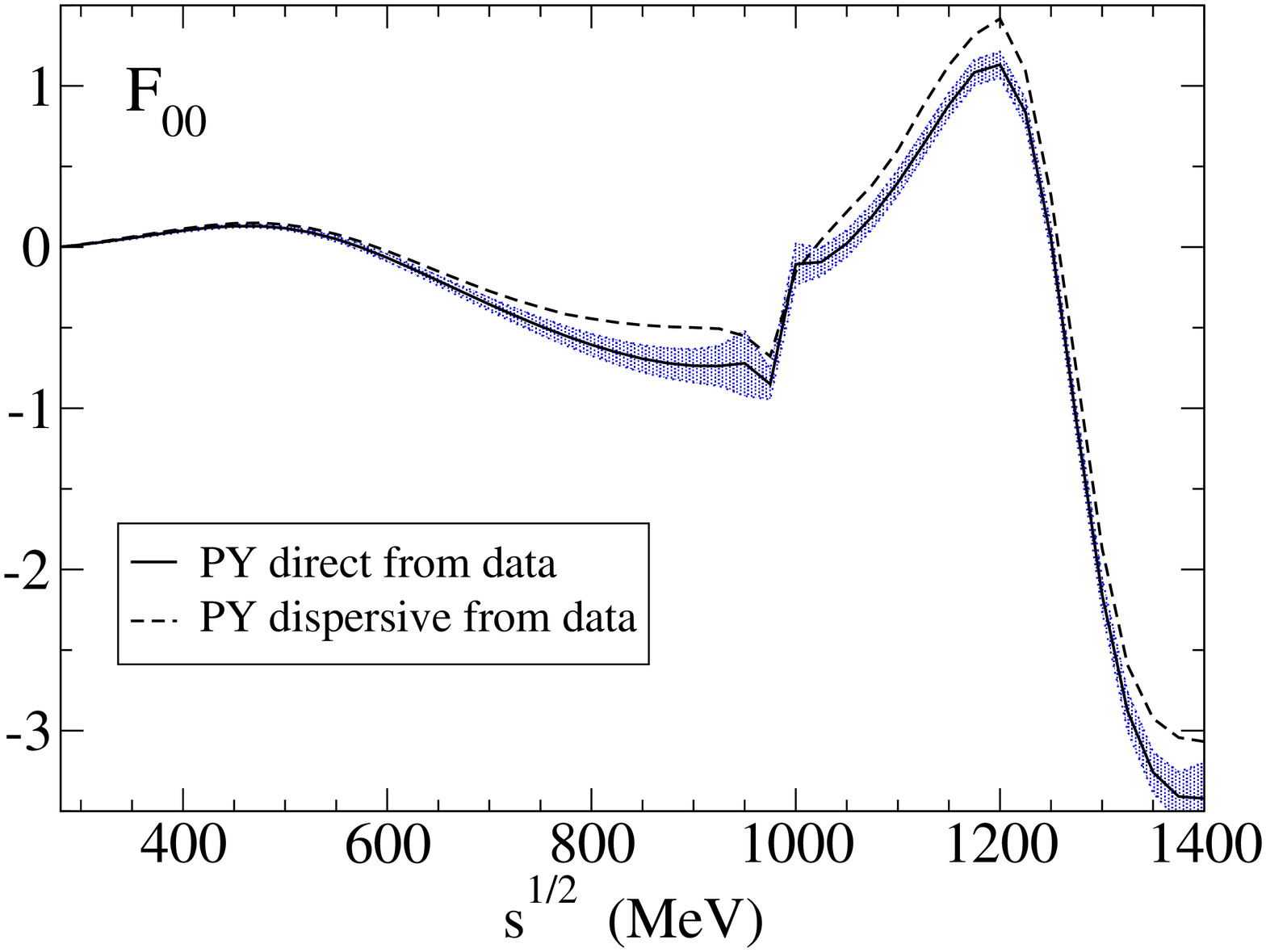,width=9.1truecm,angle=-90}}} 
\setbox6=\vbox{\hsize 15truecm\captiontype\figurasc{Figure 9. }{The combination $\real
F_{00}(s)-F_{00}(4M_{\pi}^2)$
  (continuous line) and the dispersive integral 
(broken line).
}\hb} 
\centerline{\tightboxit{\box0}}
\bigskip
\centerline{\box6}
\medskip
}\endinsert

We can also verify the dispersion relation (3.1a). 
The result is shown in Fig.~9, where a certain mismatch is observed in 
some regions. 
As we will see below, the 
matching is better for the other dispersion relations
 because of the smaller weight of the
S0 wave there.

\booksubsection{3.1.2. $\pi^0\pi^+$ scattering}
 
\noindent
 We have, with $F_{0+}(s)$ the forward $\pi^0\pi^+$ amplitude,
$$\real F_{0+}(s)-F_{0+}(4M_{\pi}^2)=
\dfrac{s(s-4M^2_\pi)}{\pi}\pepe\int_{4M_{\pi}^2}^\infty\dd s'\,
\dfrac{(2s'-4M^2_\pi)\imag F_{0+}(s')}{s'(s'-s)(s'-4M_{\pi}^2)(s'+s-4M_{\pi}^2)}.
\equn{(3.3a)}$$
In particular, at the point $s=2M^2_\pi$, this becomes
$$F_{0+}(4M_{\pi}^2)=F_{0+}(2M_{\pi}^2)+D_{0+},\quad
D_{0+}=\dfrac{8M_{\pi}^4}{\pi}
\int_{8M_{\pi}^2}^\infty\dd s\,\dfrac{\imag F_{0+}(s)}{s(s-2M_{\pi}^2)(s-4M_{\pi}^2)}.
\equn{(3.3b)}$$

The calculation is now more precise because $D_{0+}$ is dominated by the P wave, very well
known.  We find, for the dispersive evaluation,
$$
D_{0+}=(12\pm 1)\times 10^{-3}.
\equn{(3.4a)}$$
On the other hand, using directly the explicit parametrizations 
for the partial wave amplitudes we found in \sect~2 one has
$$F_{0+}(4M^2_\pi,0)-F_{0+}(2M^2_\pi,0)=(6\pm16)\times 10^{-3}.
\equn{(3.4b)}$$
For the difference,
$$F_{0+}(4M^2_\pi,0)-F_{0+}(2M^2_\pi,0)-D_{0+}=
(-6\pm17)\times10^{-3}
\equn{(3.4c)}$$
i.e., perfect agreement.

This is a good place to remark that the
 agreement of the values of $F_{0+}(2M^2_\pi,0)$ and 
 $F_{00}(2M^2_\pi,0)$ obtained with our parametrizations, and those 
found evaluating dispersion relations  (Eqs.~(3.2c), (3.4c))  provides  
a nontrivial test of the 
validity of our parametrizations even in regions below threshold, well 
beyond the region were we fitted data.

\topinsert{
\setbox0=\vbox{{\psfig{figure=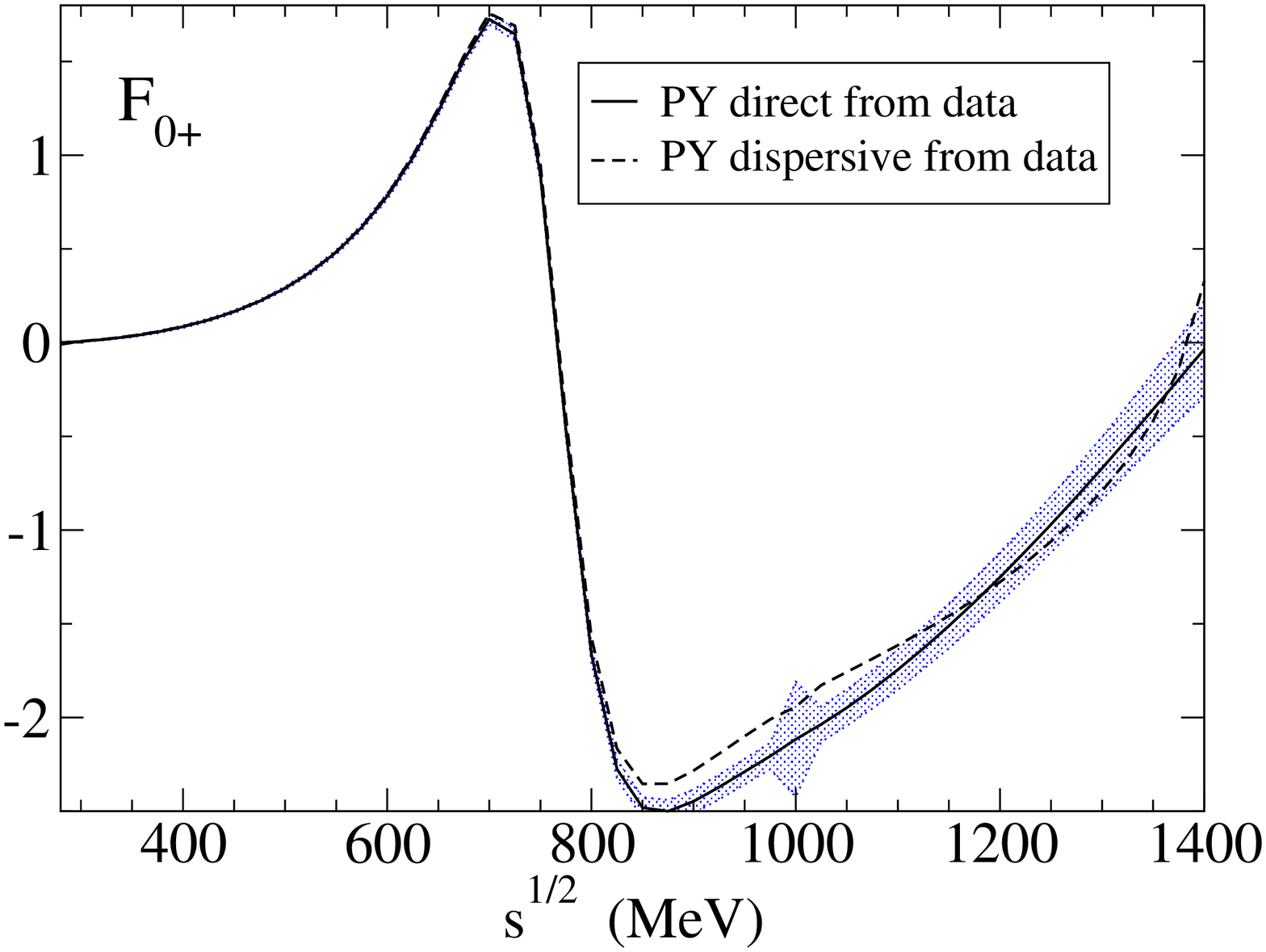,width=9.1truecm,angle=-90}}} 
\setbox6=\vbox{\hsize 15truecm\captiontype\figurasc{Figure 10. }{The combination $\real
F_{0+}(s)-F_{0+}(4M_{\pi}^2)$
  (continuous line) and the dispersive integral 
(broken line).
}\hb} 
\centerline{\tightboxit{\box0}}
\bigskip
\centerline{\box6}
\medskip
}\endinsert

The fulfillment of the dispersion relation (3.3a) is 
shown in Fig.~10 for $s^{1/2}$ below $1.4\,\gev$.
 The agreement is now  good in the
whole range; the average \chidof\ for $s^{1/2}\lsim0.925\,\gev$ is of 1.7.
The fact that the fulfillment of the dispersion relation reaches the 
energy  where the Regge formulas start being valid,  $s^{1/2}\sim1.4\,\gev$, is 
yet another 
 test of the consistency of the Regge analysis with the low energy data. 

To test the dependence of our results on 
the point at which we effect the junction 
between the phase shift analyses and the Regge formulation, 
we have repeated the calculation of the 
forward $\pi^0\pi^+$ dispersion relation 
performing this junction at 1.32~\gev, instead of doing so at 1.42~\gev. 
The fulfillment of the dispersion relation improves slightly 
(below the percent level) at low energy, $s^{1/2}<0.75\,\gev$; 
while it deteriorates a bit more for $s^{1/2}\geq0.75\,\gev$.
The net results are practically unchanged, 
with the choice $s^{1/2}=1.42\,\gev$ for the junction 
slightly favoured. 
We will use this number ($1.42\,\gev$) henceforth.

\brochuresubsection{3.2. The odd amplitude $F^{(I_t=1)}$: dispersion relation  and 
Olsson sum rule (global fit)}

\noindent
We consider first a forward dispersion relation for the amplitude
$F^{(I_t=1)}$ with isospin 1 in the $t$ 
channel, evaluated at threshold. This is known at times as the (first) 
{\sl Olsson sum rule}. 
Expressing $F^{(I_t=1)}(4M^2_\pi,0)$  in terms of the scattering lengths,  
this reads
$$2a_0^{(0)}-5a_0^{(2)}=D_{\rm Ol.},\quad
D_{\rm Ol.}\equiv 3M_\pi\int_{4M_\pi^2}^\infty \dd s\,
\dfrac{\imag F^{(I_t=1)}(s,0)}{s(s-4M_\pi^2)}.
\equn{(3.5)}$$
In terms of isospin in the $s$ channel,
$$F^{(I_t=1)}(s,t)=\tfrac{1}{3}F^{(I_s=0)}(s,t)+\tfrac{1}{2}F^{(I_s=1)}(s,t)-
\tfrac{5}{6}F^{(I_s=2)}(s,t).
$$

\topinsert{
\setbox0=\vbox{{\psfig{figure=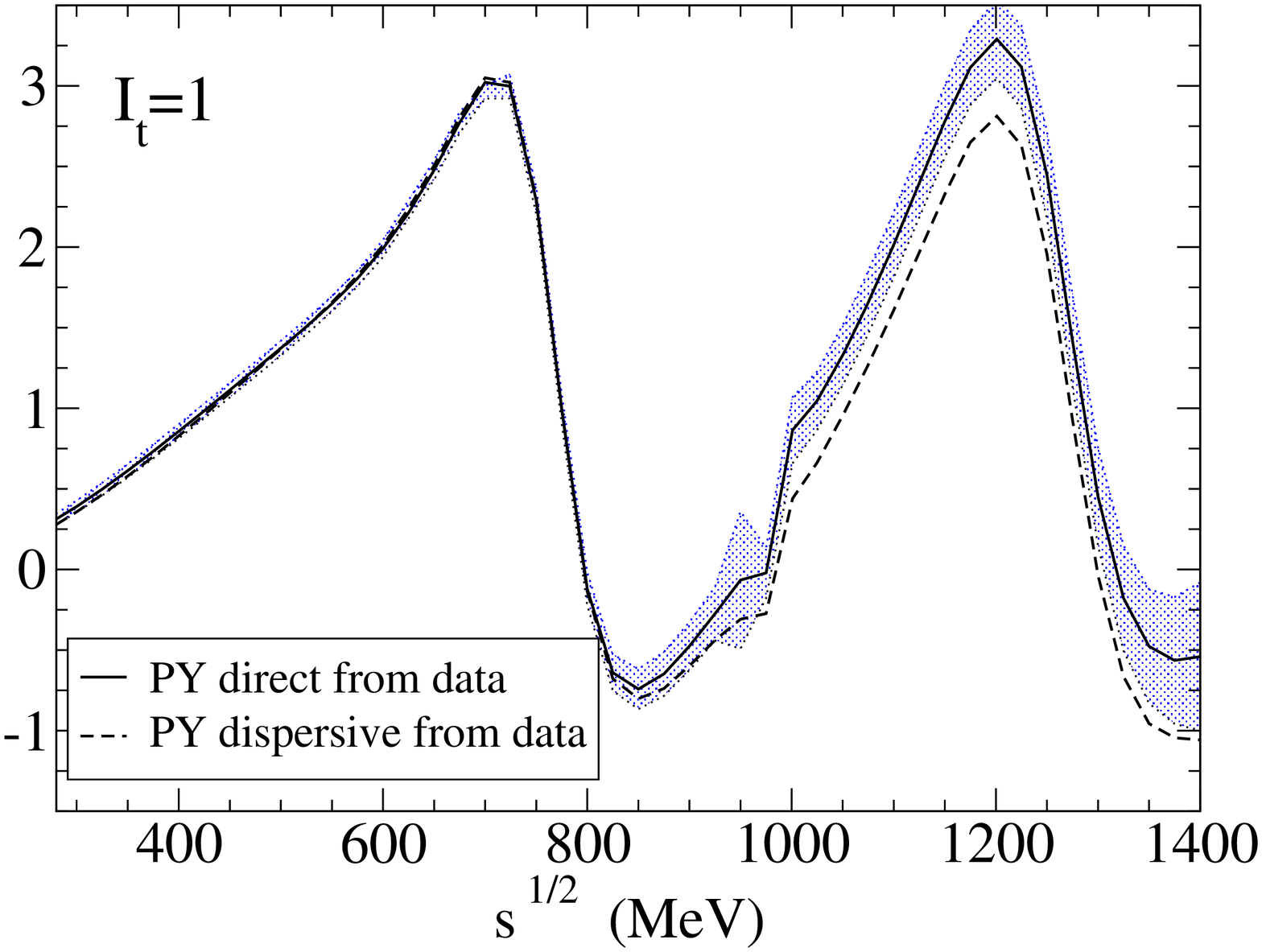,width=9.1truecm,angle=-90}}}
\setbox6=\vbox{\hsize 15truecm\captiontype\figurasc{Figure 11. }{The amplitude $\real
F^{I_t=1}(s,0)$
  (continuous line) and the dispersive integral 
(broken line).
}\hb} 
\centerline{\tightboxit{\box0}}
\bigskip
\centerline{\box6}
\medskip
}\endinsert

Substituting in the right hand side above the  scattering amplitudes
 we have just determined up to
$1.42\,\gev$, and the Regge expression  for rho  exchange of 
Appendix~B at higher energies, we find,
\smallskip
$$D_{\rm Ol.}=0.647\pm0.021.
\equn{(3.6a)}$$
(Here, and in all the numbers for scattering lengths and effective ranges, 
we will take the pion mass $M_\pi$ as unity). 
This is to be compared with what we find from the values of the $a_0^{(I)}$ we found 
in the fits of \sect~2,
$$2a_0^{(0)}-5a_0^{(2)}=0.719\pm0.072.
\equn{(3.6b)}$$

For the difference,
$$2a_0^{(0)}-5a_0^{(2)}-D_{\rm Ol.}=0.073\pm0.077,
\equn{(3.6c)}$$
thus vanishing within errors.

One can also evaluate the corresponding dispersion relation, 
$$\real F^{(I_t=1)}(s,0)=\dfrac{2s-4M^2_\pi}{\pi}\,\pepe\int_{4M^2_\pi}^\infty\dd s'\,
\dfrac{\imag F^{(I_t=1)}(s',0)}{(s'-s)(s'+s-4M^2_\pi)}, 
\equn{(3.7)}$$
evaluating $\real F^{(I_t=1)}(s,0)$ at all values of $s$, either directly using the fits of \sect~2, or 
from the dispersive integral. The agreement, as shown in \fig~11, is very good  
below 1~\gev, and reasonably good above this.

\brochuresubsection{3.3. Dispersion relations using the individual fits to S0 wave data}

\noindent 
In this Subsection we present the results of checking the forward dispersion relations 
using the individual fits for the S0 wave that we performed in \subsect~2.2.3. 
The methods are identical to those used for the solution with the global fit for this wave 
in the previous Subsections, so we will skip details and give only the results, summarized in 
Table~2. 
Here we give the {\sl average}\fnote{That is to say, the sum 
of the $\chi^2$ of each point, spaced at intervals of 25~\mev, divided by the number of points minus the
number of free parameters.}
\chidof\ for the  dispersion relations for the amplitudes that contain the S0 wave:
 $I_t=1$ and $\pi^0\pi^0$.
The values of the parameters $B_i$, ${\mu}_0$ are, of course, 
as in Table~1, but we repeat them here for
ease of reference.

We have separated in Table~2 the 
fits which produce a total  \chidof\ of less than 6, from 
those that give a number larger  than or equal to 6 when running 
the corresponding amplitudes through dispersion relations. 
We may consider that the second set is disfavoured by this test. 
Also, we may repeat some of the comments made in \subsect~2.2.3 
with regard to Solution~E of Grayer et al., and the evaluation 
using $K$ decay data alone: their 
errors are very large, due of course to the small number of points 
they fit, so that their fulfillment of dispersion relations is less 
meaningful than what looks at first sight.

\midinsert
{\medskip
\setbox0=\vbox{
\setbox1=\vbox{\petit \offinterlineskip\hrule
\halign{
&\vrule#&\strut\hfil\ #\ \hfil&\vrule#&\strut\hfil\ #\ \hfil&
\vrule#&\strut\hfil\ #\ \hfil&
\vrule#&\strut\hfil\ #\ \hfil&
#&\strut\hfil\ #\ \hfil&
\vrule#&\strut\hfil\ #\ \hfil&\vrule#&\strut\hfil\ #\ \hfil\cr
 height2mm&\omit&&\omit&&\omit&&\omit&&\omit&&\omit&&\omit&\cr 
&\hfil \hfil&&\hfil $B_0$ \hfil&&\hfil $B_1$\hfil&&\hfil ${\mu}_0$ (\mev)\hfil&&
&&\hfil $\dfrac{\chi^2}{\rm d.o.f.}(I_t=1)$\hfil&&\hfil 
 $\dfrac{\chi^2}{\rm d.o.f.}(\pi^0\pi^0)$\hfil& \cr
 height1mm&\omit&&\omit&&\omit&&\omit&&\omit&&\omit&&\omit&\cr
\noalign{\hrule} 
height1mm&\omit&&\omit&&\omit&&\omit&&\omit&&\omit&&\omit&\cr
&PY, Eq. (2.14)&&\vphantom{\Big|}$21.04\;^{\rm (a)}$&&
$6.62\;^{\rm(a)}$&&$782\pm24$&&& &\hfil$0.3$ \hfil&&
$3.5$& \cr 
\noalign{\hrule}
height1mm&\omit&&\omit&&\omit&&\omit&&\omit&&\omit&&\omit&\cr
&\vphantom{\Big|}$K\; {\rm decay\;  only}$
&&$\phantom{\Big|}18.5\pm1.7$&&$\equiv0$&&$766\pm95$&&&
&\hfil $0.2$  \hfil&&$1.8$& \cr
\noalign{\hrule}
height1mm&\omit&&\omit&&\omit&&\omit&&\omit&&\omit&&\omit&\cr
&\vphantom{\Big|}
${\displaystyle{{ K\; {\rm decay\; data}}}\atop{\displaystyle +\,{\rm Grayer,\;B}}}$&&
$22.7\pm1.6$ &&$12.3\pm3.7 $&&$858\pm15 $&
&&
&$1.0$&&$2.7$& \cr
\noalign{\hrule}
height1mm&\omit&&\omit&&\omit&&\omit&&\omit&&\omit&&\omit&\cr
&\vphantom{\Big|}
${\displaystyle{{ K\; {\rm decay\; data}}}\atop{\displaystyle +\,{\rm Grayer,\;C}}}$&&
$ 16.8\pm0.85$&&$-0.34\pm2.34 $&&$787\pm9 $&
&&
&$0.4$&&$1.0$& \cr
\noalign{\hrule}
height1mm&\omit&&\omit&&\omit&&\omit&&\omit&&\omit&&\omit&\cr
&\vphantom{\Big|}
${\displaystyle{{ K\; {\rm decay\; data}}}\atop{\displaystyle +\,{\rm Grayer,\;E}}}$&&
$21.5\pm3.6 $&&$12.5\pm7.6 $&&$1084\pm110 $&
&&
&$2.1$&&$0.5$& \cr
\noalign{\hrule}
height1mm&\omit&&\omit&&\omit&&\omit&&\omit&&\omit&&\omit&\cr
&\vphantom{\Big|}
${\displaystyle{{ K\; {\rm decay\; data}}}\atop{\displaystyle +\,{\rm Kaminski}}}$&&
$ 27.5\pm3.0$&&$21.5\pm7.4 $&&$789\pm18 $&
&&
&$0.3$&&$5.0$& \cr
\noalign{\hrule}
height1mm&\omit&&\omit&&\omit&&\omit&&\omit&&\omit&&\omit&\cr
\noalign{\hrule}
height1mm&\omit&&\omit&&\omit&&\omit&&\omit&&\omit&&\omit&\cr
&\vphantom{\Big|}
${\displaystyle{{ K\; {\rm decay\; data}}}\atop{\displaystyle +\,{\rm Grayer,\;A}}}$&&
$ 28.1\pm1.1$&&$26.4\pm2.8 $&&$866\pm6 $&
&&
&$2.0$&&$7.9$& \cr
\noalign{\hrule}
height1mm&\omit&&\omit&&\omit&&\omit&&\omit&&\omit&&\omit&\cr
&\vphantom{\Big|}
${\displaystyle{{ K\; {\rm decay\; data}}}\atop{\displaystyle +\,{{\rm EM},\;s{\rm
-channel}}}}$&&
$ 29.8\pm1.3$&&$25.1\pm3.3 $&&$811\pm7 $&
&&
&$1.0$&&$9.1$& \cr
\noalign{\hrule}
height1mm&\omit&&\omit&&\omit&&\omit&&\omit&&\omit&&\omit&\cr
&\vphantom{\Big|}
${\displaystyle{{ K\; {\rm decay\; data}}}\atop{\displaystyle +\,{{\rm EM},\;t{\rm
-channel}}}}$&&
$ 29.3\pm1.4$&&$26.9\pm3.4 $&&$829\pm6 $&
&&
&$1.2$&&$10.1$& \cr
\noalign{\hrule}
height1mm&\omit&&\omit&&\omit&&\omit&&\omit&&\omit&&\omit&\cr
&\vphantom{\Big|}
${\displaystyle{{ K\; {\rm decay\; data}}}\atop{\displaystyle +\,{\rm Protopopescu,\;VI}}}$&&
$ 27.0\pm1.7$&&$22.0\pm4.1 $&&$855\pm10 $&
&&
&$1.2$&&$5.8$& \cr
\noalign{\hrule}
height1mm&\omit&&\omit&&\omit&&\omit&&\omit&&\omit&&\omit&\cr
&\vphantom{\Big|}
${\displaystyle{{ K\; {\rm decay\; data}}}\atop{\displaystyle +\,{\rm Protopopescu,\;XII}}}$&&
$ 25.5\pm1.7$&&$18.5\pm4.1 $&&$866\pm14 $&
&&
&$1.2$&&$6.3$& \cr
\noalign{\hrule}
height1mm&\omit&&\omit&&\omit&&\omit&&\omit&&\omit&&\omit&\cr
&\vphantom{\Big|}
${\displaystyle{{ K\; {\rm decay\; data}}}\atop{\displaystyle +\,{\rm Protopopescu,\;VIII}}}$&&
$ 27.1\pm2.3$&&$23.8\pm5.0 $&&$913\pm18 $&
&&
&$1.8$&&$4.2$& \cr
\noalign{\hrule}}
\vskip.05cm}
\centerline{\box1}
\bigskip
{\noindent\petit
$^{\rm(a)}$ Errors as in \equn{(2.14b)}.\hb
PY, Eq.~(2.14): our 
{\sl global} fit of \subsect~2.2.2. The next rows
show the fits to K decay\ref{13} alone or combined
with $\pi\pi$ scattering data.
Grayer A, B, C, E: the solutions in the 
paper of Grayer et al.\ref{11a}  EM: the
solutions of Estabrooks and  Martin.\ref{11a}  Kaminski refers to the papers of 
Kami\'nski et al.\ref{11c}  Protopopescu VI, XII and VIII:  
the corresponding solutions in ref.~10.}
\medskip
\centerline{\sc Table~2}
\smallskip
\centerrule{5truecm}}
\box0
}
\endinsert

\booksection{4. Improving  the parameters with the help of dispersion relations}

\noindent
In this Section we will show how one can improve the results for the 
fits to the individual waves that we found in \sect~2: 
the fact that the dispersion relations are fulfilled with reasonable accuracy at low energy,
 and that at the $\lsim\,3\;\sigma$ level they still hold 
to higher energies, suggests that we may {\sl improve} the values of the 
parameters we have found with our fits to data requiring also 
better fulfillment of 
such dispersion relations. 
This will provide us with a parametrization of the various waves with central values 
more compatible with analyticity and $s\,-\,u$ crossing. 
This method is an alternative to that of the Roy equations to which it is, 
in principle, inferior in
that  we do not include  $s\,-\,t$ crossing 
(although we check it {\sl a posteriori});
 but it is 
clearly superior in that we do not need as input the 
values of the scattering amplitude for
$|t|$ up to
$30M^2_\pi$, where the various Regge fits  existing in the literature disagree strongly
one with another (see Appendix~B)
 and also in that, with  dispersion relations, we can 
test all energies, whereas the Roy equations are only  valid for
$s^{1/2}<\sqrt{60}\,M_\pi\sim 1.1\,\gev$
 --and, in practice, only applied up to $0.8\,\gev$.

\booksubsection{4.1. Improved parameters for the global fit of \subsect~2.2.2}

\noindent
We will  consider the displacement
 of the central values of the parameters, 
requiring fulfillment, within errors, of all 
three dispersion relations for 
$\sqrt{2}\,M_\pi\leq s^{1/2}\leq0.925\,\gev$ (note that we even 
go {\sl below} threshold),  starting with 
the global solution in Eq.~(2.14). 
We do not fit higher energies because 
 we feel that the errors in the input for some waves 
is too poorly known to give a reliable test there. 
Specifically, the P wave in the region $1.15\,\gev\lsim s^{1/2}\lsim 1.5\,\gev$
 is not at all determined by 
experiment; depending on the fit, a resonance appears, or does not appear, in that region: 
 its mass varies between 
$1.25$ and $1.6$~\gev. Something similar occurs for the 
S0 wave above 0.95~\gev. 
Thus, 
it may well be that the parametrizations we use (which, for example, 
assume no P wave resonance below 1.5~\gev) are biased. 
This could likely be  the explanation 
of the slight mismatch of dispersion relations above 1~\gev, 
 when the corresponding 
contributions begin to become important; 
especially for the {\sl real} parts of the 
scattering amplitudes.

Because  of this uncertainty with the P wave above 1.15~\gev, 
and the S0 wave above 0.95~\gev, which goes 
beyond their nominal errors, and because they are of small importance at low energy, 
we have {\sl not} varied the parameters that describe these waves here.
We have then minimized the sum of $\chi^2$'s obtained from the variation of the 
parameters of the waves, within their errors, as obtained from data, plus 
the average $\chi^2$ of the dispersion relations (that we call ``$\chidof$"). 
This average is obtained evaluating each dispersion 
relation 
at intervals of 25~\mev\ 
in $s^{1/2}$, 
from threshold up to $s^{1/2}=0.925~\mev$, dividing this by the total number of points. 
For the dispersion relations for  $\pi^0\pi^+$ and
 $\pi^0\pi^0$ scattering, we also include in the fit the 
relations (3.1b) and (3.3b), which are important in fixing the location of the 
Adler zeros for the S0, S2 waves.

According to this, we allow variation of
 the parameters of the S0 wave up to $\bar{K}K$ threshold (including the location 
of the Adler zero, $z_0$); the parameters of the P
wave  up to $s^{1/2}=1.0\,\gev$; and the parameters of S2, D0, D2 and F 
waves for all   $s^{1/2}\leq1.42\,\gev$. 
For S2 we also leave free $z_2$.
We find that the  total  variation of the parameters 
has an average $\chi^2$ of $0.38$, 
 showing  
the remarkable stability of our fits. 
The only parameters that have varied by $\sim1\,\sigma$ or a bit more
 are some of the
parameters for the S0 and D2 waves. 
For both, this hardly
affects the low energy shape, but alters them a little 
at medium and higher energies (for D2, see \fig~7). 
Given the low quality of experimental data in the two cases, 
this feature should not be surprising.

As stated above, in the present Subsection 
we take as starting point the S0 wave we obtained with our global fit in 
\subsect~2.2.2.  The new
central values of the parameters, and the scattering length and  effective range 
parameters (both in units of $M_\pi$) are listed below.

$$\eqalign{
{\rm S0};\; s^{1/2}\leq 2m_K:&\quad  B_0=17.4\pm0.5;\quad B_1=4.3\pm1.4;
\cr &\quad {\mu}_0=790\pm21\,\mev;\quad
z_0=195\,\mev\;\hbox{[Fixed]};\cr
&\quad  a_0^{(0)}=0.230\pm0.015;\quad b_0^{(0)}=0.312\pm0.014.\cr
{\rm S2};\; s^{1/2}\leq 1.0:&\quad B_0=-80.8\pm1.7;\quad B_1=-77\pm5;\quad
z_2=147\,\mev\;\hbox{[Fixed]};\cr
&\quad  a_0^{(2)}=-0.0480\pm0.0046;\quad b_0^{(2)}=-0.090\pm0.006.\cr
{\rm S2};\;  1.0\leq s^{1/2}\leq1.42:&\quad  B_0=-125\pm6;\quad B_1=-119\pm14;\quad
\epsilon=0.17\pm0.12.\cr
{\rm P};\; s^{1/2}\leq 1.05:&\quad 
B_0=1.064\pm0.11;\quad B_1=0.170\pm0.040;\quad
M_\rho=773.6\pm0.9\;\mev;\cr
&\quad a_1=(38.7\pm1.0)\times10^{-3};\quad b_1=(4.55\pm0.21)\times10^{-3}.\cr
 {\rm D0};\; s^{1/2}\leq
1.42:&\quad B_0=23.5\pm0.7;\quad B_1=24.8\pm1.0;\quad \epsilon=0.262\pm0.030;\cr
& \quad a_2^{(0)}=(18.4\pm3.0)\times10^{-4};\quad b_2^{(0)}=(-8.6\pm3.4)\times10^{-4}.
\cr  {\rm D2};\; s^{1/2}\leq 1.42:&\quad  
B_0=(2.9\pm0.2)\times10^3;\quad B_1=(7.3\pm0.8)\times10^3;\quad
B_2=(25.4\pm3.6)\times10^3;\cr
&\quad\deltav=212\pm19;\cr
&\quad a_2^{(2)}=(2.4\pm0.7)\times10^{-4};\quad b_2^{(2)}=(-2.5\pm0.6)\times10^{-4}.\cr
{\rm F};\; s^{1/2}\leq 1.42:&\quad B_0=(1.09\pm0.03)\times10^5;\quad
B_1=(1.41\pm0.04)\times10^5;
\quad a_3=(7.0\pm0.8)\times10^{-5}.\cr
&[\;s^{1/2}\leq0.925\;\gev;\quad 
\hbox{total average $\chi^2/{\rm d.o.f.}$=0.80}\;].
\cr}
\equn{(4.1a)}$$

We note that the central values and errors for all the waves, with 
the exception of the S0 and D2 waves (and a little the S2 wave), are almost unchanged
 (in some cases, they are unchanged within our
precision). The results for the S0, S2 and D2 waves are shown in Figs.~13, 14 and 7. 

This brings us to the matter of the errors. 
In general, one cannot  improve much the errors we found 
in \sect~2  
by imposing the dispersion relations since, to begin with, 
they are reasonably well satisfied by our original 
parametrizations; and, indeed, 
 the errors obtained fitting also the 
dispersion relations are 
almost identical to the ones we found in \sect~2. 
Any improvement would thus be purely nominal and would be 
marred by the strong correlations among 
the parameters of the various waves that would be introduced.

\topinsert{
\setbox0=\vbox{\hsize8truecm

\setbox1=\vbox{{\psfig{figure=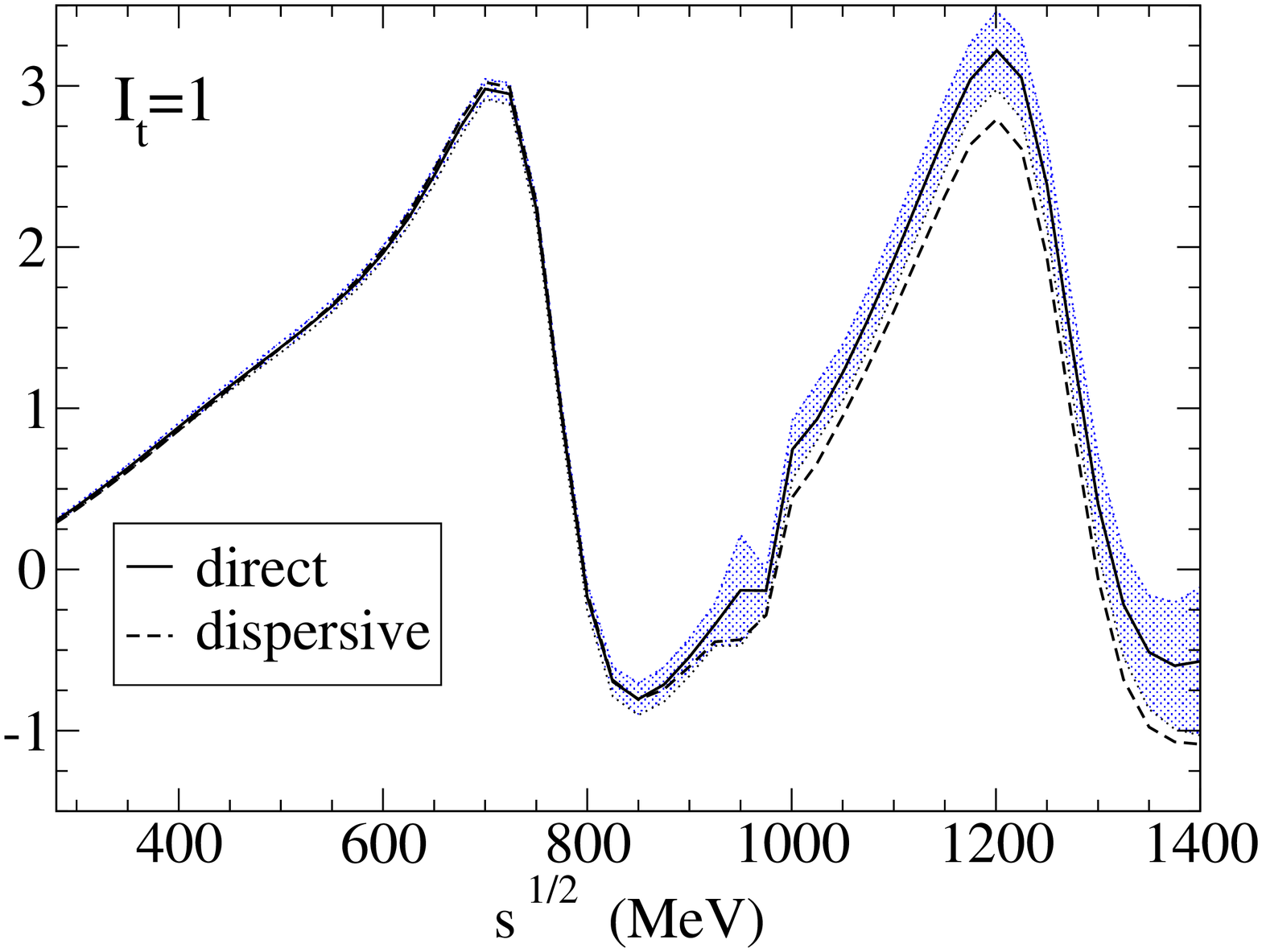,width=9.1truecm,angle=-90}}} 
\setbox2=\vbox{{\psfig{figure=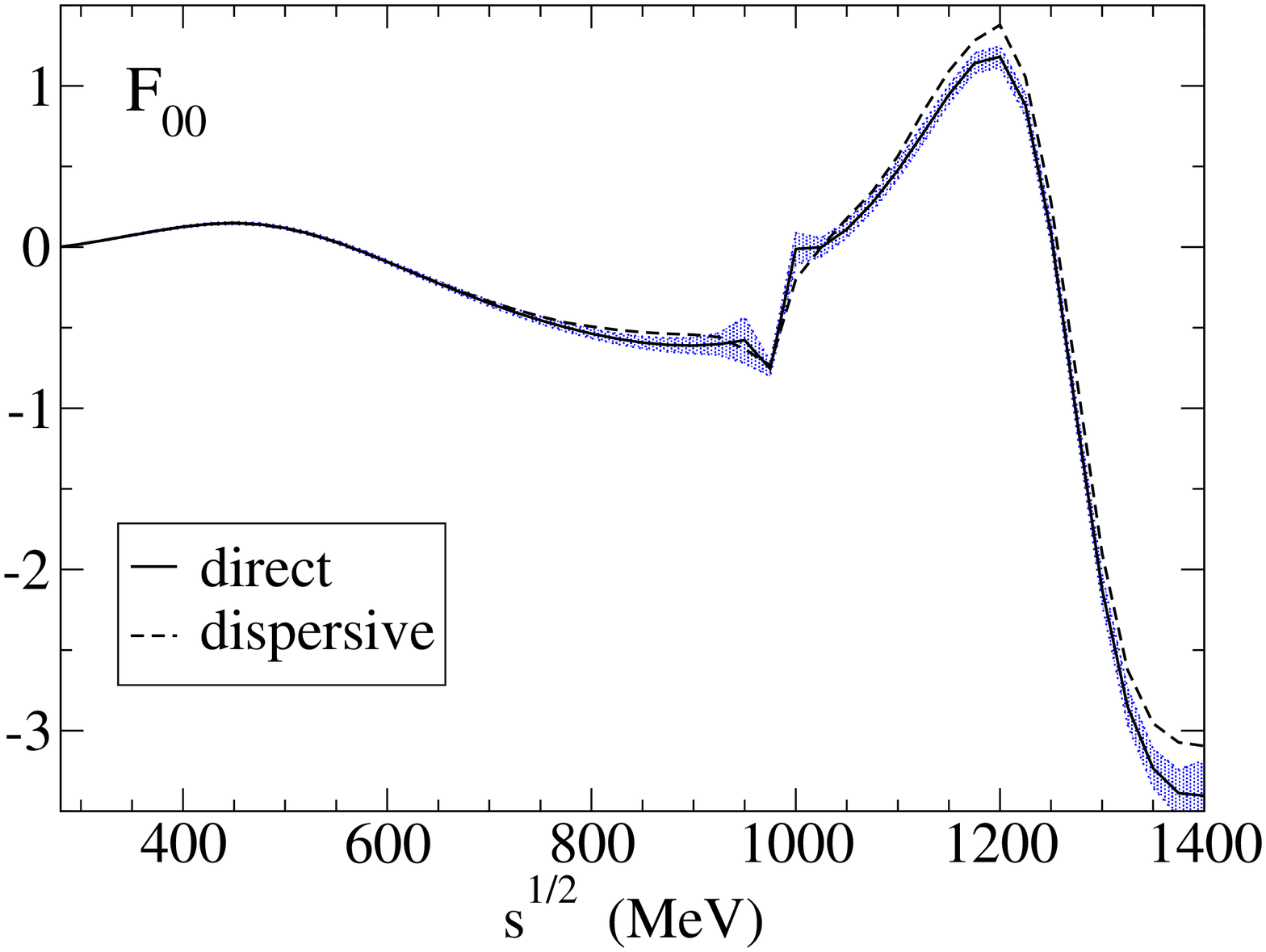,width=9.1truecm,angle=-90}}} 
\setbox3=\vbox{{\psfig{figure=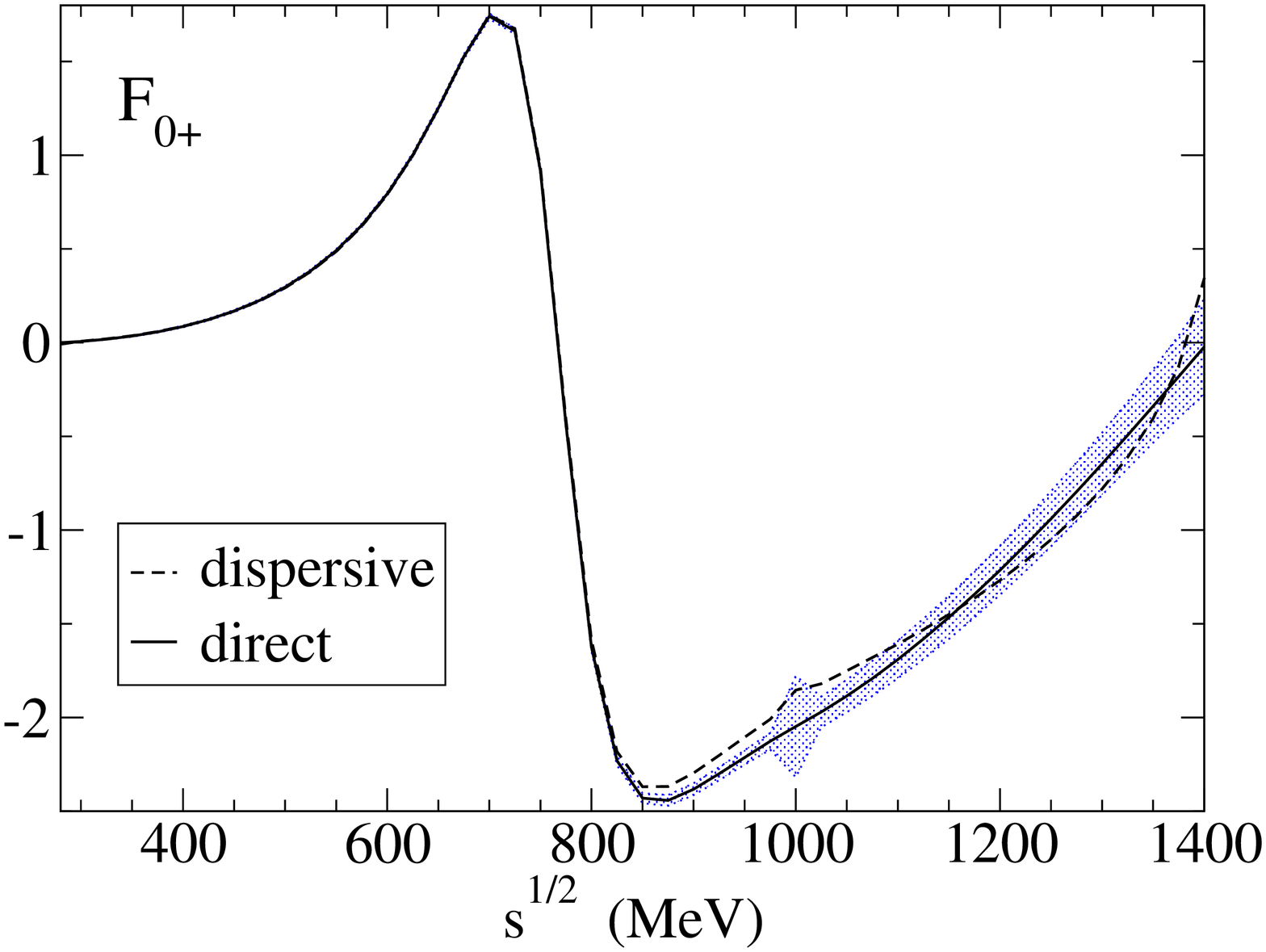,width=9.1truecm,angle=-90}}}
\centerline{\tightboxit{\box2}}
\centerline{\tightboxit{\box3}}
\centerline{\tightboxit{\box1}}}
\setbox6=\vbox{\hsize 15truecm\captiontype\figurasc{Figure 12. }{Fulfillment of 
dispersion
relations,  with the  central parameters  in (4.1a).  The error bands are also shown.}\hb}
\centerline{\ntightboxit{\box0}}
\bigskip
\centerline{\box6}
}\endinsert

The exceptions are, as stated, the S0, S2 and  D2 waves. 
For the first two when including the fulfillment of the dispersion relations in the fits, 
we have first left the Adler zeros, located  at $\tfrac{1}{2}z_0^2$ 
and
$2z_2^2$, free. We find
$$z_0=195\pm21\,\mev,\quad
z_2=147\pm7\,\mev.
\equn{(4.1b)}$$
Unfortunately, the parameters are now strongly correlated so the 
small gain obtained 
would be offset by the complications of dealing with many correlated 
errors.\fnote{There is another reason 
 for not attaching errors to $z_0$; the Adler zero is located 
at $\tfrac{1}{2}z_0^2\sim0.01\,\gev^2$, 
so near the left hand cut that our 
conformal expansion cannot be considered to be convergent there.} 
Because of this we have 
fixed the Adler zeros  at their central values as given in (4.1b) 
when evaluating the errors for the other parameters.
Then the errors are almost uncorrelated.

For the D2 wave we accept the new errors because its
 parameters vary by more than $1\,\sigma$ from those of \sect~2. 
Given the poor quality of experimental data, obvious from a look at \fig~7, 
we feel justified in trusting more the central values and errors buttressed by 
fulfillment of dispersion relations.

We consider (4.1) to provide the best central values 
(and some improved errors) for the parameters shown there. 
The various dispersion relations are fulfilled, up to 
$s^{1/2}=0.925\,\gev$, with an 
average $\chi^2$ of 0.66 (for $\pi^0\pi^0$), of 
 1.62 for the $\pi^0\pi^+$ dispersion relation, and of  0.40 for the $I_t=1$ case.  
The 
consistency of our 
$\pi\pi$ scattering amplitudes that this shows is remarkable, and may 
be seen depicted graphically for the dispersion relations in \fig~12, 
where we show the fulfillment of 
the dispersion relation even up to $s^{1/2}=1.42\,\gev$. 
Mismatch  occurs to more than one unity of \chidof\ 
 due to the 
artificial joining of our low and higher energy fits to data, and, 
above all, to 
the uncertainties of the P and, especially, the S0 wave between 1 and 1.42 \gev.
 
In fact, this mismatch is  small; 
if we take the improved values of the parameters as given in (4.1a), 
and re-calculate the various dispersion relations up to 
$s^{1/2}=1.42\,\gev$, we find  that for $\pi^0\pi^0$ scattering the dispersion relation 
is fulfilled  
with an average $\chi^2$ of 1.85, while for $\pi^0\pi^+$ scattering 
we find an average $\chi^2$ of 1.57 and  the $I_t=1$ dispersion relation is fulfilled 
to an average $\chi^2$ of 1.16. 
However, the last two numbers become smaller than unity if only we increase
the error of the {\sl slope} of the
 P wave between 1 and 1.4~\gev\ by a factor of 2, 
i.e., if we take $\lambda_1=1.1\pm0.4$ in \equn{(2.7)}.
 
Likewise, if we replaced the S0 inelasticity in (2.15b) by that in 
(2.15b$'$), the average $\chi^2$ for $\pi^0\pi^0$ would improve to 
1.35, while   
the  average $\chi^2$ for the $I_t=1$ amplitude would 
become slightly worse,  1.47. 
Doubtlessly, the incompatibility of phase and inelasticity for the S0 
wave above 1~\gev\ precludes a perfect fit.
  We plan, in a coming paper, to use
the results here to  find {\sl consistent} values for the phase shift and inelasticity for 
the S0 and P waves between 1~\gev\ and 1.42~\gev, 
by requiring consistency below the $1\,\sigma$ level of the dispersion relations 
in that energy range.

The agreement of dispersion relations up to 1.42~\gev\ is the more remarkable 
in that the dispersion relations above 0.925~\gev\ have {\sl not} 
been used to improve any wave.

For the sum rules (3.2) and (3.4)  we now find
$$\eqalign{
2a_0^{(0)}-5a_0^{(2)}&-D_{\rm Ol.}=(25\pm32)\times10^{-3},\cr
F_{00}(4M^2_\pi,0)&\,-F_{00}(2M^2_\pi,0)-D_{00}=(-15\pm9)\times10^{-3},\cr
F_{0+}(4M^2_\pi,0)&\,-F_{0+}(2M^2_\pi,0)-D_{0+}=(3\pm7)\times 10^{-3}.
\cr}
\equn{(4.2)}$$

\booksubsection{4.2. Improved parameters for the individual fits of \subsect~2.2.3}

\noindent
We now refine the parameters of the fits to data, but taking as starting 
point the numbers obtained from the individual fits to the various data sets 
as described in \subsect~2.2.3. 
Apart from this, the procedure is identical to that used in \subsect~4.1; 
in particular, we also fit the relations (3.1b) and (3.3b).
The results of the evaluations are presented in Table~3,  
where we  include the solution (4.1), and also
  what we find if requiring  the fit to only 
$K$ decay data for the S0, as given in Eq.~(2.12).

\bigskip
\setbox0=\vbox{
\setbox1=\vbox{\petit \offinterlineskip\hrule
\halign{
&
\vrule#&\strut\hfil\ #\ \hfil&
\vrule#&\strut\hfil\ #\ \hfil&
\vrule#&\strut\hfil\ #\ \hfil&
\vrule#&\strut\hfil\ #\ \hfil&
\vrule#&\strut\hfil\ #\ \hfil&
\vrule#&\strut\hfil\ #\ \hfil&
\vrule#&\strut\hfil\ #\ \hfil&
\vrule#&\strut\hfil\ #\ \hfil&
\vrule#&\strut\hfil\ #\ \hfil\cr
 height2mm&\omit&&\omit&&\omit&&\omit&&\omit&&\omit&&\omit&&\omit&&\omit&\cr 
&\hfil Improved fits: \hfil&&\hfil $B_0$ \hfil&&\hfil $B_1$\hfil&&\hfil ${\mu}_0$ (\mev)\hfil&&
\hfil $z_0$ (\mev)\hfil&&\hfil $\dfrac{\chi^2(I_t=1)}{\rm d.o.f.}$\hfil&&
\hfil $\dfrac{\chi^2(\pi^0\pi^0)}{\rm d.o.f.}$\hfil&&\hfil 
 $\dfrac{\chi^2(\pi^0\pi^+)}{\rm d.o.f.}$\hfil&&
\hfil(3.1b)\hfil& \cr
 height1mm&\omit&&\omit&&\omit&&\omit&&\omit&&\omit&&\omit&&\omit&&\omit&\cr
\noalign{\hrule} 
height1mm&\omit&&\omit&&\omit&&\omit&&\omit&&\omit&&\omit&&\omit&&\omit&\cr
&PY, Eq. (4.1)&&\vphantom{\Big|}$ 17.4\pm0.5$&&$4.3\pm1.4$&&$790\pm21$&&$198\pm21$
&&$0.40$&
&\hfil$0.66$ \hfil&&
$1.62$&&$1.6\;\sigma$& \cr 
\noalign{\hrule}
height1mm&\omit&&\omit&&\omit&&\omit&&\omit&&\omit&&\omit&&\omit&&\omit&\cr
&\vphantom{\Big|}$K\; {\rm decay\;  only}$
&&$\phantom{\Big|}16.4\pm0.9$&&$\equiv0$&&$809\pm53$&&$182\pm34$
&&$0.30$&
&\hfil $0.29$  \hfil&&$1.77$&&$1.5\;\sigma$& \cr
\noalign{\hrule}
height1mm&\omit&&\omit&&\omit&&\omit&&\omit&&\omit&&\omit&&\omit&&\omit&\cr
&\vphantom{\Big|}
${\displaystyle{{ K\; {\rm decay\; data}}}\atop{\displaystyle +\,{\rm Grayer,\;C}}}$&&
$\phantom{\Big|} 16.2\pm0.7$&&$0.5\pm1.8 $&&$788\pm9 $&&$184\pm39 $&
&\hfil $0.37 $ \hfil&
&$0.32$&&$1.74$&&$1.5\;\sigma$& \cr
\noalign{\hrule}
height1mm&\omit&&\omit&&\omit&&\omit&&\omit&&\omit&&\omit&&\omit&&\omit&\cr
\noalign{\hrule}
height1mm&\omit&&\omit&&\omit&&\omit&&\omit&&\omit&&\omit&&\omit&&\omit&\cr
&\vphantom{\Big|}
${\displaystyle{{ K\; {\rm decay\; data}}}\atop{\displaystyle +\,{\rm Grayer,\;B}}}$&&
$\phantom{\Big|}20.7\pm1.0$ &&$11.6\pm2.6 $&&$861\pm14 $&&$233\pm30 $&
&\hfil$0.37$ \hfil&
&$0.83$&&$1.6$&&$4.0\;\sigma$& \cr
\noalign{\hrule}
height1mm&\omit&&\omit&&\omit&&\omit&&\omit&&\omit&&\omit&&\omit&&\omit&\cr
&\vphantom{\Big|}
${\displaystyle{{ K\; {\rm decay\; data}}}\atop{\displaystyle +\,{\rm Grayer,\;E}}}$&&
$ 20.2\pm2.2$&&$8.4\pm5.2 $&&$982\pm95 $&&$272\pm50 $&
&\hfil $0.60 $ \hfil&
&$0.09$&&$1.4$&&$6.0\;\sigma$& \cr
\noalign{\hrule}
height1mm&\omit&&\omit&&\omit&&\omit&&\omit&&\omit&&\omit&&\omit&&\omit&\cr
&\vphantom{\Big|}
${\displaystyle{{ K\; {\rm decay\; data}}}\atop{\displaystyle +\,{\rm Kaminski}}}$&&
$ 20.8\pm1.4$&&$13.6\pm3.7 $&&$798\pm17 $&&$245\pm39 $&
&\hfil $0.43 $ \hfil&
&$1.08$&&$1.36$&&$4.5\;\sigma$& \cr
\noalign{\hrule}}
\vskip.05cm}
\centerline{\box1}
\bigskip
{\noindent\petit
PY,~Eq.~(4.1):  our 
{\sl global} fit, improved with forward dispersion relations, as explained in \subsect~4.1. 
Grayer A, B, C, E means that we take, as experimental low energy data for the S0 wave,  
  the solutions in  
Grayer et al.,\ref{11a}  fitted as shown in Table~1. 
Kaminski means we have used the data of Kami\'nski et al.\ref{11c}
In  these fits we have improved the parameters 
requiring fulfillment of dispersion relations. 
Although errors are given for the Adler zero, we have fixed it at its central 
value when evaluating other errors. 
We have included the fulfillment of the sum rule (3.1b) in the last column, but not 
 (3.3b) and (3.5), which are verified within $1\;\sigma$ by all 
solutions.
}
\smallskip
\centerline{\sc Table~3}
\smallskip
\centerrule{5truecm}}
\box0

\topinsert{
\setbox0=\vbox{{\psfig{figure=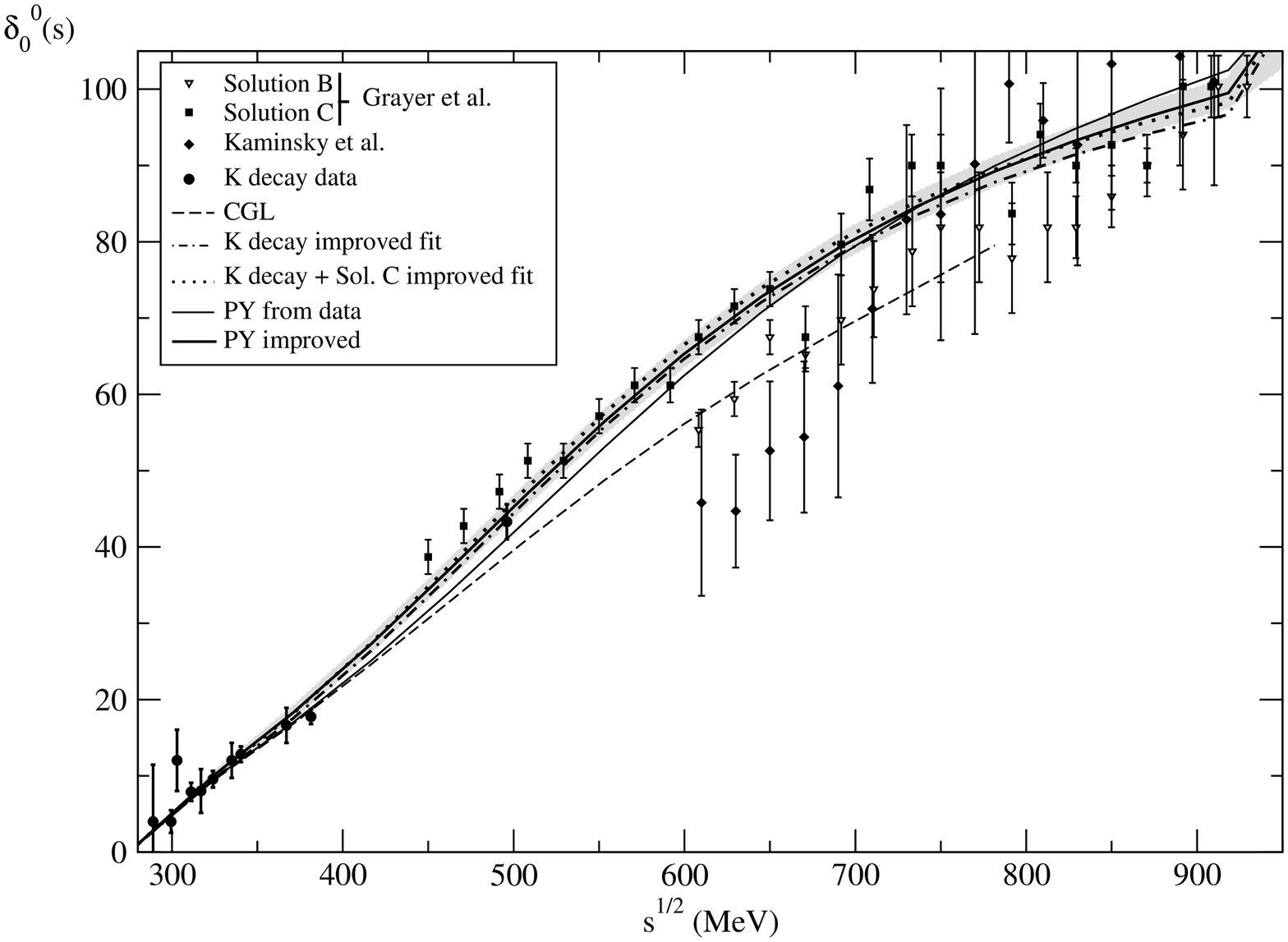,width=12.4truecm,angle=-90}}}
\setbox6=\vbox{\hsize 15.0truecm\captiontype\figurasc{Figure 13. }{
The S0 phase shift 
 corresponding to Eq.~(4.1) (PY, thick continuous
line  
and error band), the unimproved solution of Eq.~(2.14) (thin continuous line),
 and the 
{\sl improved} solutions ``$K$ decay only" and
 ``Grayer~C" of Table~3 (difficult to see as they
fall almost on top of  PY). 
The solution CGL\ref{2} (lowest discontinuous line) is also shown.}} 
\centerline{\tightboxit{\box0}}
\medskip
\centerline{\box6}
\medskip
}\endinsert

The  following comments are in order. 
First of all, we have the remarkable convergence of 
the first three solutions in Table~3; 
and even the last three solutions approach our evaluation, PY, Eq.~(4.1). 
This convergence  is not limited to the S0 wave: 
the parameters for all waves other
 than the S0 agree, within $\lsim1\,\sigma$, 
for all solutions in Table~3, with the numbers given for the PY solution in Eq.~(4.1).
For this reason we do not give in detail, for each individual fit,
 the improved solutions for waves other than the S0 wave.
Of course, the values of the $\chi^2$'s given in Table~3 have been 
evaluated with the corresponding {\sl improved} solutions 
for all waves.

Secondly, the fulfillment of dispersion relations is similar 
for all solutions in Table~3. 
However, the solutions based on Grayer~B, Grayer~E, and that based on the data of Kami\'nski et al., 
fail to satisfy the sum rule 
(3.1b) by a large amount even though they 
have large errors; see the last column in Table~3. 
We should, therefore, consider the three solutions PY, 
that based on $K$ decay data only, and the solution 
based on Grayer~C, to be clearly favoured by 
this consistency test.
This is satisfactory in that the solution~C was obtained by Grayer et al.\ref{11a} 
from solution~B by {\sl including} absorption corrections.

Thirdly, we may consider the solution based on $K$ decay data only, and with a single 
parameter $B_0$, to be in the nature of a first approximation, 
and the introduction of the parameter $B_1$ as producing the more accurate solutions 
denoted by 
Grayer~C and PY.  
In this respect, the coincidence of the common parameters of the first three solutions 
in Table~3 within $\lsim2\,\sigma$ is satisfactory. And, indeed, this 
coincidence is even more pronounced. In \fig~13 we show our starting, global fit from 
Eq.~(2.14) together with the improved solution PY,~Eq.~(4.1) 
and the improved solution corresponding to Grayer~C: 
they are all three contained inside the error band of  PY,~Eq.~(4.1), 
with which the solution Grayer~C practically overlaps. 
We also show, in \fig~14, the S2 wave, before and after improvement.

\topinsert{
\setbox0=\vbox{{\psfig{figure=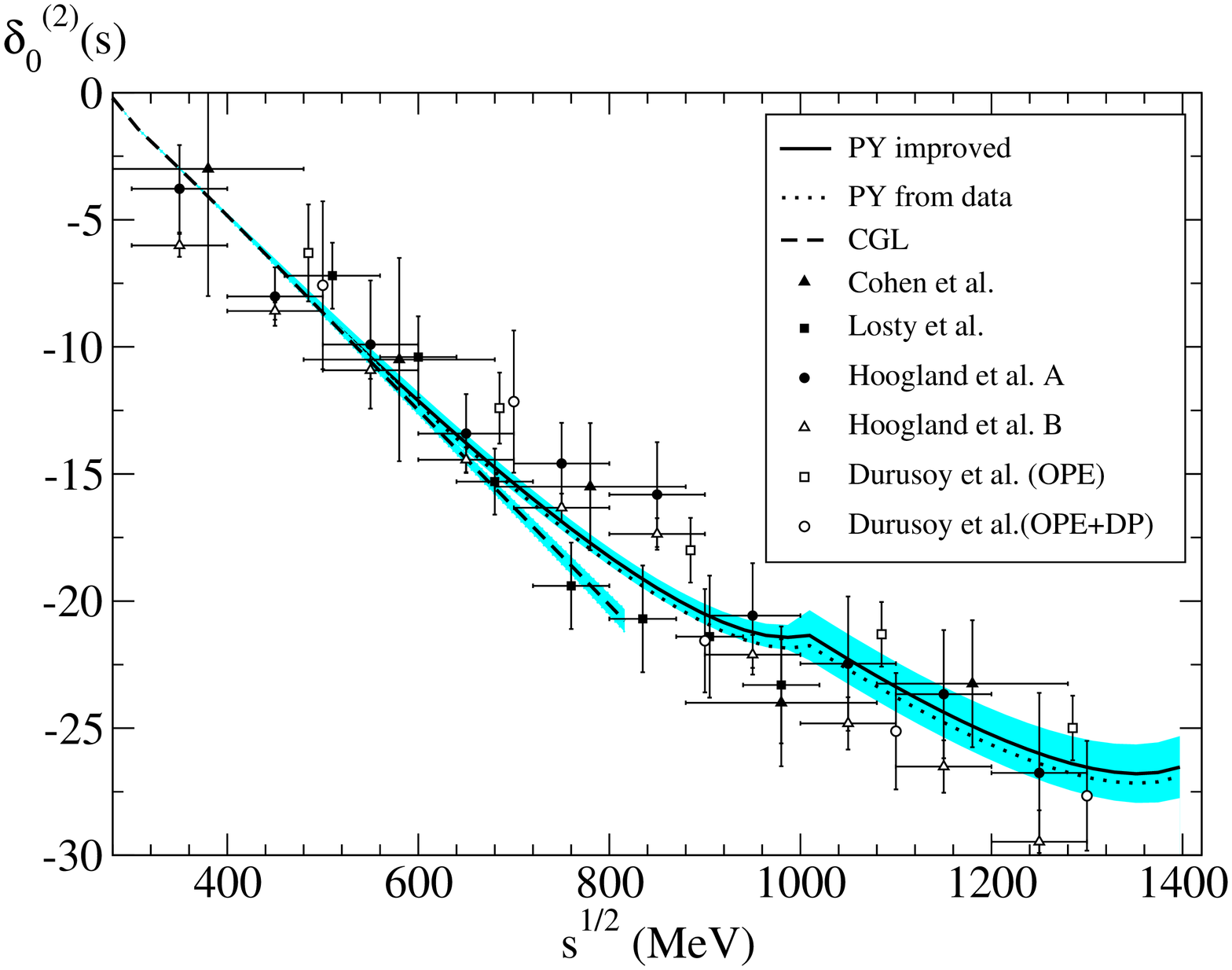,width=11.2truecm,angle=-90}}}
\setbox6=\vbox{\hsize 15truecm\captiontype\figurasc{Figure 14. }{
Phase shift for the S2 wave,  
  Eq.~(4.1) (PY, thick continuous
line, and error band), our unimproved fit, Eq.~(2.8) (dotted line) and 
 the solution CGL,\ref{2} and  error band  (dashed line). 
We  also show the data from Durusoy et al.\ref{12} and solution B in 
Hoogland et al.,\ref{12} not included in the fit.}} 
\centerline{\tightboxit{\box0}}
\medskip
\centerline{\box6}
\medskip
}\endinsert

Finally, and in spite of the virtual coincidence between the 
two improved solutions  Grayer~C and PY,~Eq.~(4.1), we consider the last 
to be preferred: it incorporates data from various experiments and 
is thus less likely to be biased by systematic errors.
Because of all this, we feel confident in considering our solution 
in Eq.~(4.1) to be fully validated, and the method 
we have used to be well tested. 
We will then henceforth accept and work with the solution given in Eq.~(4.1) 
for the improved S0 wave below 0.95~\gev.

\booksection{5. Consistency tests of $s\, -\, t\,-\,u$ crossing: 
two sum rules}

\noindent
In this Section we discuss two  sum rules that follow from crossing 
symmetry; 
in next Section we will consider Froissart--Gribov sum rules, 
which can also be viewed as checks of $s\, -\, t\,-\,u$ crossing. The sum rules we 
discuss now  relate high ($s^{1/2}\geq1.42\,\gev$) and low energy, with the low energy 
given by the P, D, F waves in the region  $s^{1/2}\leq1.42\,\gev$
 and the high energy 
is dominated  by, 
respectively, the rho and Pomeron Regge trajectories. 
The interest in checking them is that they were given by the authors in refs.~1,~2 
 (following 
Pennington\ref{15}) as the reason for the  incorrect Regge parameters. 
Contrarily to the assertions in these references, however,
 we will here check, once again,
 that there is perfect consistency provided one uses standard Regge behaviour for
energies  above 1.42 \gev, and accurate representations of the 
experimental partial waves below  1.42 \gev, such as the ones found here in
 \sects~2,~4 and collected in 
 Appendix~A. 
We should perhaps remark here that the contributions of the S0, S2 waves 
cancel in both sum rules, hence 
we do not even need to worry 
about which solution to use for the S0 wave.

The first sum rule is obtained by profiting from the threshold behaviour to 
write an unsubtracted forward dispersion relation for the 
quantity $F^{(I_s=1)}(s,0)/(s-4M^2_\pi)$. 
One gets
$$\dfrac{6 M_\pi}{\pi}a_1=\dfrac{1}{\pi}\int_{M^2_\pi}^\infty\dd s\,
\dfrac{\imag F^{(I_s=1)}(s,0)}{(s-4M^2_\pi)^2}+\dfrac{1}{\pi}\sum_IC^{(su)}_{1I}
\int_{M^2_\pi}^\infty\dd s\,
\dfrac{\imag F^{(I)}(s,0)}{s^2},
\equn{(5.1)}$$
 which is known at times as the (second) Olsson sum rule; see e.g. 
the textbook of Martin et al.\fnote{Martin, B. R., Morgan, D., and and Shaw, G.   {\sl 
Pion-Pion Interactions in Particle Physics}, Academic Press, New~York (1976).}
The index $I$ refers to isospin in the $s$ channel and $C^{(su)}_{1I}$ are the $s-u$ 
crossing matrix elements.
Canceling $a_1$ with the Froissart--Gribov expression  for this quantity (cf.~\sect~6), 
and substituting the $C^{(su)}_{1I}$, we find the result
$$\eqalign{
I\equiv&\, \int_{M^2_\pi}^\infty\dd s\,
\dfrac{\imag F^{(I_t=1)}(s,4M^2_\pi)-\imag F^{(I_t=1)}(s,0)}{s^2}-
 \int_{M^2_\pi}^\infty\dd s\,\dfrac{8M^2_\pi[s-2M^2_\pi]}{s^2(s-4M^2_\pi)^2}
\imag F^{(I_s=1)}(s,0)=0.\cr
}
\equn{(5.2)}$$

The contributions of the S waves cancel in (5.2), so only the P, D,  F and G waves 
contribute. 
At high energy the integrals are dominated by rho exchange.
We take the improved central values of the 
parameters for the different waves as given in \equn{(4.1)}.
In  units of the pion mass we get
$$I=(-0.12\pm1.27)\times10^{-5},
\equn{(5.3)}$$
that is to say, perfect consistency.

The second sum rule we discuss is that given in Eqs.~{(B.6), (B.7)} of ref.~1.
 It reads,
$$J\equiv\int_{4M^2_\pi}^\infty\dd s\,\Bigg\{
\dfrac{4\imag F'^{(0)}(s,0)-10\imag F'^{(2)}(s,0)}{s^2(s-4M^2_\pi)^2}
-6(3s-4m^2_\pi)\,\dfrac{\imag F'^{(1)}(s,0)-\imag F^{(1)}(s,0)}{s^2(s-4M^2_\pi)^3}
\Bigg\}=0.
\equn{(5.4)}
$$
Here $F'^{(I)}(s,t)=\partial F^{(I)}(s,t)/\partial\cos\theta$, 
and the  upper indices  refer to isospin in the $s$ channel.
We  get, again with the improved 
parameters and with $M_\pi=1$,
$$J=(-0.2\pm4.2)\times10^{-3}. 
\equn{(5.5)}$$ 

These are not the only crossing sum rules that our $\pi\pi$
 amplitude verifies; the coincidence of the 
values for the parameters $a_1$, $b_1$, $b_2^{(I)}$ obtained from direct fits to data 
in \sect~2 with those from the Froissart--Gribov projection, that 
involves simultaneously $s,\,u$ and $t$ crossing,  
 are  highly nontrivial ones. We will see this in next Section.

\booksection{6.  Low energy parameters for waves with $l\geq1$ from the 
Froissart--Gribov\hb projection, and 
a new sum rule for $b_1$}
\vskip-0.5truecm
\booksubsection{6.1. The Froissart--Gribov representation for $a_1,\,b_1$}

\noindent
The quantities $a_1$, $b_1$ may be evaluated in terms of the $I_t=1$ amplitude 
using the Froissart--Gribov representation (for more details on the 
 Froissart--Gribov representation, see~refs.~6,~16 and work quoted there). 
For, e.g. $a_l$ with $l=$odd, we have
$$2a_l^{(I=1)}=\dfrac{\sqrt{\pi}\,\gammav(l+1)}{4M_{\pi}\gammav(l+3/2)}
\int_{4M_{\pi}^2}^\infty\dd s\,\dfrac{\imag F^{(I_t=1)}(s,4M_{\pi}^2)}{s^{l+1}}.
\equn{(6.1)}$$
We will use the parametrizations in (4.1) for the imaginary 
parts of the scattering amplitudes in {\sl all}
 the Froissart--Gribov integrals, and the Regge expressions in
Appendix~B here at high energy.  We find the following results, in units of
   $M_\pi$:

$$\matrix{
&
\hbox{F.--G.}&\hbox{Eq. (4.1)}\cr
10^3\times a_1:\;&37.7\pm1.3&38.7\pm1.0.\cr
}
\equn{(6.2)}$$
We here compare the result obtained with (6.1), denoted by F.-G., and the value  
for $a_1$ which we found in our improved fits, as  given in \equn{(4.1a)}. 
Because the two determinations are 
essentially independent, we can compose the errors to get a  precise and reliable
value for 
$a_1$:
$$a_1=(38.4\pm0.8)\times 10^{-3}\,M_{\pi}^{-3}.
\equn{(6.3)}$$

For the quantity $b_1$ we have,  with the same conventions as before,
$$\matrix{
&
\hbox{F.--G.}&\hbox{Eq. (4.1)}\cr
10^3\times b_1:\;&4.69\pm0.98&4.55\pm0.21.\cr
}
\equn{(6.4)}$$
The Froissart--Gribov integral here is dominated by the rho Regge pole, as the low energy 
contributions cancel almost completely.  
The agreement between these two determinations of $b_1$ is, therefore,
 a highly nontrivial test of the
consistency  of the high and low energy parts of our pion-pion scattering amplitude, 
unfortunately not very precise because of the large error in 
the Froissart--Gribov determination. 
We will discuss this further at the end of Appendix~B.

The F wave scattering length may be similarly 
evaluated; we find
$$a_3=(6.3\pm0.4)\times10^{-5}\;M_\pi^{-7}.
\equn{(6.5)}$$

\brochuresubsection{6.2.  The 
Froissart--Gribov 
projection for even amplitudes: the $a_2^{(I)},\,b_2^{(I)}$ parameters}

\noindent
We first  
calculate the two combinations of scattering lengths
 $a(0+)=\tfrac{2}{3}[a_2^{(0)}-a_2^{(2)}]$ and 
$a(00)= \tfrac{2}{3}[a_2^{(0)}+2a_2^{(2)}]$. 
They correspond to the $s-$channel amplitudes
$$F_{\pi^0\pi^+}=\tfrac{1}{2}F^{(I_s=1)}+\tfrac{1}{2}F^{(I_s=2)},\quad
F_{\pi^0\pi^0}=\tfrac{1}{3}F^{(I_s=0)}+\tfrac{2}{3}F^{(I_s=2)}
$$
for which (as mentioned before) errors are minimized.

The dominant high energy part in the Froissart--Gribov 
representation is given now by the Pomeranchuk 
trajectory  and its importance 
is smaller than previously because the integrals converge faster.
We  find, in units of $M_\pi$,
\smallskip
$$a(0+)=(10.61\pm0.14)\times10^{-4}\,M^{-5}_\pi;\quad
a(00)=(16.17\pm0.75)\times10^{-4}\,M^{-5}_\pi.
\equn{(6.6)}$$
For the  effective range parameters,  
\smallskip
$$
b(0+)=(-0.183\pm0.061)\times10^{-4}\,M^{-7}_\pi;\quad
 b(00)=(-7.96\pm0.57)\times10^{-4}\,M^{-7}_\pi.
\equn{(6.7)}$$

\booksubsection{6.3. A new sum rule for $b_1$}

\noindent
The Froissart--Gribov representation for the effective range $b_1$ 
 depends  strongly on the Regge parameters for  rho exchange, 
and is affected by large errors. 
Here we will present a sum rule that, contrarily, depends almost entirely on 
the low energy scattering amplitudes and is much more precise. 
It is obtained in a way similar to that used for the first 
crossing sum rule in  Section~5. 
We now write a dispersion relation for the quantity
$$\dfrac{\partial}{\partial s}\,\left(\dfrac{F^{(I_s=1)}(s,0)}{s-4M^2_\pi}\right),$$
which we evaluate at threshold. 
Taking into account that
$$\dfrac{\partial}{\partial s}\,\left(\dfrac{F^{(I_s=1)}(s,0)}{s-4M^2_\pi}
\right)_{s=4M^2_\pi}=\dfrac{3M_\pi}{2\pi}b_1,$$
we obtain a  fastly convergent relation for $b_1$:

$$\eqalign{M_\pi b_1=&\,
=\tfrac{2}{3}\int_{4M^2_\pi}^\infty\dd s\,\Bigg\{
\tfrac{1}{3}\left[\dfrac{1}{(s-4M^2_\pi)^3}-\dfrac{1}{s^3}\right]\imag F^{(I_t=0)}(s,0)
+\tfrac{1}{2}\left[\dfrac{1}{(s-4M^2_\pi)^3}+\dfrac{1}{s^3}\right]\imag F^{(I_t=1)}(s,0)\cr
-&\,\tfrac{5}{6}\left[\dfrac{1}{(s-4M^2_\pi)^3}-
\dfrac{1}{s^3}\right]\imag F^{(I_t=2)}(s,0) \Bigg\}.
\cr}
\equn{(6.8)}$$
Most of the contribution to $b_1$ comes from the S0 and P waves at low energy, while  
all other contributions (in particular, the Regge contributions) 
are substantially smaller than $10^{-3}$.  Adding all pieces we find
$$b_1=(4.99\pm0.21)\times10^{-3}\;M_{\pi}^{-5},
\equn{(6.9)}$$
a value reasonably compatible with what we  found  in 
(4.1), $b_1=(4.55\pm0.21)\times10^{-3}\;M_{\pi}^{-5}$
 (because of correlations, the distance is actually below
$1\,\sigma$). We can combine both and find a precise estimate,
$$b_1=(4.75\pm0.16)\times10^{-3}\;M_{\pi}^{-5}.
\equn{(6.10)}$$

\booksection{7. Comparison with the low energy parameters of CGL, DFGS and KLL}

\noindent
We   
present in Table~4 a global comparison of the low energy parameters
 as given here, that we
denote by 
 PY, as
well as   recent  evaluations, that use  the Roy equations, by
Colangelo, Gasser and Leutwyler\ref{2} (CGL),
 by Descotes et al.\ref{8}, that
we denote by DFGS, and by Kami\'nski, Le\'sniak and
Loiseau\ref{8}, denoted by KLL. 
Besides scattering lengths and effective range parameters, we give the quantities
 $[a_0^{(0)}-a_0^{(2)}]^2$ and $\delta_0^{(0)}(m^2_K)-\delta_0^{(2)}(m^2_K)$,
relevant, respectively, for pionic atom decays and CP violating kaon decays.

\midinsert{
\setbox0=\vbox{\petit
\setbox1=\vbox{ \offinterlineskip\hrule
\halign{
&\vrule#&\strut\hfil\ #\ \hfil&\vrule#&\strut\hfil\ #\ \hfil&
\vrule#&\strut\hfil\ #\ \hfil&
\vrule#&\strut\hfil\ #\ \hfil&\vrule#&\strut\hfil\ #\ \hfil\cr
 height2mm&\omit&&\omit&&\omit&&\omit&&\omit&\cr 
&\hfil \hfil&&\hfil DFGS \hfil&&KLL&
&\hfil CGL\hfil&
&\hfil PY\hfil& \cr
 height1mm&\omit&&\omit&&\omit&&\omit&&\omit&\cr
\noalign{\hrule} 
height1mm&\omit&&\omit&&\omit&&\omit&&\omit&\cr
&$a_0^{(0)}$&&\vphantom{\Big|}$0.228\pm0.032$&&$0.224\pm0.013$&
&\hfil$0.220\pm0.005$ \hfil&&
$0.230\pm0.015$& \cr 
\noalign{\hrule}
height1mm&\omit&&\omit&&\omit&&\omit&&\omit&\cr
&$a_0^{(2)}$&&\vphantom{\Big|}$-0.0382\pm0.0038$&&$-0.0343\pm0.0036$&
&\hfil$-0.0444\pm0.0010$ \hfil&&
$-0.0480\pm0.0046$& \cr 
\noalign{\hrule}
height1mm&\omit&&\omit&&\omit&&\omit&&\omit&\cr
&$\big[a_0^{(0)}-a_0^{(2)}\big]^2$&&\vphantom{\Big|}$0.071\pm0.018 $&&$0.067\pm0.007 $&
&\hfil$0.070\pm0.003 $ \hfil&&
$0.077\pm0.008$& \cr 
\noalign{\hrule}
height1mm&\omit&&\omit&&\omit&&\omit&&\omit&\cr
&$\delta_0^{(0)}(m^2_K)-\delta_0^{(2)}(m^2_K)$&&\vphantom{\Big|}&
&&
&\hfil$47.7\pm1.5\degrees$ \hfil&&
$52.9\pm1.6\degrees$& \cr 
\noalign{\hrule}
height1mm&\omit&&\omit&&\omit&&\omit&&\omit&\cr
&$b_0^{(0)}$&&\vphantom{\Big|}&&$0.252\pm0.011$&
&\hfil$0.280\pm0.001$ \hfil&&
$0.312\pm0.014$& \cr 
\noalign{\hrule}
height1mm&\omit&&\omit&&\omit&&\omit&&\omit&\cr
&$b_0^{(2)}$&&\vphantom{\Big|}&&$-0.075\pm0.015$&
&\hfil$-0.080\pm0.001$ \hfil&&
$-0.090\pm0.006$& \cr 
\noalign{\hrule} 
height1mm&\omit&&\omit&&\omit&&\omit&&\omit&\cr
&$a_1$&&\vphantom{\Big|}&&$39.6\pm2.4$&
&\hfil${\displaystyle 37.9\pm0.5}$ \hfil&&
$38.4\pm0.8\quad(\times\,10^{-3})$& \cr 
\noalign{\hrule}
height1mm&\omit&&\omit&&\omit&&\omit&&\omit&\cr
&\vphantom{\Big|}$b_1$ 
\phantom{\big|}&&\phantom{\Big|}&&$2.83\pm0.67$&
&\hfil ${\displaystyle5.67\pm0.13}
\vphantom{\big|}$  
\hfil&&\hfil 
$4.75\pm0.16\quad(\times\,10^{-3})$& \cr
\noalign{\hrule} 
height1mm&\omit&&\omit&&\omit&&\omit&&\omit&\cr
&$a_2^{(0)}$&&\vphantom{\Big|}&
&\hfil  \hfil&&$17.5\pm0.3$&&$18.70\pm0.41$
$\quad(\times\,10^{-4})$& \cr
\noalign{\hrule} 
height1mm&\omit&&\omit&&\omit&&\omit&&\omit&\cr
&\vphantom{\Big|}$a_2^{(2)}$&&&
&\hfil  \hfil&&$1.70\pm0.13$&&$2.78\pm0.37$
$\quad(\times\,10^{-4})$& \cr
\noalign{\hrule}
height1mm&\omit&&\omit&&\omit&&\omit&&\omit&\cr
&\vphantom{\Big|}$a(0+)$&&&
&\hfil  \hfil&&$\vphantom{\big|}$&&
$10.61\pm0.14$
$\quad(\times\,10^{-4})$& \cr
\noalign{\hrule}
height1mm&\omit&&\omit&&\omit&&\omit&&\omit&\cr
&\vphantom{\Big|}$a(00)$&&&
&\hfil  \hfil&&$\vphantom{\big|}$&&
$16.17\pm0.75$
$\quad(\times\,10^{-4})$& \cr
\noalign{\hrule}
height1mm&\omit&&\omit&&\omit&&\omit&&\omit&\cr
&$b_2^{(0)}$&&\vphantom{\Big|}&
&\hfil  \hfil&&$-3.55\pm0.14$&&$-4.16\pm0.30$
$\quad(\times\,10^{-4})$& \cr
\noalign{\hrule} 
height1mm&\omit&&\omit&&\omit&&\omit&&\omit&\cr
&\vphantom{\Big|}$b_2^{(2)}$&&&
&\hfil  \hfil&&$-3.26\pm0.12$&&$-3.89\pm0.28$
$\quad(\times\,10^{-4})$& \cr
\noalign{\hrule}
height1mm&\omit&&\omit&&\omit&&\omit&&\omit&\cr
&\vphantom{\Big|}$b(0+)$&&$ $&
&\hfil  \hfil&
&$\vphantom{\big|}$&&$-0.183\pm0.061$
$\quad(\times\,10^{-4})$& \cr
\noalign{\hrule}
height1mm&\omit&&\omit&&\omit&&\omit&&\omit&\cr
&\vphantom{\Big|}$b(00)$&&$ $&
&\hfil  \hfil&
&$\vphantom{\big|}$&&$-7.96\pm0.57$
$\quad(\times\,10^{-4})$& \cr
\noalign{\hrule}
height1mm&\omit&&\omit&&\omit&&\omit&&\omit&\cr
&\vphantom{\Big|}$a_3$&&$ $&
&\hfil  \hfil&
&$5.6\pm0.2$&&$6.3\pm0.4$
$\quad(\times\,10^{-5})$& \cr
\noalign{\hrule}}
\vskip.05cm}
\centerline{\box1}
\bigskip
{\noindent\petit
Units of $M_\pi$. The numbers  in the CGL column are   as given by 
CGL in  Table~2 and elsewhere in their text. \hb
In PY, the values for the D, F waves parameters are from the 
Froissart--Gribov representation. 
The rest are from the fits, improved with dispersion relations, except for $a_1$ and
$b_1$ that have been taken from Eqs.~(6.3) and (6.10). }
\medskip
\centerline{\sc Table~4}
\smallskip
\centerrule{5truecm}}
\box0
}\endinsert

The  DFGS solution is compatible, within its errors, 
both with CGL and PY. 
As to the KLL solution, its values for  $a_0^{(I)}$,   $b_0^{(I)}$ 
are compatible with those of DFGS (for the first), somewhat less so 
with the numbers of PY and with what CGL find.
The central value for $a_1$ of KLL is too high, although it is compatible within its 
errors with other determinations. 
The value of $b_1$, however, is  $3\,\sigma$ below the next lowest one 
in the Table~4.  
The reason could be that
Kami\'nski, Le\'sniak and
Loiseau use some approximate 
calculation techniques, like  effective, separable potentials 
and Pad\'e approximants.

\topinsert{
\setbox0=\vbox{\hsize15truecm
\setbox1=\vbox{{\psfig{figure=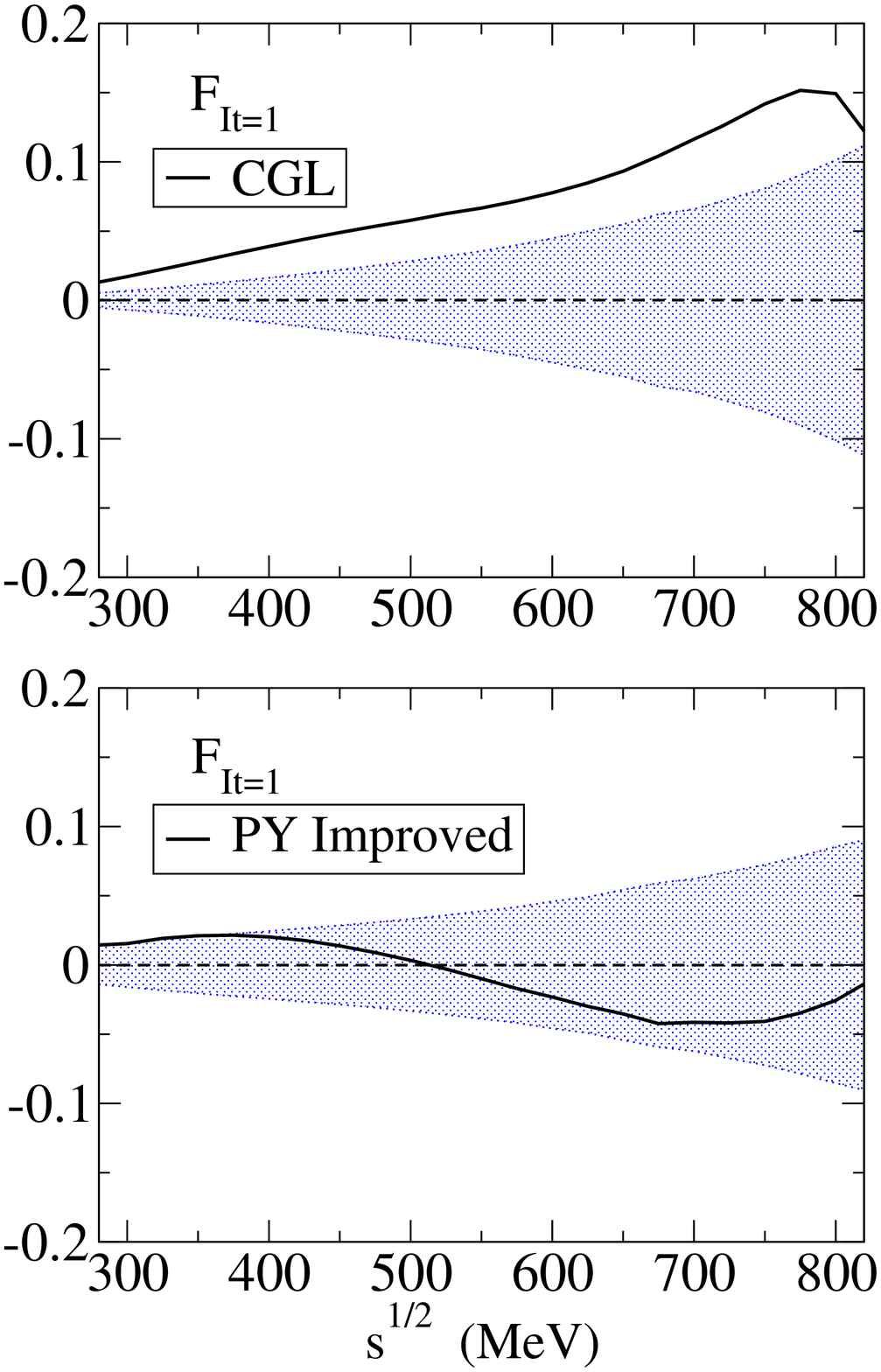,width=6.2truecm,angle=-0}}}
\setbox2=\vbox{{\psfig{figure=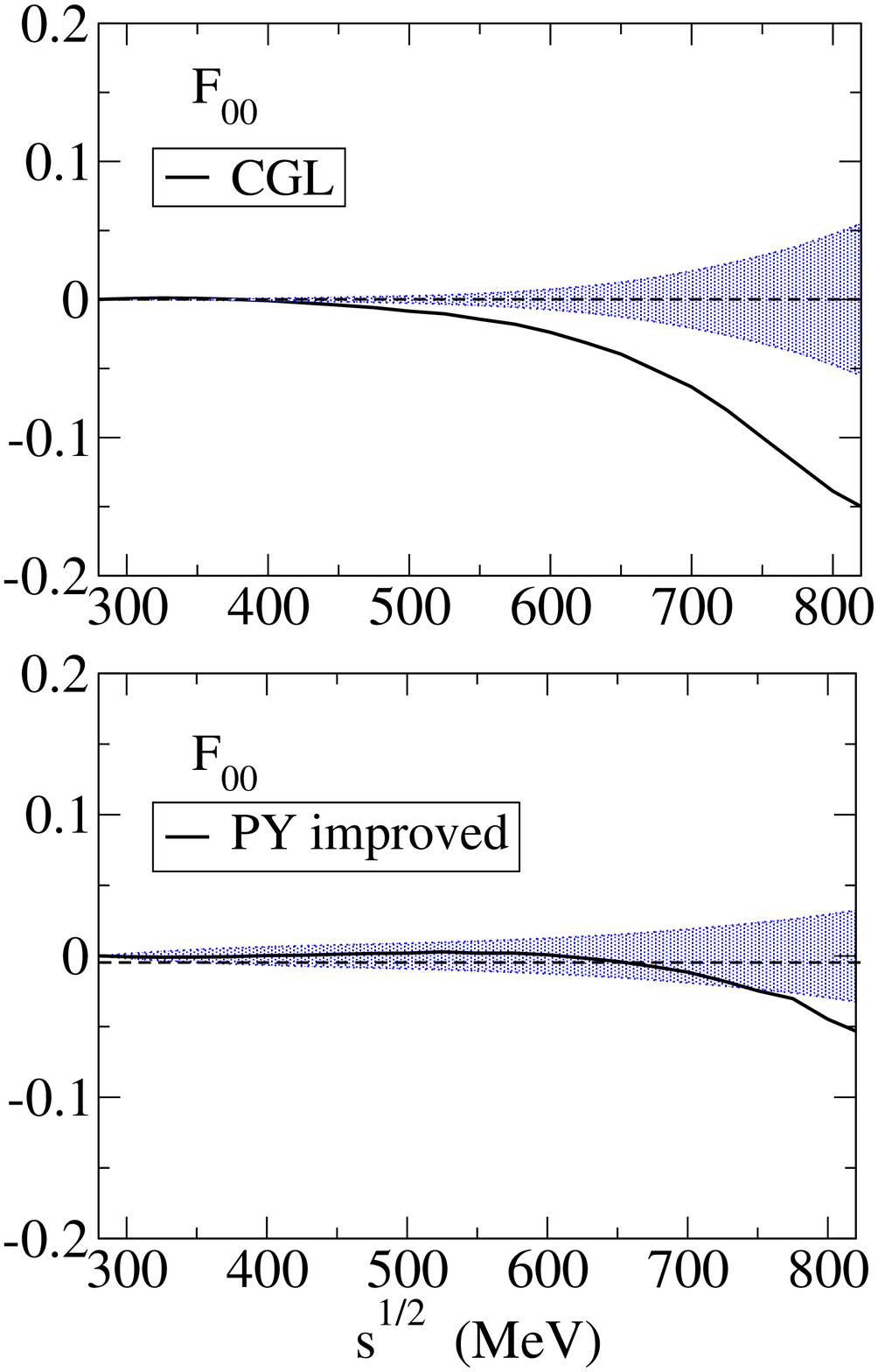,width=6.2truecm,angle=-0}}}
\setbox3=\vbox{{\psfig{figure=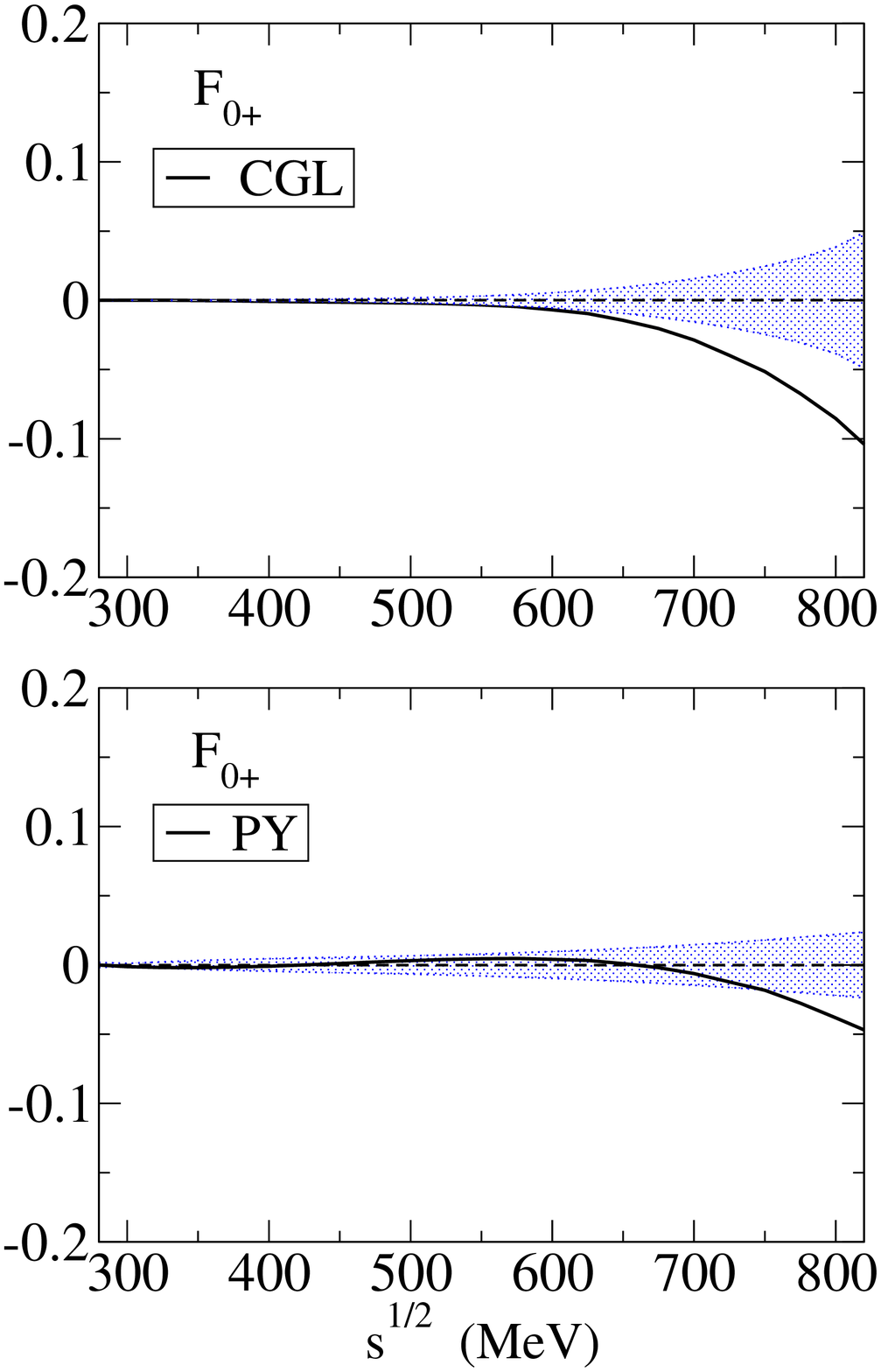,width=6.2truecm,angle=-0}}} 
\line{\tightboxit{\box1}\hfil\tightboxit{\box2}}
\setbox6=\vbox{\hsize 6truecm\captiontype\figurasc{Figure 15. }{Consistency of 
dispersion
relations for the 
$\pi\pi$ amplitudes 
of ref.~2~(CGL) and for our amplitudes, 
with the  parameters in (4.1),
denoted by PY.  We plot the differences $\deltav_i$, given in 
Eqs.~(7.1), between the results of the 
calculation of the real parts directly 
with the various parametrizations given, or from the 
dispersive formulas. Perfect consistency 
would occur if the continuous curves coincided with the dotted lines.
The error bands are also shown. 
The progressive deterioration of the CGL results as the energy
 increases is apparent
here.}\hb
\vskip0.5truecm
\phantom{x}}
\line{\tightboxit{\box3}\hfil{\box6}}}
\centerline{{\box0}}
}\endinsert

The calculation of CGL\ref{2} is different from 
 the others, since  
 CGL {\sl impose} ch.p.t. to two loops. 
However, CGL use Roy equations, which 
require phenomenological input for energies above 0.8~\gev. 
As stated in the Introduction (and discussed in more detail in refs.~6,~7) 
some of this input is not accurate, which may cause biases in the final results. 
In this respect, we find that the predictions of CGL for 
the scattering lengths $a_0^{(I)},\,a_1$ agree very well 
with what we find from experiment. This means that our results do {\sl not}
challenge the standard chiral counting of ch.p.t. However, the solution
given by CGL deviates from our results for quantities like $b_1$,
$a_2^{(I)}$, $b_2^{(I)}$ or the S0 and S2 phase shifts above 500 MeV
(and in particular the value 
used as {\it input} by CGL at the matching point
 $\delta_0^{(0)}(800MeV)=82.3\pm3.4^{\rm o}$).

That the CGL solution 
deteriorates as the energy increases  is quite transparent if we 
compare the fulfillment of dispersion
relations with the parameters of CGL for the S0, S2 and P waves al low energy, 
or with our parameters here.
This is depicted in \fig~15, where we show the mismatch between the real part and the
dispersive 
evaluations, that is to say, the differences $\deltav_i$, 
$$\deltav_{1}\equiv\real
F^{(I_t=1)}(s,0)-\dfrac{2s-4M^2_\pi}{\pi}\pepe\int_{4M^2_\pi}^\infty\dd s'\,
\dfrac{\imag F^{(I_t=1)}(s',0)}{(s'-s)(s'+s-4M^2_\pi)},
\equn{(7.1a)}$$
$$\deltav_{00}\equiv\real F_{00}(s)-F_{00}(4M_{\pi}^2)-
\dfrac{s(s-4M_{\pi}^2)}{\pi}\pepe\int_{4M_{\pi}^2}^\infty\dd s'\,
\dfrac{(2s'-4M^2_\pi)\imag F_{00}(s')}{s'(s'-s)(s'-4M_{\pi}^2)(s'+s-4M_{\pi}^2)},
\equn{(7.1b)}$$
and
$$\deltav_{0+}\equiv\real F_{0+}(s)-F_{0+}(4M_{\pi}^2)-
\dfrac{s(s-4M^2_\pi)}{\pi}\pepe\int_{4M_{\pi}^2}^\infty\dd s'\,
\dfrac{(2s'-4M^2_\pi)\imag F_{0+}(s')}{s'(s'-s)(s'-4M_{\pi}^2)(s'+s-4M_{\pi}^2)}.
\equn{(7.1c)}$$
These quantities would vanish,  $\deltav_i=0$, 
if the dispersion relations were exactly satisfied.
 
We  include in the comparison of \fig~14 the errors; in the case of CGL, 
these errors are as follow from the parametrizations given by these authors in ref.~2, 
for $s^{1/2}\lsim0.8\,\gev$. At higher energies they are taken from experiment via our 
parametrizations.
By comparison, we show the same quantities for our 
best results in the present paper, that is to say, with amplitudes  
improved by use of dispersion relations, \equn{(4.1)}. 
In both cases we have taken the Regge parameters from 
Appendix~B here.

\booksection{8. Summary of conclusions}

\noindent
In the previous Sections we have given a representation of the 
$\pi\pi$ scattering amplitudes obtained fitting experimental data below 
1.42~\gev, supplemented by standard Regge formulas above this energy. 
We have shown that our representations satisfy reasonably well 
 forward dispersion relations at low energy, as well as crossing sum rules. 
We have shown that requiring fulfillment of the dispersion relations 
up to  $s^{1/2}=0.925\,\gev$  leads to a refinement of 
the central values and errors for the 
parameters for the various waves, giving a set of these quantities
  such that dispersion
relations are now satisfied, 
 within acceptable errors, at all energies. 

In particular, for the S0 wave, 
we have analyzed the results found 
starting from different sets of data. 
If we eliminate from those sets the ones that are less consistent 
with the dispersion relations, it is seen that the remaining determinations converge to a solution,
essentially unique, 
when improved by requiring fulfillment of the 
dispersion relations.
 We have, therefore, obtained a complete set of $\pi\pi$ 
scattering amplitudes that are {\sl consistent}, 
 with theoretical requirements as well as 
with experiment: they are collected in  
~Appendix~A.

After this, we use these scattering amplitudes to evaluate low energy parameters 
for P, D0, D2, and F waves in a reliable manner, 
clearly improving on previous work.\ref{1,2,8}
These parameters may then be used to test chiral
 perturbation theory to one and two loops,
or to find quantities relevant for pionium decays or 
CP 
violating kaon decays:
$$[a_0^{(0)}-a_0^{(2)}]^2=0.077\pm0.008\,M^{-2}_\pi,\quad
\delta_0^{(0)}(m^2_K)-\delta_0^{(2)}(m^2_K)=52.9\pm1.6^{\rm o}.$$

We may remark here that our errors are, typically, a factor 2 to 3 times larger than 
those in CGL,\ref{2} {\sl at low energy}. 
Given that the scattering amplitudes of this reference show mismatches 
between high and low energy,
 for many observables
by several standard deviations (cf., for example, \fig~14), and that our 
results are of higher precision at intermediate energies for some quantities 
(compare, for example, the error bars in Figs.~15), we do not
consider our results to be inferior. But  
one certainly can ask the question whether it would be possible to improve on {\sl our} 
precision. The answer is, no in the sense that our amplitudes agree, within errors, 
with theoretical requirements and with data. 
A sizable improvement would require substantially improved experimental data, 
certainly  
 for the S0, S2 waves at low energy, and for the S0 and P waves above $\sim1\,\gev$.
One may think that imposing chiral perturbation theory could lead to 
decreasing the errors of  the 
$\pi\pi$ scattering amplitudes. 
However, the matter is complicated and 
 will be left for a future publication.

\vfill\eject

\booksection{Appendix A: Summary of low energy ($s^{1/2}\leq1.42\,\gev$) partial waves}

\noindent
In this Appendix we collect the best values for the parametrizations of the various
partial waves. 
We give the values obtained after improving with the help of dispersion 
relations, \equn{(4.1)}. 
The values of the parameters and 
errors obtained from raw fits to  data are as given in \sect~2.

\booksubsection{A.1. The S wave with isospin zero below $950$ \mev}

\noindent
We  impose the Adler zero at $s=\tfrac{1}{2}z_0^2$, with 
$z_0$ fixed,
 and a ``resonance" with mass
${\mu}_0$, a free parameter. 
After improving the global fit with dispersion relations we have
$$\eqalign{
\cot\delta_0^{(0)}(s)=&\,\dfrac{s^{1/2}}{2k}\,\dfrac{M_{\pi}^2}{s-\tfrac{1}{2}z_0^2}\,
\dfrac{{\mu}^2_0-s}{{\mu}^2_0}\,
\left\{B_0+B_1\dfrac{\sqrt{s}-\sqrt{s_0-s}}{\sqrt{s}+\sqrt{s_0-s}}\right\};\cr
{B}_0=&\,17.4\pm0.5,\quad {B}_1=4.3\pm1.4,\quad
{\mu}_0=790\pm21\,\mev;\quad z_0=195\pm21\;\mev\;\hbox{[Fixed]}.\cr}
\equn{(A.1)}$$
This fit we take to be valid for $s^{1/2}\lsim0.95\,\gev$. 
Note that we have fixed the value of $z_0$ when evaluating the 
errors of the other parameters.

This gives
$$a_0^{(0)}=(0.230\pm0.015)\; M_{\pi}^{-1};\quad
b_0^{(0)}=(0.312\pm0.014)\;
M_{\pi}^{-3};\quad\delta_0^{(0)}(m_K)=44.4\degrees\pm1.5\degrees.
\equn{(A.2)}$$

\booksubsection{A.2. The $I=0$ S wave between $950\,\mev$ and $1420\,\mev$}

 \noindent
We write
$$\cot\delta_0^{(0)}(s)=c_0\,\dfrac{(s-M^2_s)(M^2_{f}-s)}{M^2_{f} s^{1/2}}\,
\dfrac{|k_2|}{k^2_2},\quad k_2=\dfrac{\sqrt{s-4m^2_K}}{2};
\equn{(A.3)}$$
$$\eta_0^{(0)}=1-\left(\epsilon_1\dfrac{k_2}{s^{1/2}}+\epsilon_2\dfrac{k_2^2}{s}\right)
\,\dfrac{M'^2-s}{s}.
\equn{(A.4)}$$ 
Then,
$$\eqalign{
c_0=&\,1.3\pm0.5,\quad M_f=1320\pm50\,\mev;\quad M_s=920\,\mev\;({\rm fixed});\cr
\epsilon_1=&\,6.4\pm0.5,\quad \epsilon_2=-16.8\pm1.6;\quad M'=1500\,\mev\;({\rm fixed}).\cr
}
\equn{(A.5)}$$
It should be clear that this fit is not very reliable since there are several incompatible
sets 
of experimental data. Thus, the errors  are 
purely nominal. 
If using $\pi\pi\to\bar{K}K$ data to fit the inelasticity we would have 
obtained different values for the $\epsilon_i$:
$$\epsilon_1=2.4\pm0.2,\quad \epsilon_2=-5.5\pm0.8.
\equn{(A.5')}$$
This fit has not been improved with dispersion relations.
For dispersion relations below 0.925~\gev, it is irrelevant 
whether we use (A.5) or (A.$5'$).

\booksubsection{A.3. Parametrization of the S wave for  $I=2$ below 1~\gev}

\noindent
We fit only the low energy data, 
$s^{1/2}<1.0\,\gev$, taking the inelastic threshold at $s_0=1.05^2\gev^2$,
$$\cot\delta_0^{(2)}(s)=\dfrac{s^{1/2}}{2k}\,\dfrac{M_{\pi}^2}{s-2z_2^2}\,
\left\{B_0+B_1\dfrac{\sqrt{s}-\sqrt{s_0-s}}{\sqrt{s}+\sqrt{s_0-s}}\right\},
\quad s^{1/2}\leq1.0\;\gev.
\equn{(A.6)}$$
The central values and errors are improved with dispersion relations; we  get 
$$\eqalign{
B_0=&\,-80.8\pm1.7,\quad B_1=-77\pm5,\quad
z_2=147\pm7\;\mev\quad\hbox{[Fixed]};\cr
 a_0^{(2)}=&\,(-0.0480\pm0.0046)\times M_\pi^{-1};\quad 
b_0^{(2)}=(-0.090\pm0.006)\times M_\pi^{-3};\quad
\delta_0^{(2)}(m^2_K)=-8.5\pm0.3\degrees.\cr
}
\equn{(A.7)}$$

\booksubsection{A.4. Parametrization of the S wave for  $I=2$ above 1~\gev}

\noindent
At high energy, we may fit data from $s^{1/2}=0.91\,\gev$
 up to $s^{1/2}=1.42\,\gev$. We require junction with the 
low energy fit at 1~\gev, neglect inelasticity 
below $1.45\,\gev$, and   write
$$\eqalign{
\cot\delta_0^{(2)}(s)=&\,\dfrac{s^{1/2}}{2k}\,\dfrac{M_{\pi}^2}{s-2z_2^2}\,
\left\{B_0+B_1\dfrac{\sqrt{s}-\sqrt{s_0-s}}{\sqrt{s}+\sqrt{s_0-s}}\right\};\cr
s_0^{1/2}=1.45\;\gev;&\quad z_2=M_\pi\;\hbox{(fixed)}.\cr
}
\equn{(A.8a)}$$
After improving with dispersion relations we then find  
$$B_0=-125\pm6,\quad B_1=-119\pm14;\qquad s>(1.0\;\gev)^2.
\equn{(A.8b)}$$
The inelasticity  may be described by the 
empirical fit
$$\eta_0^{(2)}(s)=1-\epsilon(1-\hat{s}/s)^{3/2},\quad 
\epsilon=0.17\pm0.12\quad(\hat{s}^{1/2}=1.05\;\gev).
\equn{(A.8c)}$$
These formulas are expected to hold 
from 1~\gev\ up to $1.42$~\gev.

\booksubsection{A.5. The P  wave below $1$ \gev}

\noindent
 We have
$$\cot\delta_1(s)=\dfrac{s^{1/2}}{2k^3}
(M^2_\rho-s)\left\{B_0+B_1\dfrac{\sqrt{s}-\sqrt{s_0-s}}{\sqrt{s}+\sqrt{s_0-s}}
\right\};\quad s_0^{1/2}=1.05\;\gev.
\equn{(A.9a)}$$
The best result, from (4.1), is 
\smallskip
$$\eqalign{
B_0=&\,1.064\pm0.011,\quad B_1=0.170\pm0.040,\quad M_{\rho}=773.6\pm0.9;\cr 
a_1=&\,(38.7\pm1.0)\times10^{-3}M_{\pi}^{-3},\quad 
b_1=(4.55\pm0.21)\times10^{-3}M_{\pi}^{-5}. }
\equn{(A.9b)}$$
Slightly better values for $a_1$, $b_1$, improved with sum rules, may be found in Table~1.

\booksubsection{A.6. The P wave for $1\gev\leq s^{1/2}\leq 1.42\gev$}

\noindent
For the P wave between 1 \gev\ and 1.42 \gev\ we use an empirical formula, 
obtained  with a linear fit to the phase and inelasticity of 
Protopopescu et al.\ref{10} and Hyams et al.:\ref{11}
$$\eqalign{
\delta_1(s)=&\,\lambda_0+\lambda_1(\sqrt{s/\hat{s}}-1),\quad
 \eta_1(s)=1-\epsilon(\sqrt{s/\hat{s}}-1);\cr
\epsilon=&\,(0.30\pm0.15),\quad \lambda_0=2.69\pm0.01,\quad\lambda_1=1.1\pm0.2.\cr }
\equn{(A.10)}$$
Here $\quad\hat{s}=1\,{\gev}^2$.
The value of $\lambda_0$ ensures the 
agreement of the phase with the value given by (A.9) at $s=\hat{s}\equiv1\,\gev^2$.
 This fit is good; 
however, it should be remarked that 
there are other sets of experimental data for this wave disagreeing
 with the one used here for $s^{1/2}>1.15\,\gev$. 
Hence (A.10) may be biased  beyond its nominal errors.

This fit has not been improved with dispersion relations.

\booksubsection{A.7. Parametrization of the D wave for  $I=0$}

\noindent
We   fit data on $\delta_2^{(0)}$ altogether neglecting inelasticity,
 which we will then add by hand.
The data are scanty, and of poor quality.  
 We  impose in the fit the scattering length,
 as obtained from the Froissart--Gribov 
representation, and the experimental width of the $f_2$.
Moreover, we improve the central values with dispersion relations.

We write
$$\cot\delta_2^{(0)}(s)=\dfrac{s^{1/2}}{2k^5}\,(M^2_{f_2}-s)\,{{M_\pi}^2}\,
\Big\{B_0+B_1w(s)\Big\},\quad
\equn{(A.11a)}$$
and
$$w(s)=\dfrac{\sqrt{s}-\sqrt{s_0-s}}{\sqrt{s}+\sqrt{s_0-s}},\quad
 s_0^{1/2}=1450\,\mev;\quad
M_{f_2}=1275.4\,\mev\quad\hbox{[Fixed]}.$$ 
We find,
$$ B_0=23.5\pm0.7,\quad B_1=24.8\pm1.0.
\equn{(A.11b)}$$
 
We take into account the inelasticity by writing
$$\eta_2^{(0)}(s)=\cases{1,\qquad  s< 4m_K^2;\cr
1-\epsilon\,\dfrac{k_2(s)}{k_2(M^2_{f_2})},\;
\quad \epsilon=0.262\pm0.030,\quad s> 4m_K^2.\cr} 
\equn{(A.11c)}$$
Here $ k_2=\sqrt{s/4-m^2_K}.$

The fit returns the values
$$\eqalign{
a_2^{(0)}=&\,(18.4\pm3.0)\times10^{-4}\,\times M_{\pi}^{-5},\quad
b_2^{(0)}=(-8.6\pm3.4)\times10^{-4}\,\times M_{\pi}^{-7}.\cr
}
$$
Better values for $a_2^{(0)},\,b_2^{(0)}$ are given in Table~1.

\booksubsection{A.8. Parametrization of the D wave for  $I=2$}

\noindent
For isospin equal 2, there are no resonances in the D wave. 
If we want a parametrization that 
applies down to threshold, we must incorporate the  
zero of the corresponding phase shift. So 
we write
$$\cot\delta_2^{(2)}(s)=
\dfrac{s^{1/2}}{2k^5}\,\Big\{B_0+B_1 w(s)+B_2 w(s)^2\Big\}\,
\dfrac{{M_\pi}^4 s}{4({M_\pi}^2+\deltav^2)-s},\quad
s^{1/2}\leq1.05\;\gev.
\equn{(A.12a)}$$
with $\deltav$ a free parameter and
$$w(s)=\dfrac{\sqrt{s}-\sqrt{s_0-s}}{\sqrt{s}+\sqrt{s_0-s}},\quad
 s_0^{1/2}=1450\,\mev.$$ 
 Moreover, we impose  the 
value for the scattering length 
that follows from the Froissart--Gribov representation.
We assume negligible inelasticity and find, after improvement with 
dispersion relations, 
$$B_0=(2.9\pm0.2)\times10^3,\quad B_1=(7.3\pm0.8)\times10^3,
\quad B_2=(25.4\pm3.6)\times10^3,\quad
\deltav=212\pm19\,\mev.
\equn{(A.12b)}$$

The fit returns reasonable values for the scattering length and 
 effective range parameter,
$b_2^{(2)}$: 
$$a_2^{(2)}=(2.4\pm0.7)\times10^{-4}\,{M_\pi}^{-5};\quad
b_2^{(2)}=(-2.5\pm0.6)\times10^{-4}\,{M_\pi}^{-7}.
\equn{(A.13)}$$
Better values for $a_2^{(2)},\,b_2^{(2)}$ are given in Table~1.

\booksubsection{A.9. The F wave}

\noindent
For the  
 F wave below $s^{1/2}=1.42\,\gev$ we  fit the  phase 
shifts of Protopopescu et al.,\ref{10} and those
 of Hyams et al.,\ref{11b} plus the scattering length as given 
by the Froissart--Gribov representation. After honing the central values 
of the parameters with the help of
dispersion relations we find 
$$\eqalign{
\cot\delta_3(s)=&\,\dfrac{s^{1/2}}{2k^7}\,M^6_\pi\,
\left\{B_0+B_1\dfrac{\sqrt{s}-\sqrt{s_0-s}}{\sqrt{s}+\sqrt{s_0-s}}\right\};
\cr
 B_0=&\,(1.09\pm0.03)\times 10^5,\quad
B_1=(1.41\pm0.04)\times 10^5.
\cr}
\equn{(A.14)}$$
This implies $a_3=(7.0\pm0.8)\times10^{-5}\,M^{-7}_\pi$, in agreement (within errors) 
with the result of the Froissart--Gribov calculation 
(Table~4).

The contribution of the F wave to all our sum rules is very small 
(but not always negligible); the interest 
of calculating it lies in that it provides a test (by its very smallness)
 of the convergence of the partial wave expansions.

\booksubsection{A.10. The G waves}

\noindent 
For the G0 wave, we take its imaginary part to be given by the 
tail of the $f_4(2050)$ resonance, with its 
properties as given in the Particle Data Tables:
$$\eqalign{
\imag \hat{f}_4^{(0)}(s)=&\,\left(\dfrac{k(s)}{k(M^2_{f_4})}\right)^{18}
{\rm BR}\dfrac{M^2_{f_4}\gammav^2}{(s-M^2_{f_4})^2+M^2_{f_4}
\gammav^2[k(s)/k(M^2_{f_4})]^{18}};\cr
s^{1/2}\geq 1\;\gev;\qquad{\rm BR}=&\,0.17\pm0.02,\quad
 M_{f_4}=2025\pm8\;\mev,\quad \gammav=194\pm13\;\mev.\cr}
\equn{(A.15)}$$

For the wave G2, we can write, neglecting its eventual inelasticity,
$$\cot\delta_4^{(2)}(s)=\dfrac{s^{1/2}M^8_\pi}{2k^9}\,B,\quad 
B=(-9.1\pm3.3)\times 10^6;\quad
s^{1/2}\geq 1\;\gev.
\equn{(A.16)}$$

It should be noted that  the expressions  for 
the G0, G2 waves, are little more than order of magnitude estimates. 
Moreover, at low energies they certainly fail; 
below 1~\gev, expressions in terms of the scattering length 
approximation, with
$$ a_4^{(0)}=(8.0\pm0.4)\times10^{-6}\,M_{\pi}^{-9},\quad 
a_4^{(2)}=(4.5\pm0.2)\times10^{-6}\,M_{\pi}^{-9},$$
  are more appropriate.
If, in a given calculation, the contribution of either of the two G waves is 
important, it means that the calculation will have a large error. 
We have checked in a representative number of our calculations 
that the contributions of the G0, G2 waves are totally negligible.

\booksection{Appendix B: The   Regge amplitude ($s^{1/2}\geq 1.42\,\gev$)}

\noindent
In the calculations of the various sum rules and
 dispersion relations we require the imaginary 
part of the scattering amplitudes at high energy. 
For this we use 
 Regge formulas, applicable when $s\gg\lambdav^2$, with $\lambdav\simeq0.3\,\gev$ the QCD parameter, 
and 
$s\gg|t|$. 
 Specifically, we  use them above $s^{1/2}=1.42\,\gev$ and 
expect them to be valid at least for $|t|\lsim4M^2_\pi\simeq0.08\,\gev^2$. 
These formulas are obtained from $NN$, $\pi N$ scattering, 
using factorization, and from direct fits to $\pi\pi$ data.
We will not include refinements that take into account 
the logarithmic increase of total cross sections 
at ultra-high energies (that may be found in ref.~5), superfluous for applications to $\pi\pi$ 
scattering. 
We  give 
here the relevant formulas, for ease of reference, for Pomeron and $P'$ Regge poles,  
and will only discuss in some detail (and even improve slightly)
 the parameters of the rho residue.

We consider three Regge poles: Pomeron ($P$), the $P'$ (associated with the 
$f_2(1270)$ resonance) and the rho. For the first two,
$$\eqalign{ 
\imag F^{(I_t=0)}_{\pi\pi}(s,t)&\,\simeqsub_{{s\to\infty}\atop{t\,{\rm fixed}}}
P(s,t)+P'(s,t),\cr
P(s,t)=&\,\beta_P\,\alpha_P(t)\,\dfrac{1+\alpha_P(t)}{2}\,
\ee^{bt}(s/\hat{s})^{\alpha_P(t)},\cr
P'(s,t)=&\,\beta_{P'}\,
\dfrac{\alpha_{P'}(t)[1+\alpha_{P'}(t)]}{\alpha_{P'}(0)[1+\alpha_{P'}(0)]}\,
\ee^{bt}(s/\hat{s})^{\alpha_{P'}(t)},\quad
\alpha_{P'}(t)=\alpha_\rho(t).\cr
}
\equn{(B.1)}$$

For exchange of unit isospin, we write  
$$\eqalign{
\imag F^{(I_t=1)}_{\pi\pi}(s,t)&\,\simeqsub_{{s\to\infty}\atop{t\,{\rm fixed}}}
\rho(s,t);\quad\rho(s,t)=\beta_\rho(t)\,(s/\hat{s})^{\alpha_\rho(t)}.\cr
}
\equn{(B.2a)}$$
In the present paper we have taken a formula slightly different from 
that in ref.~5, viz.,
$$\beta_\rho(t)=\beta_\rho(0)\,\dfrac{1+\alpha_\rho(t)}{1+\alpha_\rho(0)}\,
\big[1+d_\rho t\big]\ee^{bt}\,
\simeqsub_{t\to0}\,
\beta_\rho(0)\big[1+c_\rho t\big],\quad c_\rho=b+d_\rho+
\dfrac{\alpha'_\rho(0)}{1+\alpha_\rho(0)}.
\equn{(B.2b)}$$ 
In ref.~5, 
where we took an expression from Rarita et al.,\ref{17} solution 1a, 
assuming the $t$ dependence to be equal for $\pi\pi$ and $\pi N$ scattering 
(see below, Eq.~(B.7)). 
Our expression now, (B.2b), 
is simpler. 
As we will show below, we can fix its parameters (for small values of $t$) 
without assuming the equality 
of  $\pi\pi$ and $\pi N$.

For exchange of isospin two,
$$\imag F^{(I_t=2)}_{\pi\pi}(s,t)\simeqsub_{{s\to\infty}\atop{t\,{\rm
fixed}}}R_2(s,t)\equiv
\beta_2\,\ee^{bt}(s/\hat{s})^{\alpha_\rho(t)+\alpha_\rho(0)-1}.
\equn{(B.3)}$$
This last amplitude corresponds to double rho exchange; the expression we 
use for it at $t\neq0$ is rather arbitrary, since little is known about it.
Fortunately, it has almost no influence in the calculations.

Following the analysis of ref.~5, the trajectories are taken as follows: 
$$\eqalign{
\alpha_P(t)\simeqsub_{t\sim0}\alpha_P(0)+\alpha'_Pt,\quad
\alpha_\rho(t)\simeqsub_{t\sim0}\alpha_\rho(0)+\alpha'_{\rho}\,t+
\tfrac{1}{2}\alpha''_{\rho}\,t^2,\cr
}
\equn{(B.4)}$$
and one has
$$\eqalign{
\alpha_\rho(0)=&\,0.52\pm0.02,\quad\alpha'_\rho=
0.90\, {\gev}^{-2},\quad \alpha''_{\rho}=-0.3\;{\gev}^{-4};\cr
\alpha_P(0)=&\,1,\quad\alpha'_P=(0.20\pm0.10)\, {\gev}^{-2}.\cr
}
\equn{(B.5)}$$
Moreover, in the present paper, we have taken the following values 
for the parameters of the residues:
$$\eqalign{
\beta_P=&\,2.54\pm0.03,\quad
\beta_{P'}=1.05\pm0.02,\quad
\beta_2=0.2\pm0.2,\quad b=(2.4\pm0.2)\;{\gev}^{-2};\cr
\beta_\rho\equiv&\,\beta_\rho(0)=1.02\pm0.11,\quad d_\rho=(2.4\pm0.9)\;{\gev}^{-2}.
\cr
}
\equn{(B.6)}$$
The parameters for $P$, $P'$ are like in ref.~5. Those for the rho trajectory are as improved below.

The Pomeron and $P'$ 
parameters are well fixed from factorization and (besides $\pi\pi$ cross sections) 
by $\pi N$, $NN$ data, down to kinetic energies $\simeq1\,\gev$;
 but some extra words will be added on
the rho residue, $\beta_\rho(t)$. 
This quantity was obtained in ref.~5 in a mixed manner: 
$\beta_\rho=\beta_\rho(0)$ 
was found fitting high energy $\pi\pi$ data, improving the 
error 
with the help of the sum rule (5.2). 
The $t$ dependence of  $\beta_\rho(t)$ 
was taken as in solution ~1a of Rarita et al.,\ref{17} 
assuming it to be identical for $\pi\pi$ and $\pi N$ scattering. 
Thus, we wrote
$$
\beta_\rho(t)=\beta_\rho(0)\,\Big[(1.5+1)\ee^{bt}-1.5\Big]
\dfrac{1+\alpha_\rho(t)}{1+\alpha_\rho(0)}\,\simeqsub_{t\to0}\,
\beta_\rho(0)\big[1+c_\rho t\big],\quad c_\rho=2.5\,b+
\dfrac{\alpha'_\rho(0)}{1+\alpha_\rho(0)}.
\equn{(B.7)}$$
This, in fact, is similar to (B.2b),~(B.6) for small values of 
$|t|$, as we will check below.

There are  other possibilities for the $t$ dependence of the 
rho residue. We could have chosen solution~1 of Rarita et al.;\ref{17}
or we could have  taken  a Veneziano-type 
formula with a diffractive factor included, writing
$$\eqalign{
\rho(s,t)=&\,\beta_\rho(t)(s/\hat{s})^{\alpha_\rho(t)};\cr
\beta_\rho(t)=&\,\beta_\rho(0)\,
\dfrac{\gammav(1-\alpha_\rho(t))\sin\pi\alpha_\rho(t)}
{\gammav(1-\alpha_\rho(0))\sin\pi\alpha_\rho(0)}\,\ee^{bt}\,
\,\simeqsub_{t\to0}\,
\beta_\rho(0)\big[1+c_\rho t\big],\quad c_\rho\simeq b+\gammae\,\alpha'_\rho(0).
\cr} 
\equn{(B.8)}$$ 
Another possibility is afforded by the $t$ dependence obtained by 
 Froggatt and Petersen\ref{18} from an analysis 
of $\pi^+\pi^-$ dispersion relations at fixed $t$.
These authors take 
$$\beta_\rho(t)=\beta_\rho(0)\dfrac{\sin \pi\alpha_\rho(t)}{\sin \pi\alpha_\rho(0)}
\big[1-
t/t_\rho\big]\ee^{b_{\rm FP}t},
\equn{(B.9)}$$ 
with $t^{-1}_\rho=-2.2\,\gev^{-2},\;
b_{\rm FP}=0.8\,\gev^{-2}$. 
This gives a slope for the residue of about half the value of Solution~1a of 
Rarita et al.

\topinsert{
\setbox0=\vbox{\hsize14truecm{\epsfxsize 11.2truecm\epsfbox{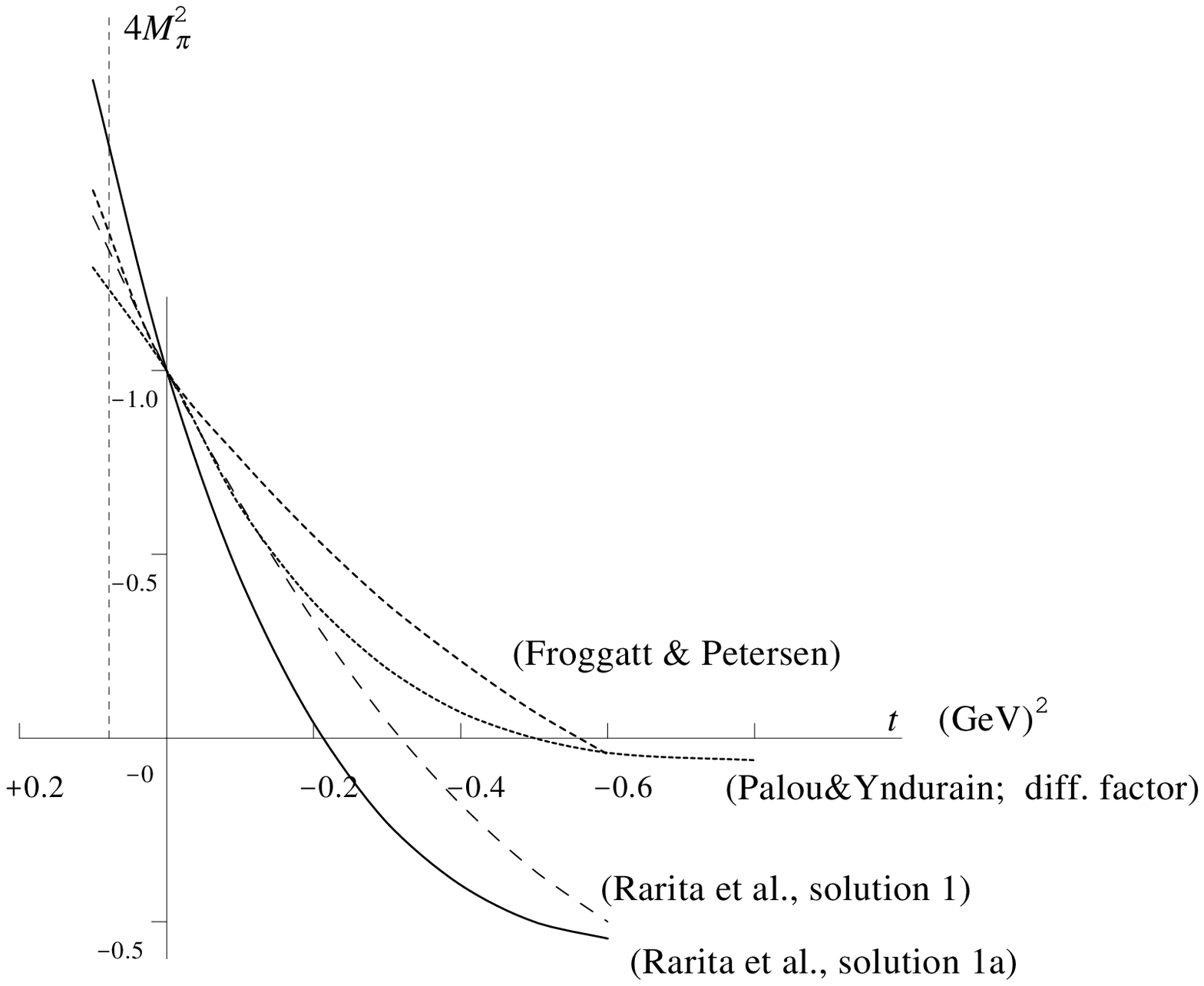}}}
\setbox6=\vbox{\hsize 14truecm\captiontype\figurasc{Figure 16. }{The $t$-dependence 
of the  
different Regge residues, $\beta(t)/\beta(0)$  (refs.~17,~18):
 { Rarita et al., solution 1}
and { solution 1a}:
the  fits from Rarita et al. (assuming equal slope for $\pi N$ and $\pi\pi$). 
{ Froggatt \& Petersen}: from fits to  $\pi\pi$ dispersion relations 
at fixed $t$.  { Palou \& Yndurain; diff. factor}:
 actually, from the Veneziano model, 
 including a diffraction factor $\ee^{bt},\,b=2.4\,\gev$, as in \equn{(B.7).}\hb 
The residue given in (B.6) would lie between ``Palou \& Yndurain; diff. factor" and
``Rarita et al.,   solution~1a" (closer to the second).}}
\centerline{\tightboxit{\box0}}
\bigskip
\centerline{\box6} 
}\endinsert 

The values of $\beta_\rho(t)$, both $\beta_\rho(0)$ and $t$ dependence,
 are given by Regge
fits  with less precision than other quantities; a few determinations 
(including the ones mentioned here) are shown in
\fig~16. The variation from one estimate to another in the range necessary for Roy 
equations, $-t\simeq0.6\,\gev^2$, is very large, and even 
at $|t|\simeq0.1\,\gev^2$ 
there are noticeable differences.

 Fortunately,  we can at least check
the correctness ({\sl within errors}) of the values of 
$\beta_\rho(t)$ given in ref.~5 at
$t=0$ and $t=4M^2_\pi\simeq0.08\,\gev^2$, and even improve them. 
In fact, apart from the sum rule (5.2), the fulfillment of the dispersion 
relation (3.5) fixes with good accuracy $\beta_\rho(0)$, and  combinations of both values 
 $\beta_\rho(0)$  and  $\beta_\rho(4M^2_\pi)$ 
enter in the evaluations, using the Froissart--Gribov representation, 
of two quantities, $a_1$, $b_1$,  that can also be determined by other methods, 
independent of the Regge assumptions. 
The ensuing equality of the two estimates each 
for $a_1$, $b_1$ given in Eqs.~(6.2) and (6.4) 
fixes  $c_\rho$,  as we will see below.
 Thus, if, for example, we had taken the
Veneziano-type
$t$ dependence of Palou and Yndur\'ain\ref{18}(i.e., as in (B.8) without the exponential
 factor $\ee^{bt}$) we would have obtained 
$b_1\sim3\times10^{-3}\;M_{\pi}^{-5}$, quite incompatible with 
determinations from the pion form factor or the sum rule (6.8). 
Even if we include a diffraction factor as in (B.8), 
 the values of $a_1$, $b_1$ following from the 
Froissart--Gribov representation 
vary by  more than $1\,\sigma$.

 One can in principle  fix 
the  parameters for $\pi\pi$ scattering from 
other processes profiting from factorization, which allows one to improve the 
poor precision of the  $\pi\pi$ data.
This is all right for the $P$, $P'$ poles because here we can use 
  data from 
$\pi N$ and $\bar{p}p+pp$, scattering which are very precise and to which 
only these two poles, plus the rho pole for $\pi N$, contribute. 
The situation is different for the coupling of the rho pole to $\pi\pi$. 
Direct fits to $\pi\pi$ data are not precise and, to use  factorization, 
one has to incorporate  (besides $\pi N$) 
proton, antiproton and
neutron  scattering data. The last presents important systematic errors because the data  
have to be extracted from scattering on deuterons. 
The large number of trajectories that contribute to these processes also make the 
analysis unreliable; for example, some existing 
fits have very likely their rho trajectory contaminated by 
pion exchange, which is not 
negligible at the lower energy range. 
Because of this, we will now give a 
new 
determination  of $\beta_\rho(0)$ and $d_\rho$  {\sl independent} 
 of Regge fits.
 
To obtain  $\beta_\rho(0)$, we rewrite the dispersion relation (3.7) as
$$\eqalign{
\real F^{(I_t=1)}(s,0)-&\,\dfrac{2s-4M^2_\pi}{\pi}
\pepe\int_{4M^2_\pi}^{1.42^2{\gev}^2}\dd s'\,
\dfrac{\imag F^{(I_t=1)}(s',0)}{(s'-s)(s'+s-4M^2_\pi)}\cr
=&\,
\dfrac{2s-4M^2_\pi}{\pi}\int_{1.42^2{\gev}^2}^{\infty}\dd s'\,
\dfrac{\imag F^{(I_t=1)}(s',0)}{(s'-s)(s'+s-4M^2_\pi)}.\cr
}
\equn{(B.10)}$$
We  evaluate the left hand side for values of $s^{1/2}$ from threshold to 
925~\mev, at intervals of 25~\mev, using low energy experimental
data, via the  parametrizations\fnote{Of course, {\sl before} 
improving with dispersion relations, or 
the reasoning would be somewhat circular. 
Alternatively, we could
 fit simultaneously the parameters for phase shifts and inelasticities and 
the Regge parameters of the rho, 
using all three dispersion relations; the method, 
quite complicated, would not give results substantially different from what we get making 
independent fits.}  given in \sect~2. 
In this region,
$s^{1/2}\leq0.925\,\gev$, the uncertainties due to the middle energy S0 and P waves
are not very important.  If we write in the right hand side of (B.10) 
$$\imag F^{(I_t=1)}(s',0)=\beta_\rho(0)(s'/\hat{s})^{\alpha_\rho(0)},
\quad
s\geq1.42^2\;{\gev}^2, 
$$
and fix $\alpha_\rho(0)=0.52$ from factorization or deep 
inelastic scattering, we may fit $\beta_\rho(0)$ and find a very 
precise number,
$$\beta_\rho(0)=1.06\pm0.10.
\equn{(B.11)}$$
This is   quite compatible with the value $\beta_\rho(0)=0.94\pm0.14$ that we 
obtained from 
fits to $\pi\pi$ cross sections in ref.~5. 
The result (B.11) is robust; if we 
{\sl double} the 
errors of the parameters for the P and S0 
waves between 1 and 1.42~\gev, where (as discussed in the main text) one has larger
uncertainties, and repeat the fit, we get the same value.

As  for $c_\rho$, we can use the values of $a_1$, $b_1$ obtained  
from the analysis of the pion form factor,\ref{9} together with the Froissart--Gribov 
representation of the same quantities.
We rewrite Eq.~(6.1) as 
$$\eqalign{
a_1-&\,\dfrac{\sqrt{\pi}\,\gammav(2)}{4M_{\pi}\gammav(1+3/2)}
\int_{8M_{\pi}^2}^{1.42^2{\gev}^2}\dd s\,\dfrac{\imag F^{(I_t=1)}(s,4M_{\pi}^2)}{s^{3}}
\cr
=&\,
\dfrac{\sqrt{\pi}\,\gammav(2)}{8M_{\pi}\gammav(1+3/2)}
\int^{\infty}_{1.42^2{\gev}^2}\dd s\,\dfrac{\imag
F^{(I_t=1)}(s,4M_{\pi}^2)}{s^{3}}.\cr }
\equn{(B.12a)}$$
For the $b_l$, $l=$odd,
$$\eqalign{b_l-&\,\dfrac{\sqrt{\pi}\,\gammav(l+1)}{4M_{\pi}\gammav(l+3/2)}
\int_{4M_{\pi}^2}^{1.42^2\,{\gev}^2}\dd s\,
\left\{\dfrac{4\imag F^{(I_t=1)'}(s,4M_{\pi}^2)}{(s-4M_{\pi}^2)s^{l+1}}-
\dfrac{(l+1)\imag F(s,4M_{\pi}^2)}{s^{l+2}}\right\}\cr
=&\,\dfrac{\sqrt{\pi}\,\gammav(l+1)}{4M_{\pi}\gammav(l+3/2)}
\int_{1.42^2\,{\gev}^2}^\infty\dd s\,
\left\{\dfrac{4\imag F^{(I_t=1)'}(s,4M_{\pi}^2)}{(s-4M_{\pi}^2)s^{l+1}}-
\dfrac{(l+1)\imag F(s,4M_{\pi}^2)}{s^{l+2}}\right\};\cr
\imag F^{(I_t=1)'}&(s,t)\equiv\partial \imag
F^{(I_t=1)}(s,t)/\partial{\cos\theta}.\cr
}
\equn{(B.12b)}$$
The left hand sides in (B.12) are evaluated using also the experimental phase shifts and
inelasticities  as fitted in \sect~2 of the present paper 
and we take, in the right hand side of (B.12a),
$$\imag F^{(I_t=1)}(s,4M_{\pi}^2)=
\beta_\rho(4M_{\pi}^2)(s/\hat{s})^{\alpha_\rho(4M_{\pi}^2)},\quad 
s\geq1.42^2\;{\gev}^2,
$$
$\beta_\rho(t)\simeq\beta_\rho(0)[1+c_\rho t].$
An analogous formula may be written  for 
the right hand side of (B.12b). The sum rule for $b_1$ gives the more precise 
constraint for $c_\rho$ for two reasons: first, 
it depends directly on the derivative of $\imag F^{(I_t=1)}(s,t)$ with 
respect to $t$ ($\cos\theta$), hence 
$c_\rho$ appears explicitly; and, secondly, 
the low energy contributions to the left hand side in 
(B.12b) cancel to a large extent, so 
$b_1$ is given almost exclusively by the high energy, 
Regge formulas; cf.~ref.~6.

We can fit, with these sum rules for $a_1$, $b_1$, plus the dispersion 
relation for $I_t=1$, the parameters $\beta_\rho(0)$, $c_\rho$ 
simultaneously. As was to be expected, the 
value of $\beta_\rho(0)$ is unchanged and, for $c_\rho$, we find  the values
$$c_\rho=\cases{
(6.0\pm2.8)\;{\gev}^{-2}\quad {\rm from}\; a_1\cr
(4.7\pm0.8)\;{\gev}^{-2}\quad {\rm from}\; b_1.\cr
}
\equn{(B.13)}$$
This is compatible with what one has from solution~1a of Rarita et al.,\ref{17}
 $c_\rho=(6.6\pm0.5)\,{\gev}^{-2}$ in the sense that they produce  similar values for 
$\beta_\rho(4M^2_\pi)$: the decrease in $c_\rho$  is compensated (within errors) 
by the increase in $\beta_\rho(0)$, Eq.~(B.1), compared with what we used in 
ref.~5. 
  Our results for $c_\rho$ thus  justify the use of the formulas of  Solution 1a of
Rarita et al.,\ref{17} for {\sl small} values of $|t|\lsim0.1\,\gev$, 
as was done in ref.~6, and the assumption that 
(in that range) they are also valid
 for $\pi\pi$ scattering, within errors.
 
We can  also include in the fit the information from
$\pi\pi$ scattering data (as discussed in ref.~5), since it is compatible with 
what we found now,
 and the sum rule (5.2), evaluated with the parameters
 found in \sect~2 here for the low energy piece. In
this case we find what we consider the best results, 
$$\beta_\rho(0)=1.02\pm0.11,\quad c_\rho=(5.4\pm0.9)\;{\gev}^{-2}.
\equn{(B.14a)}$$
This implies 
$$d_\rho=(2.4\pm0.9)\;{\gev}^{-2}.
\equn{(B.14b)}$$
Eqs.~(B.14) provide us with the 
central values  given in (B.6) for $d_\rho$, $\beta_\rho(0)$.

\booksection{Appendix C: On experimental phase shifts in the range $1.4\,\gev\simeq
s^{1/2}\simeq 2\,\gev$}

\noindent 
As is well known known, as soon as the 
center of mass kinetic energy  in a hadronic reaction, $E_{\rm kin,c.m.}$, 
increases beyond 1 \gev, inelastic processes become important and, 
for $E_{\rm kin,c.m.}\gsim1.3\,\gev$, they  dominate elastic ones. 
This is easily understandable in the QCD, ladder version of the
 Regge picture, as discussed in
ref.~6; and indeed, it is verified experimentally in
 the hadronic processes  
$\pi N$, $K N$, and $NN$ where, for   $E_{\rm kin,c.m.}>1.3\,\gev$, the 
elastic cross section is smaller than the inelastic one and, for 
  $E_{\rm kin,c.m.}>1.5\,\gev$, the elastic cross section is less than or about one half 
of the inelastic one. 
There is no reason to imagine that $\pi\pi$ scattering would follow a different pattern. 
In this case (large inelasticity), and again as mentioned in refs.~6,~7,
 it can be proved theoretically\ref{19} that 
there is not a unique solution to the phase shift analysis: 
some sets of $\eta$s and $\delta$s may fit the data; but so would others. 

In spite of this, the Cern-Munich experiments\fnote{Hyams et al. and 
 Grayer et al., refs.~11a,b.} have produced a
set of phase shifts and  inelasticities which go up to $s^{1/2}\simeq2\,\gev$. 
These have been used as input in certain 
theoretical analyses, notably in those of refs.~1,~2. 
Unfortunately, such phase shifts and inelasticities are very likely to diverge 
more and more from reality 
as   
$s^{1/2}$ becomes larger and larger than $1.3\,\gev\simeq 2M_\pi+1\,\gev$. 
This is suggested, besides the theoretical reasons
just  mentioned, by the fact that the Cern-Munich phase shifts and inelasticities
 contradict a number of
physical properties  they should fulfill: we will here mention a few.

 As we have remarked above, one would 
expect dominant inelastic cross sections above $s^{1/2}\sim1.5\,\gev$; but the
 Cern--Munich elastic cross sections are larger or comparable to the
inelastic ones up to 
$s^{1/2}=2\,\gev$. 
What is worse, the $\pi^+\pi^-$ {\sl inelastic} cross section of  Cern--Munich, alone of all 
hadronic cross sections, {\sl decreases} 
as the kinetic energy  increases between 1 \gev\ and 1.7 \gev, 
 as shown for example in Fig.~7 in the paper 
of Hyams et al.\ref{11a} 
As happens for $\gamma N$, $\pi N$, $KN$ or $NN$ scattering, 
and indeed as is also seen in other 
experimental analyses for $\pi\pi$ scattering,\ref{4} 
we  expect a levelling off of the total cross section for 
$E_{\rm kin,c.m.}>1.3\,\gev$: but 
 the Cern--Munich total cross section for $\pi^+\pi^-$ 
scattering
decreases roughly like $1/s$ up to 2 \gev. 
In fact, the Cern--Munich {\sl elastic} cross section agrees well with that 
found in other experiments (such as those in ref.~4), 
but their {\sl total} cross section  is smaller by a factor $\sim2$ 
than those in refs.~4, at the higher part of the energy range: the Cern--Munich 
 results  certainly
 misrepresent the {\sl total} cross section.
From all this it follows
that the Cern--Munich phases and inelasticities
are not reliable in that energy region. 

Secondly, the combination of $\delta$ and $\eta$ for both P and S0 
waves at  energy  $\gsim1.8\,\gev$ is incompatible with QCD results for the 
electromagnetic  form factor of the pion, and also with 
Regge behavior, which requires all phases to go to a multiple of $\pi$ at high energy, 
and the real part of the partial wave amplitudes for isospin 0,~1, to be {\sl positive}.
The phase $\delta_1$ is also incompatible with QCD results for the 
electromagnetic  form factor of the pion. 
In fact, the phase of this form factor behaves like  
$$\delta_{\rm Form\;factor}(s)\simeq\pi\left(1+\dfrac{1}{\log s/\hat{t}}\right),\quad s\gg \lambdav^2;\quad
\hat{t}\sim\lambdav^2; 
$$
$\lambdav$ is the QCD parameter (see, e.g.,~ref.~20). 
One may take this to hold for $s\gsim3\,\gev^2$ ($s^{1/2}>1.6\,\gev$).
If one had negligible inelasticity for the P wave 
somewhere in the region $1.6\,\gev\lsim s^{1/2}\lsim 2\,\gev$, as the 
Cern--Munich data seem to imply, the form factor and 
partial wave would have the same phase at such energies, and thus 
the same behaviour  should hold 
for $\delta_1(s)$, $\delta_{\rm Form\;factor}(s)$. 
But the phase which the Cern--Munich experiment gives  clearly contradict this behaviour 
around $1.8$ \gev; there the Cern--Munich phase 
$\delta_1$ stays consistently {\sl below}
$\pi$, while, as just shown, it should be {\sl above}.

Thirdly, for the process $\pi^\pm\pi^\pm\to\pi^\pm\pi^\pm$, 
we would expect large inelasticity as soon as the production  
$\pi^\pm\pi^\pm\to\rho^\pm\rho^\pm$ is energetically possible, 
and therefore large inelasticity for the S2, D2 waves for $s^{1/2}\gg 2M_\rho$. 
This, in particular, occurs in any theoretical model. 
It is therefore not possible to 
extrapolate these phases above 
$\sim1.45$ \gev as being elastic. 
In fact, the extrapolation used in refs.~1,~2 for the D2 wave 
is clearly incorrect above $\sim1.4\,\gev$ as 
the corresponding $|\delta_2^{(2)}(s)|$ grows linearly with $s$ 
while, from Regge theory (and also from low energy 
fits, see \fig~7), one expects it to go to zero.
Thus, besides the general problem for the cross sections 
we have individual problems 
for each of the S0, S2, P and D2 phases.

 Finally we would like to mention that both the  Regge picture and the values of
 the experimental cross sections for all hadronic processes 
indicate that the number of waves that contribute 
effectively to the imaginary part (say) of the scattering amplitudes 
grows with the kinetic energy as $E_{\rm kin}/\lambdav$, for
 $E_{\rm kin}$ upwards of 1 \gev.
We thus expect 2 to 3 waves (for fixed isospin) 
at  $E_{\rm kin}\sim1\,\gev$, and almost double this, 4 or 5 
waves, at $E_{\rm kin}\sim1.7\,\gev$. 
In fact, for $\pi\pi$ scattering  at this energy,  the contribution to the 
total cross section of the F wave is as large as that of the P wave, 
the D0 wave is as large as the S0 wave and the contribution of the 
 D2 wave is actually larger than that of 
the S2 wave. The partial wave series with only two waves per isospin channel 
is not convergent, and the approximations, like those of refs.~1,~2,~11,
 that neglect higher waves  at 
such energies have yet another reason for being irrealistic.

Because of all this, it follows that use of the Cern--Munich phases and inelasticities  
 must  lead to a rather distorted imaginary part of the
 $\pi\pi$  scattering amplitude above $s^{1/2}\sim1.4\,\gev$.
It is thus  not surprising that authors like those 
in refs.~1,~2,~15, who 
fix their Regge parameters by balancing them above $2\,\gev$ with Cern--Munich 
phase shift analyses below $2\,\gev$, get  incorrect
 values for the first.\fnote{We would
 like to emphasize
that what has been said should not be taken as implying  criticism of the 
 Cern--Munich experiment which, for $s^{1/2}\lsim1.4\,\gev$, 
 produced what are probably the best determinations of phase shifts and inelasticities. 
Above $1.4$ \gev, they did what they could: it is for 
theorists to realize that this was not enough to produce acceptable phase shifts 
and inelasticities at these higher energies.}

\vfill\eject
\brochuresection{Acknowledgments}
\noindent
We are grateful to CICYT, Spain, and to INTAS, for partial financial support.

We are also grateful to  G.~Colangelo and H.~Leutwyler, whose questions about the 
rho residue, $\beta_\rho(t)$, prompted us to give the alternate derivation and 
improvement of its values in Appendix~B.

\brochuresection{References}
\item{1 }{Ananthanarayan, B., et al., {\sl Phys. Rep.}, {\bf 353}, 207,  (2001).}
\item{2 }{Colangelo, G., Gasser, J.,  and Leutwyler, H.,
 {\sl Nucl. Phys.} {\bf B603},  125, (2001).}
\item{3 }{Roy, S. M., {\sl Phys. Letters}, {\bf 36B}, 353,  (1971).}
\item{4 }{Biswas, N. N., et al., {\sl Phys. Rev. Letters}, 
{\bf 18}, 273 (1967) [$\pi^-\pi^-$, $\pi^+\pi^-$ and $\pi^0\pi^-$];
 Cohen, D. et al., {\sl Phys. Rev.}
{\bf D7}, 661  (1973) [$\pi^-\pi^-$];
 Robertson, W. J.,
Walker, W. D., and Davis, J. L., {\sl Phys. Rev.} {\bf D7}, 2554  (1973)  [$\pi^+\pi^-$]; 
Hoogland, W., et al.  {\sl Nucl. Phys.}, {\bf B126}, 109 (1977) [$\pi^-\pi^-$];
Hanlon, J., et al,  {\sl Phys. Rev. Letters}, 
{\bf 37}, 967 (1976) [$\pi^+\pi^-$]; Abramowicz, H., et al. {\sl Nucl. Phys.}, 
{\bf B166}, 62 (1980) [$\pi^+\pi^-$]. These  references cover the region
between  1.35 and 16 \gev, and agree within errors in the regions where they overlap 
(with the exception of $\pi^-\pi^-$ below 2.3 \gev, discussion in ref.~5).}
\item{5 }{Pel\'aez, J. R., and Yndur\'ain, F. J.,  {\sl Phys. Rev.} {\bf D69}, 114001 (2004).}
\item{6 }{Pel\'aez, J. R., and Yndur\'ain, F. J., {\sl Phys. Rev.} {\bf D68}, 074005 (2003).}
\item{7 }{Yndur\'ain, F. J., FTUAM 03-14; revised August, 2004
(hep-ph/0310206).}
\item{8 }{Descotes, S., Fuchs, N. H.,  Girlanda, L., and   Stern, J., {Eur. Phys. J. C}, 
{\bf 24}, 469, (2002); 
Kami\'nski, R., Le\'sniak, L., and Loiseau, B. {\sl Phys. Letters} {\bf B551},
241 (2003).}
\item{9 }{de Troc\'oniz, J. F., and Yndur\'ain, F. J., {\sl Phys. Rev.},  {\bf D65}, 093001,
 (2002),  
and hep-ph/0402285. 
When quoting numbers, we will quote from this last paper.}
\item{10 }{Protopopescu, S. D., et al., {\sl Phys Rev.} {\bf D7}, 1279, (1973).}
\item{11 }{The Cern--Munich experiments ($\pi\pi$ scattering). (a): 
Hyams, B., et al., {\sl Nucl. Phys.} {\bf B64}, 134, (1973); 
Grayer, G., et al.,  {\sl Nucl. Phys.}  {\bf B75}, 189, (1974). 
See also the analysis of the 
same experimental data in
Estabrooks, P., and Martin, A. D., {\sl Nucl. Physics}, {\bf B79}, 301,  (1974). 
(b): Hyams, B., et al., {\sl Nucl. Phys.} {\bf B100}, 205, (1975). 
(c): Kami\'nski, R., Lesniak, L, and Rybicki, K., {\sl Z. Phys.} {\bf C74}, 79 (1997) and 
{\sl Eur. Phys. J. direct} {\bf C4}, 4 (2002). $\pi\pi\to\bar{K}K$ scattering: (d) 
Wetzel,~W., et al., {\sl Nucl. Phys.} {\bf B115}, 208 (1976);
Polychromatos,~V.~A., et al.,  {\sl Phys. Rev.} {\bf D19}, 1317 (1979);
Cohen, ~D. et al., {\sl Phys. Rev.} {\bf D22}, 2595 (1980); 
Etkin,~E. et al.,  {\sl Phys. Rev.} {\bf D25}, 1786 (1982).}
\item{12 }{Losty, M.~J., et al.  {\sl Nucl. Phys.}, {\bf B69}, 185 (1974); 
Hoogland, W., et al. 
{\sl Nucl. Phys.}, {\bf B126}, 109 (1977); 
Durusoy,~N.~B., et al., {\sl Phys. Lett.} {\bf B45}, 517 (19730.}
\item{13 }{Rosselet, L., et al. {\sl Phys. Rev.} {\bf D15}, 574  (1977); 
Pislak, S.,  et al.  {\sl
Phys. Rev. Lett.}, {\bf 87}, 221801 (2001).}
\item{14 }{For the more recent determination, 
see Aloisio, A., et al., {\sl Phys. Letters}, {\bf B538}, 21  (2002); an older one is
Pascual, P., and Yndur\'ain, F. J., {\sl Nucl. Phys.} {\bf B83}, 362, (1974).}
\item{15 }{Pennington, M. R., {\sl Ann. Phys.} (N.Y.), {\bf 92}, 164, (1975).}
\item{16 }{Palou, F. P., S\'anchez-G\'omez, J. L., and Yndur\'ain, F. J., 
{\sl Z. Phys.}, {\bf A274}, 161, (1975).}
\item{17 }{Rarita, W., et al., {\sl Phys. Rev.} {\bf 165}, 1615, (1968).}
\item{18 }{Froggatt,~C.~D.,
and Petersen,~J.~L., {\sl Nucl. Phys.} {\bf B129}, 89 (1977); 
Palou, F. P., and Yndur\'ain,~F.~J.,  {\sl Nuovo Cimento} {\bf 19A}, 245 (1974).}
\item{19 }{Atkinson, D., et al., {\sl Nucl. Phys.} {\bf B7}, 375  (1968) and  
 {\sl Nucl. Phys.}, {\bf B23}, 397 (1970).}
\item{20 }{Yndur\'ain, F. J., {\sl Phys. 
Letters}, {\bf B578}, 99 (2004) and (E), {\sl Phys. Letters}, {\bf B586}, 439 (2004).}

\bye